\documentclass[showpacs,twocolumn,aps,prx,longbibliography,superscriptaddress,notitlepage]{revtex4-2}
\pdfoutput=1
\usepackage{booktabs}
\usepackage{graphicx}
\usepackage{tikz}
\usepackage{commath,amsmath,amssymb}
\usepackage{mathrsfs,dsfont}
\usepackage[font=small,labelfont=bf]{caption}
\usepackage{braket}
\usepackage[algo2e,ruled]{algorithm2e}
\usepackage{subcaption}
\usepackage{pifont}
\usepackage{fontawesome}

\newcommand{\xmark}{\ding{55}}
\SetKwComment{Comment}{/* }{ */}
\newcommand\lseq[1]{\stackrel{{\rm LS}}{=}}

\newcommand{\Tr}{{\rm Tr}}
\newcommand{\rk}{{\rm rank}}
\DeclareRobustCommand{\rchi}{{\mathpalette\irchi\relax}}
\newcommand{\irchi}[2]{\raisebox{\depth}{$#1\chi$}}
\newcommand*{\vertbar}{\rule[-0.2 ex]{0.5pt}{3.0 ex}}
\DeclareMathOperator*{\argmin}{arg\,min}
\newcommand{\Dt}{\Delta t}
\newcommand{\Ko}{\mathcal{K}_{\Delta t}}

\newcommand{\bbC}{\mathbb{C}}
\newcommand{\bG}{\mathbf{G}}
\newcommand{\bX}{\mathbf{X}}
\newcommand{\bK}{\mathbf{K}}

\newcommand{\bW}{\mathbf{W}}
\newcommand{\bZ}{\mathbf{Z}}
\newcommand{\vV}{\mathbb{V}}
\newcommand{\GOi}{\Gamma_{O_i}}
\newcommand{\bDel}{\boldsymbol{\Delta}}

\usepackage[colorlinks=true,urlcolor=blue]{hyperref}
\usepackage[capitalise]{cleveref}

\begin{document}
\title{Efficient Measurement-Driven Eigenenergy Estimation with Classical Shadows}

\author{Yizhi Shen}
\email{yizhis@lbl.gov}
\thanks{Equal contributions}
\affiliation{Applied Mathematics and Computational Research Division, Lawrence Berkeley National Laboratory, Berkeley, CA 94720, USA}

\author{Alex Buzali}
\thanks{Equal contributions}
\affiliation{Department of Physics, Harvard University, 17 Oxford Street, Cambridge, MA 02138, USA}

\author{Hong-Ye Hu}
\affiliation{Department of Physics, Harvard University, 17 Oxford Street, Cambridge, MA 02138, USA}

\author{Katherine Klymko}
\affiliation{NERSC, Lawrence Berkeley National Laboratory, Berkeley, California 94720, USA}

\author{Daan Camps}
\affiliation{NERSC, Lawrence Berkeley National Laboratory, Berkeley, California 94720, USA}

\author{Susanne F. Yelin}
\affiliation{Department of Physics, Harvard University, 17 Oxford Street, Cambridge, MA 02138, USA}

\author{Roel~{Van~Beeumen}}
\affiliation{Applied Mathematics and Computational Research Division, Lawrence Berkeley National Laboratory, Berkeley, CA 94720, USA}

\begin{abstract}
Quantum algorithms exploiting real-time evolution under a target Hamiltonian have demonstrated remarkable efficiency in extracting key spectral information. However, the broader potential of these methods, particularly beyond ground state calculations, is underexplored. In this work, we introduce the framework of multi-observable dynamic mode decomposition (MODMD), which combines the observable dynamic mode decomposition, a measurement-driven eigensolver tailored for near-term implementation, with classical shadow tomography. MODMD leverages random scrambling in the classical shadow technique to construct, with exponentially reduced resource requirements, a signal subspace that encodes rich spectral information. Notably, we replace typical Hadamard-test circuits with a protocol designed to predict low-rank observables, thus marking a new application of classical shadow tomography for predicting many low-rank observables. We establish theoretical guarantees on the spectral approximation from MODMD, taking into account distinct sources of error. In the ideal case, we prove that the spectral error scales as $\exp(- \Delta E  t_{\rm max})$, where $\Delta E$ is the Hamiltonian spectral gap and $t_{\rm max}$ is the maximal simulation time. This analysis provides a rigorous justification of the rapid convergence observed across simulations. To demonstrate the utility of our framework, we consider its application to fundamental tasks, such as determining the low-lying, \textit{i.e.}, ground or excited, energies of representative many-body systems. Our work paves the path for efficient designs of measurement-driven algorithms on near-term and early fault-tolerant quantum devices.
\end{abstract}

\maketitle

\section{Introduction}
\label{sec:introduction}

Quantum algorithms based on real-time evolution~\cite{parrish2019quantum, stair2020multireference, klymko2022real, shen2022real, cortes2022quantum, StairCortes2023,ding2022even,ding2023simultaneous,shen2023estimating,Wang2023quantumalgorithm,ding2024quantummultipleeigenvaluegaussian,Zhang2024measurement,Motta_2024,maskara2023programmablesimulationsmoleculesmaterials} have gained increasing popularity due to the unitarity of real-time dynamics and their native implementation on quantum hardware. In particular, real-time eigensolvers have demonstrated impressive efficacy in extracting the ground state information of physical many-body systems. This success raises a natural question: can the algorithms perform highly accurate excited state calculations?

The real-time approaches were initially conceptualized from a subspace diagonalization perspective, building on classical techniques that recover eigenvalues through time correlation functions -- an idea explored for decades~\cite{Wall95}. Yet their quantum computational counterparts have only seen active development in the past few years: several related methods have since emerged under different names in the literature, including quantum filter diagonalization (QFD)~\cite{parrish2019quantum,stair2020multireference}, variational quantum phase estimation (VQPE)~\cite{klymko2022real,shen2022real}, and quantum Krylov (QK)~\cite{cortes2022quantum} approaches. While these subspace methods, in theory, have the potential to extract spectral information beyond the ground state, they typically involve solving ill-conditioned generalized eigenvalue problems \cite{epperly2022theory} and can hence suffer from any perturbative error, such as prevalent hardware noise. Alternatively, distinct recent approaches \cite{ding2022even,Li2023,ding2023simultaneous,Wang2023quantumalgorithm,ding2024quantummultipleeigenvaluegaussian} draw inspirations from a signal processing perspective, where real-time evolution is primarily employed to generate time series that resonate with the target eigenmodes. The signal processing approaches tend to be resilient to perturbative noise due to their robust regularization. On the other hand, they often rely on state preparation such that the initial state undergoing the Hamiltonian dynamics must overlap dominantly with the eigenstates of interest. Without this key assumption, accurately resolving multiple eigenenergies may demand substantial quantum or classical resources.

On the other hand, quantum phase estimation (QPE) and its variants~\cite{Kitaev2002ClassicalAQ,Nielsen_Chuang_2010,Griffiths1996,Higgins2007,Berry2009,Zwierz2010,Giovannetti2011} stand as the canonical approach to access eigenphases and prepare eigenstates coherently on a quantum computer. The original QPE protocol uses a register of ancilla qubits, a sequence of controlled time evolution acting on system qubits, and the inverse quantum Fourier transform. In particular, QPE provides both energy and eigenstate information in each run, achieving Heisenberg-limited scaling with a total runtime inversely proportional to the desired accuracy. The inverse quantum Fourier transform can be performed semiclassically through single-qubit measurements and controlled rotations~\cite{Griffiths1996} whereas the ancilla overhead can be reduced to a single qubit~\cite{Higgins2007,Berry2009}, making QPE an especially attractive candidate for fault-tolerant devices.

 Despite its optimal asymptotic scaling, QPE imposes a requirement on the circuit depth or maximal runtime: unless the initial state is an exact eigenstate, the depth must increase inversely with the energy accuracy, a lower bound intrinsic to the protocol. All known QPE variants share this maximal runtime scaling, with the preconstant unchanged even as the initial state approaches an exact eigenstate~\cite{ding2022even}. This requirement can constitute a serious bottleneck for implementing QPE on near-term and early fault-tolerant hardware, where circuit depth continues to be a primary constraint.

In this work, we introduce a quantum-classical hybrid framework that combines time evolution and scrambling on quantum hardware with classical data post-processing to access the ground and low-lying excited state energies of many-body systems. Our framework prepares a single, simple initial state on the quantum computer with overall resource demands comparable to the leading ground state subspace algorithms. Specifically,
it features a shallower circuit depth~\cite{ding2024espritalgorithmhighnoise} than QPE when the initial state is close to an eigenstate. Though exceeding the Heisenberg limit, the empirical total runtime remains modest for retrieving the lowest eigenenergies at physically relevant accuracies. 
Furthermore, the classical post-processing in our approach is \textit{ansatz-free} and therefore circumvents the challenges of complex optimization landscapes commonly encountered in parametric methods, particularly as the number of target eigenenergies increases~\cite{McClean2018,Arrasmith_2022,Qi2023}.


The main technical tools that we leverage are classical shadow tomography~\cite{Preskill_shadow,hu_shadow2022,toolbox,PhysRevLett.133.020602,2022arXiv221206084V,PhysRevLett.127.110504,2023arXiv231100695L} and observable dynamic mode decomposition (ODMD)~\cite{shen2023estimating}. We develop a simple, single-ancilla shadow protocol to collect real-time signals associated with many observables, and utilize ODMD to unravel these signals into single-energy modes. Since oscillating time signals with distinct power spectra are linearly independent in the space of all sinusoidal functions, they form the basis of some \emph{signal subspace}, which can be systematically expanded to accommodate the desired frequencies. Our approach, termed the multi-observable dynamic mode decomposition (MODMD), can hence be viewed as a unifying framework that enjoys the strength of both subspace and signal processing algorithms.

\begin{figure*}[t!]
    \centering
    \includegraphics[width=0.99\linewidth]{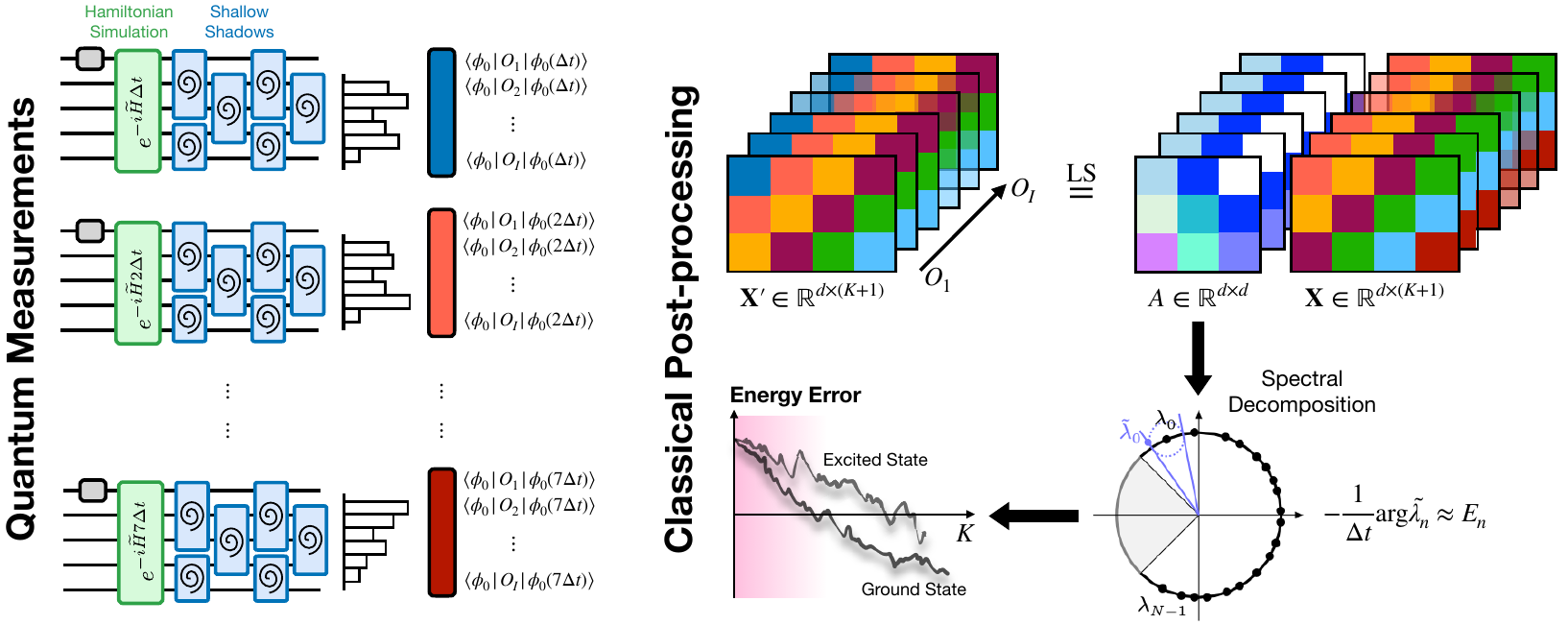}
    \caption{ MODMD for eigenenergy estimation. MODMD collects the expectations, $\Tr \left[ \rho (k\Dt) \Gamma_{O} \right] = \Re \bra{\phi_0} O e^{-i H k\Dt}\ket{\phi_0}$, with respect to a simple reference state $\ket{\phi_0}$. This data can be measured efficiently on a quantum processor through Hamiltonian simulations combined with shadow tomographic techniques, where the shadow circuits can be shallow with a depth logarithmic in the number of qubits. MODMD then constructs a pair of block Hankel matrices $(\mathbf{X},\mathbf{X}^{\prime})$, and computes the DMD system matrix $A$ that adopts a block companion structure. The eigenvalues $\tilde{\lambda}_n$ of $A$ converge to the true eigenphases $\lambda_n$ as the size of $\mathbf{X}$ and $\mathbf{X}^{\prime}$ increases. The low-lying energies $E_n$ are estimated as angles of $\tilde{\lambda}_{n}$, $\Tilde{E}_n = -\frac{1}{\Dt}\text{arg}(\tilde{\lambda}_n)$.}
\label{fig:overview}
\end{figure*}

Compared to alternative hybrid eigensolvers, our real-time framework has the following notable advantages: it (1) achieves near-exponential convergence of the eigenenergy estimates surpassing the conventional Fourier limit, (2) provides extensive knowledge of a many-body system including eigenstate properties and dynamical responses, and (3) significantly saves quantum resources in terms of the evolution time while showing stability against perturbative noise. These features make our algorithm promising for implementations on emerging quantum platforms, such as near-term analog quantum simulators and early fault-tolerant quantum computers. 

The manuscript is organized as follows. In \cref{sec:methods}, we first overview key concepts underlying recent advances in real-time eigensolvers. We then establish in \cref{sec:MODMD framework} the building blocks of the real-time MODMD framework and detail our core algorithm for eigenenergy and eigenstate estimation. Theoretical guarantees on its convergence and preliminary error analysis are presented within \cref{sec:theory}. Finally, we numerically demonstrate our algorithm in \cref{sec:applications} by focusing on many-body examples from condensed matter physics and quantum chemistry.


\section{Preliminaries}
\label{sec:methods}

In this work, we develop an approach that gives highly accurate eigenenergy estimates beyond the ground state. Specifically, we propose a real-time framework combining the observable dynamic mode decomposition (ODMD) and classical shadow tomography, where we fully leverage the synergy between these two algorithmic components.
We will show that we can design a simple shadow protocol to predict the real-time expectation value $\braket{\phi_0|Oe^{-iHt}|\phi_0}$ for the problem Hamiltonian $H$ and a reference state $\ket{\phi_0}$. In particular, classical shadows enable the simultaneous prediction of expectations for many observables $O$ of our choice. These expectations generate a multivariate time series, whose characteristic frequencies can be efficiently extracted by ODMD (see \cref{fig:overview} for a summary).

From here on, we will use $\mathcal{H} \subset \mathbb{C}^{N}$ to denote the physical Hilbert space of many-body quantum  states and $\mathbb{L}(\mathcal{H})$ to denote the Liouville space, \textit{i.e.}, the space of linear operators acting on $\mathcal{H}$ (we work with finite-dimensional and thus bounded linear operators for simplicity). Additional essential notations are defined self-consistently in \cref{tab:notations} and throughout the main text.

\subsection{Real-time quantum eigensolvers}
\label{subsec:existing approaches}

In this subsection, we first highlight the capabilities of real-time evolution in determining the eigenenergies of a target  Hamiltonian $H \in \mathbb{C}^{N \times N}$. We consider the spectral decomposition, $H = \sum_{n=0}^{N-1} E_n \ket{\psi_n} \bra{\psi_n}$, of the Hamiltonian with ordered energies $E_0 \leq E_1 \leq \ldots \leq E_{N-1}$. The real-time approaches commonly require evaluation of the expectation value~\cite{parrish2019quantum,stair2020multireference,klymko2022real,cortes2022quantum,ding2022even,Li2023,ding2023simultaneous,shen2023estimating,Wang2023quantumalgorithm,ding2024quantummultipleeigenvaluegaussian,Zhang2024measurement,Motta_2024},
\begin{align}
    \braket{\phi_0|e^{-i H t}|\phi_0} = \sum_{n=0}^{N-1} \abs{\alpha_n}^2 e^{- i E_n t},
    \label{eq:overlap}
\end{align}
where $\ket{\phi_0} \in \mathcal{H}$ is a reference state, $\alpha_n = \braket{\psi_n|\phi_0}$ is the overlap between the reference state and eigenstate $\ket{\psi_n}$, and $t = k\Dt$ is typically an integer multiple of some time step $\Dt$. The expectation can be efficiently sampled from repeated quantum measurements through the Hadamard test~\cite{Cleve:1997dh} or mirror fidelity estimation~\cite{mirror2019,cortes2022quantum}.

\begin{table}[t!]
\caption{\label{tab:notations} Glossary of key notations.}
\begin{ruledtabular}
\begin{tabular}{ll}
\textbf{Simulation protocol} \\
Hilbert space & $\mathcal{H},~{\rm dim}\mathcal{H} = N = 2^{L}$ \\
Liouville space & $\mathbb{L}(\mathcal{H}),~{\rm dim}\mathbb{L}(\mathcal{H}) = N^2$ \\
Hamiltonian in Pauli basis & $H = \sum_{\nu =1}^{M} \kappa_{\nu} P_{\nu}$ \\
Hermitian observables & $O_i = \sum_{\nu =1}^{M_i} \kappa_{i,\nu} P_{i,\nu}$ \\
Hamiltonian eigenstates & $\{\ket{\psi_n}\}_{n=0}^{N-1}$ \\
Pure quantum state & $\rho_{\phi} = \ket{\phi}\bra{\phi}$ \\
Multi-observable signal & $\{ \braket{\phi_0|O_i e^{-iHt}|\phi_0} \}_{i=1}^{I}$ \\
Classical shadow dataset & $\{\hat{\sigma}_{q}=U_q^{\dagger}|b_q\rangle\langle b_q|U_q \}_{q=1}^{Q}$ \\

\textbf{Algorithmic controls} \\
Time step & $\Dt$ \\
Number of distinct observables & $I$ \\
Shape of data slices per observable & $(d,K),~\frac{K}{d} = {\rm constant}$ \\
Systematic error (\emph{e.g.}, shot noise) & $\epsilon_{1}(\Dt,I, \displaystyle \max_{1 \leq i \leq I} \lVert \vec{\kappa}_i \rVert_1 ,Q)$ \\
Algorithmic error from eigensolver & $\epsilon(\ket{\phi_0}, \Dt, I, d, \epsilon_1)$ \\

\end{tabular}
\end{ruledtabular}
\end{table}

There are several approaches for obtaining energy estimates from data of the type \cref{eq:overlap}. Quantum subspace diagonalization~\cite{parrish2019quantum,stair2020multireference,klymko2022real,cortes2022quantum,Motta_2024,Zhang2024measurement} approximates the extremal eigenenergies by forming a projected eigenvalue problem. In particular, the quantum subspace can be built through successive applications of unitary time evolutions, where increasing powers $k = \frac{t}{\Dt}$ of the time-evolution operator $e^{-iH\Dt}$ generate additional matrix elements for the projected problem. Earlier work on quantum filter diagonalization (QFD)~\cite{parrish2019quantum,stair2020multireference} introduced multi-reference variants of this approach, improving coverage of spectrum at the expense of a quadratic overhead in the number of initial states. More recent approaches, such as variational quantum phase estimation (VQPE)~\cite{klymko2022real,shen2022real} and quantum Krylov (QK) algorithms~\cite{cortes2022quantum,StairCortes2023}, have focused on single-reference protocols that remain effective even with relatively coarse time step -- going beyond the perturbative regime where Taylor expansions of $e^{-iHt}$ suffice. While VQPE and QK differ in circuit implementations, they share the same underlying philosophy of treating real-time evolution as the generator of subspace. With a single reference state $\ket{\phi_0}$, the methods often struggle to find excited states, though convergence can be potentially facilitated by preparation of multiple references with the aforementioned quadratic cost. Moreover, the real-time states $\{e^{-iH k\Dt}\ket{\phi_0}\}_{k \geq 0}$ lose orthogonality as a subspace basis, which can thereby lead to ill-conditioning and susceptibility to noise.


Alternatively, signal processing phase estimation algorithms~\cite{ding2022even,ding2023simultaneous,Li2023,ding2024quantummultipleeigenvaluegaussian,Wang2023quantumalgorithm} capitalize on Fourier or harmonic analysis to resolve the eigenfrequencies or eigenenergies of interest. These approaches mitigate the impacts of noise by minimizing customized objective functions, which can sharpen into robust optima as the number of time steps increases. Despite a resilience to noise, the optimization landscape depends critically on the choice of the reference state $\ket{\phi_0}$. For example, consider the time correlations of \cref{eq:overlap} to be a spectral density over the unit circle, where the squared overlaps $\lvert \alpha_n \rvert^2$ define the normalized spectral weights. Techniques such as peak finding on the spectral density may not yield unique solution when the reference state has roughly uniform eigenstate overlaps. To ensure accurate energy estimation, more sophisticated reference state preparations and post-processing designs need to be accounted for. Among such approaches, quantum complex exponential least squares (QCELS)~\cite{ding2022even,ding2023simultaneous} algorithms fit measured time correlations to a linear combination of complex exponentials upon minimizing the residual error, thereby extracting the eigenphases with high resolution. Building on this, quantum multiple eigenvalue Gaussian filtered search (QMEGS)~\cite{ding2024quantummultipleeigenvaluegaussian} checks a grid of candidate spectral centers in the frequency domain -- interpreted as Fourier transform of the time-domain signal -- to localize peaks resonant with the eigenenergies. Both methods are contingent on a reference state that overlaps dominantly with the target eigenstates. While QCELS achieves high accuracy at a cost of classical post-processing that scales exponentially in the number of estimated eigenenergies, QMEGS offers a scalable route but suffers from spectral broadening due to decoherence and other noise sources.

To access both ground and excited state properties, we aim for a simple, accurate, and robust real-time protocol. To this end, we will introduce a \textit{quantum signal subspace} approach that utilizes signals of the form $\braket{\phi_0|O e^{-iHt} |\phi_0}$ for general operators $O$. Importantly, we seek to measure many such operator expectations simultaneously
with a cost logarithmic in the number of observables. This can be precisely achieved via classical shadows, which we explore in the following subsection.

\begin{table*}
\begin{ruledtabular}
\begin{tabular}{lllll}
Real-time methods & Measurement cost & Algorithmic convergence & Target overlaps & System properties \\
\colrule
\addlinespace[2pt]

\textbf{Single-observable signal} \\
\addlinespace[2pt]

Subspace~\cite{parrish2019quantum,stair2020multireference,klymko2022real,cortes2022quantum,Motta_2024,Zhang2024measurement} & $\displaystyle \mathcal{O}\big( \epsilon_{1}^{-2} \big)$  & $\tilde{\mathcal{O}}(\epsilon_{1}^{-2}) $ in low-$\epsilon_1$ limit & non-vanishing & \textcolor{green}{\checkmark} eigenstates \\
 & & unstable in high-$\epsilon_1$ limit & & \\
 
\textit{Example algorithms} & & & & \\
\multicolumn{5}{l}{\quad $\bullet$

QFD~\cite{parrish2019quantum,stair2020multireference}: multi-reference; uniform timegrid (with small $\Dt$) }  \\

\multicolumn{5}{l}{\quad 
$\bullet$
VQPE~\cite{klymko2022real,shen2022real}: single-reference (primary focus); flexible timegrid (with large $\Dt$); noise-thresholded convergence } \\

\multicolumn{5}{l}{\quad 
$\bullet$
QK~\cite{cortes2022quantum}: single-reference (primary focus); uniform timegrid; signal sampling with multi-fidelity estimation protocol } \\


 
\addlinespace[8 pt]

Signal processing~\cite{ding2022even,ding2023simultaneous,ding2024quantummultipleeigenvaluegaussian,Wang2023quantumalgorithm} & $\displaystyle \mathcal{O}\big( \epsilon_{1}^{-2} \big)$ 
& $\displaystyle \tilde{\mathcal{O}} \big( \epsilon^{-1} \big)$ & dominant & \textcolor{red}{\xmark} eigenstates \\

\addlinespace[1pt]

\textit{Example algorithms} & & & & \\
\multicolumn{5}{l}{\quad $\bullet$
(MM)QCELS~\cite{ding2022even,ding2023simultaneous}: single-reference; adaptive or randomly sampled timegrid } \\

\multicolumn{5}{l}{\quad
$\bullet$
QMEGS~\cite{ding2024quantummultipleeigenvaluegaussian}: single-reference; randomly sampled timegrid; improved classical post-processing
} \\

\addlinespace[4pt]

\colrule
\addlinespace[2pt]

\textbf{Multi-observable signal} \\
\addlinespace[5pt] 

Shadow spectroscopy~\cite{chan2023algorithmic} & $\displaystyle \mathcal{O} \big(\log(I) 3^{{\rm loc}[O]} \epsilon_{1}^{-2}  \big)$ & uncertain & non-vanishing & \textcolor{green}{\checkmark} dynamics \\
& & possibly ambiguous energy gaps  & & \\
\addlinespace[5pt]

Signal subspace (this work) & $\displaystyle \mathcal{O} \big(  \log(I) \epsilon_{1}^{-2} \big)$ & $\tilde{\mathcal{O}} \big( \epsilon^{-4/3} \big)$ in some cases & non-vanishing & \textcolor{green}{\checkmark} eigenstates \\
 & & stable with respect to $\epsilon_1$~\cite{Moitra2015,Li2020_Esprit,ding2024espritalgorithmhighnoise} & & \hspace{0.2525 cm} dynamics \\
\addlinespace[1pt]

\end{tabular}
\end{ruledtabular}
\caption{Landscape of real-time hybrid eigensolver leveraging quantum measurements of $\braket{\phi_0|Oe^{-iHt}|\phi_0}$ and classical post-processing schemes. We consider estimating a target set of Hamiltonian eigenergies, $\{E_{n}: n \in \mathcal{N}_{\rm eig} \subset [N] \}$, where $\lvert \mathcal{N}_{\rm eig} \rvert = N_{\rm eig}$ for $N_{\rm eig} \ll N$. On quantum hardware, real-time evolution of a reference state $\ket{\phi_0}$ is performed and different expectation values are measured with an uncertainty of $\epsilon_1$. The classical post-processing then extracts from these measurements our eigenenergy estimates with an algorithmic error of $\epsilon$. Here $\tilde{O}(\cdot)$ denotes asymptotic upper bound with polylogarithmic factors neglected.}
\label{tab:compare_methods}
\end{table*}

\subsection{Efficient measurement with classical shadows}
\label{subsubsec:scrambling with classical shadows}
Classical shadow tomography~\cite{Preskill_shadow,hu_shadow2022,HYH_shadow,Akhtar2023scalableflexible,hu2024demonstrationrobustefficientquantum} embodies a powerful suite for efficiently measuring expectations of many observables simultaneously:
\begin{align}
    \Tr\left[ \rho_{\phi} \Gamma_i \right] = \braket{\phi|\Gamma_i|\phi},\qquad \{\Gamma_i\}_{i=1}^{I},
    \label{eq:shadow_trace}
\end{align}
where $\Gamma_i \in \mathbb{L}(\mathcal{H})$ has an efficient representation on a classical computer. Here $\rho_{\phi} = \ket{\phi} \bra{\phi} \in \mathbb{L}(\mathcal{H})$ is a pure state, though a similar trace evaluation applies to mixed states. Classical shadow tomography consists of two key steps: (1) random quench evolution using $U\in \mathcal{U}$ from a unitary ensemble $\mathcal{U}$, and (2) computational basis measurement. Upon each measurement, the quantum state collapses to a bitstring $|b_q \rangle$. After repetitive simulations, one obtains a classical \emph{shadow dataset}, $\{U_q^{\dagger}|b_q \rangle\langle b_q|U_q\}_{q=1}^{Q}$, which can be viewed as a classical sketch of the quantum state. It is known~\cite{Preskill_shadow} that $Q=\mathcal{O}(\log(I) \max_i\norm{\Gamma_i}_{\text{sh}} \epsilon_1^{-2})$ shadows can predict all the $I$ expectation values given by \cref{eq:shadow_trace} to uncertainty $\epsilon_1$ with high probability. Here, the shadow norm, $\norm{\Gamma}_{\text{sh}}$, depends on both the unitary ensemble $\mathcal{U}$ and the operator $\Gamma$. For example, when the ensemble $\mathcal{U}$ is the $L$-qubit random Clifford unitaries, \emph{i.e.}, $\mathcal{U}=\text{Cl}(2^L)$, and the operator $\Gamma=\Gamma^{\dagger}$ is Hermitian, the shadow norm is $\norm{\Gamma}_{\text{sh}}=3\Tr[\Gamma^2]$. This is especially powerful if $\Gamma$ is \emph{low-rank}, meaning that its operator rank stays independent of the system size. Because $\Tr[\Gamma^2] \leq \text{rank}(\Gamma) \lVert \Gamma \rVert_2^2$, one can use classical shadows to predict exponentially many low-rank expectations simultaneously even for large systems. In \cref{sec:MODMD framework}, we demonstrate how to transfer the measurement of many expectations $\{ \Re \langle \phi_0|O_i e^{-iHt}|\phi_0\rangle\}_{i}$ to the task of predicting low-rank Hermitian operators.

Overall, the framework presented within this work is distinctive in various crucial aspects. First, our algorithm can directly estimate the individual eigenenergies and the associated energy gaps. Conversely, resolving the energy levels from estimated excitation gaps is nearly infeasible. Second, instead of the standard usage of classical shadows in predicting many local Pauli observables, we introduce a novel application of classical shadows by replacing the Hadamard-test-like circuits with those that predict many \textit{low-rank} observables, expanding their primary utility beyond entanglement witnesses in quantum information science. Last, our post-processing scheme yields robust and accurate eigenenergy estimates, achieving an exponential error reduction in the low-noise regime. This surpasses an algebraic error decay in the conventional Fourier limit. Although recent work~\cite{chan2023algorithmic} has begun exploring the use of shadow techniques in spectroscopic calculations of energy gaps, our approach distinguishes itself in the sense we have discussed above.

\cref{tab:compare_methods} presents a comprehensive comparison of state-of-the-art real-time methods for extracting valuable spectral information, highlighting the efficiency and accuracy of our eigensolver. In \cref{sec:MODMD framework} we will outline the basic construction of our measurement-driven approach, which deploys classical shadows to evaluate the real-time expectations.

\section{MODMD framework}
\label{sec:MODMD framework}

Here we present a novel perspective to the problem of eigenenergy estimation, pushing the limits of convergence and robustness via a quantum signal subspace approach. Our quantum signal space is composed of time correlation functions of the form,
\begin{align}
    \braket{\phi_0|O e^{-i H t}|\phi_0}, \quad O = O^{\dagger} \in \mathbb{C}^{N \times N},
    \label{eq:overlap_with_O}
\end{align}
which captures the system quantum dynamics. Since the expectation value oscillates over time, it can be uniquely expressed in the natural basis $e^{-iE_n t}$. Representing real-time data in this eigenfrequency basis establishes a clear notion of linear independence in the space of signals. By evaluating the expectation values in \cref{eq:overlap_with_O} for multiple independent operators $\{O_1, O_2, \ldots, O_I \}$, we thus manage to construct a signal subspace from which we can extract spectral information significantly better. We will refer to the vector of expectation values,
\begin{align}
    \vec{s}(t) = \begin{bmatrix}
    \braket{\phi_0|O_1 e^{-iHt}|\phi_0} \\
    \braket{\phi_0|O_2 e^{-iHt}|\phi_0} \\
    \vdots \\
    \braket{\phi_0|O_I e^{-iHt}|\phi_0}
    \end{bmatrix},\label{eq:multi-observable}
\end{align}
as a \emph{multi-observable} signal associated with $\{ O_i \}_{i=1}^{I}$. The central ingredient of our real-time framework is hence the efficient collection of a multi-observable signal $\vec{s}(t)$, whose dimensionality $I$ reflects independence or richness of the underlying spectral information. In particular, the state overlap $\braket{\phi_0|e^{-iHt} |\phi_0}$ in \cref{eq:overlap} can be viewed as a simple `one-dimensional' signal which oscillates over time.

To efficiently measure the multi-observable signal $\vec{s}(t)$, we leverage classical shadow tomography, specifically the shallow shadows recently demonstrated on hardware~\cite{hu2024demonstrationrobustefficientquantum}. We will show that one can evaluate all the $I$ expectation values in \cref{eq:multi-observable} up to a small error simultaneously, with each expressed as the expectation value of a low-rank observable. Before elaborating on the favorable exponential cost reduction, we first introduce our basic measurement-driven framework.

\subsection{Basic construction}

\subsubsection{Primer: ODMD}

In recent work~\cite{shen2023estimating} we examined the observable dynamic mode decomposition (ODMD) as a powerful extension of the classical DMD formalism. ODMD exploits quantum resources to efficiently measure the expectations of time evolution operators, $s(t) = \Re \braket{ \phi_0 | e^{-iHt}|\phi_0}$, rather than directly tracking the dynamics of the full quantum state $e^{-iHt} \ket{\phi_0} \in \mathcal{H}$. The extremal (both maximum and minimum) eigenphases $e^{-iE_{n} t}$ and, thereby, the eigenenergies $E_n$ can be inferred via the least-squares (LS) solution to the following system of linear homogeneous equations,
\begin{align}
    \underbrace{\begin{bmatrix}
        s_1 & s_2 & \cdots & s_{K+1} \\
        s_2 & s_3 & \cdots & s_{K+2}  \\
        \vdots & \vdots & \ddots & \vdots \\
        s_{d} & s_{d+1} & \cdots & s_{K+d} \\
    \end{bmatrix}}_{\mathbf{X}' \in \mathbb{R}^{d \times (K+1)}}  \lseq{}  & A \underbrace{\begin{bmatrix}
        s_0 & s_1 & \cdots & s_K \\
        s_1 & s_2 & \cdots & s_{K+1}  \\
        \vdots & \vdots & \ddots & \vdots \\
        s_{d-1} & s_{d} & \cdots & s_{K+d-1} \\
    \end{bmatrix}}_{\mathbf{X} \in \mathbb{R}^{d \times (K+1)}},
    \label{eq:ODMD_sys}
\end{align}
where $s_k = s(k\Delta t)$ is sampled at a regular time step $\Delta t$, and $(\mathbf{X}, \mathbf{X}')$ are time-shifted data matrices containing the overlap evaluations. We note that the Hankel structure of $\mathbf{X}$ and $\mathbf{X}'$ immediately implies a compact companion structure of the \textit{system matrix}, $A \in \mathbb{R}^{d \times d}$. The extremal-phase eigenvalues $\Tilde{\lambda} \in \mathbb{C}$ of the system matrix $A$ converge rapidly to the extremal eigenphases $\lambda_n = e^{-iE_n \Delta t}$ of the target Hamiltonian as we increase the dimensions $d$ and $K$ of the data matrices.

The ODMD algorithm excels in extracting the ground state information. As we move towards the interior of the spectrum, its convergence often slows down progressively or even stagnates. To achieve compact and accurate excited state estimation -- a task more intricate than ground state estimation -- we take advantage of classical shadows to generate extensive real-time data with minimal quantum resource overhead.

\subsubsection{MODMD with classical shadows}

Now we generalize the ODMD approach from the time evolution operator $e^{-iHt}$ to an operator pool of arbitrary size. For the signal collection process, we invoke the idea of classical shadow tomography originally introduced~\cite{Preskill_shadow} to extract arbitrary many-body properties from random projective measurements. Similar to the Hadamard test, we introduce a single ancilla qubit to control elementary operations on the system registers, demonstrating for the first time that this setup can in fact efficiently predict the multi-observable signal $\vec{s}(t)$ in \cref{eq:multi-observable}.

With the ancilla initialized in  $\ket{0_{a}} \in \mathcal{H}_{a}$, we first create a superposition state $\ket{\Phi(t)} \in \mathcal{H}_{a} \otimes \mathcal{H}$ given by,
\begin{align}
    \ket{\Phi(t)} = \frac{\ket{0,\phi_\perp} + \ket{1,\phi_0(t)}}{\sqrt{2}},
    \label{eq:Phi_t}
\end{align}
where $\ket{1,\phi_0(t)} = \ket{1_{a}}\otimes \ket{\phi_0(t)}$ denotes the product of the ancillary state and the time-evolved reference state $\ket{\phi_0(t)} = e^{-iHt} \ket{\phi_0}$, and $\ket{\phi_\perp} \in \mathcal{H}$ is any residual state. 


For the composite state $\rho(t) := \ket{\Phi(t)} \bra{\Phi(t)} \in \mathbb{L}(\mathcal{H}_{a} \otimes \mathcal{H})$ and the operator $\Gamma = \ket{1,\phi_0} \bra{0,\phi_\perp} + \ket{0,\phi_\perp} \bra{1,\phi_0} \in \mathbb{L}(\mathcal{H}_{a} \otimes \mathcal{H})$, we recognize the pivotal relation that
\begin{align}
    \Re \braket{\phi_0|\phi_0
    (t)} = \Tr\left[ \rho(t) \Gamma \right],
    \label{eq:tr_s}
\end{align}
which establishes a fundamental connection between the expectation in \cref{eq:overlap} and the trace in \cref{eq:shadow_trace}. Notably, $\Gamma = \Gamma^{\dagger}$ is Hermitian, and its classical simulability entirely depends on that of $\ket{\phi_0}$ and $\ket{\phi_{\perp}}$. Following \cref{eq:tr_s}, it is straightforward to show that the density operator $\rho(t)$ in fact encodes the time correlation function $\braket{\phi_0|O|\phi_0(t)}$ of any Hermitian operator $O = O^{\dagger} \in \mathbb{L}(\mathcal{H})$, since
\begin{align}
    \Re \braket{\phi_0|O|\phi_0(t)} = \Tr \left[ \rho(t) \Gamma_O \right]
     = \Re \sum_{n=0}^{N-1} c_n e^{-i E_n t}, \label{eq:o_shadow}
\end{align}
where $c_n = \sum_{m=0}^{N-1} \alpha_m^{\ast} \alpha_n \braket{ m | O | n}$ and
\begin{align}
    \Gamma_O = ({\rm Id}_{a} \otimes O) \ket{1, \phi_0} \bra{0,\phi_\perp} + {\rm h.c.},
\end{align}
with the identity ${\rm Id}_{a}$ acting on the ancilla Hilbert space. The classical simulability of $\Gamma_{O}$ now depends on that of $\ket{\phi_0}$, $\ket{\phi_{\perp}}$, and $O$. In particular, $\Gamma_{O}$ has an efficient classical representation if the two states and the operator are sparse in the computational and Pauli basis, respectively. For instance, if $\ket{\phi_0}$ and $\ket{\phi_\perp}$ are simple computational basis states, as in the context of quantum chemistry, $\Gamma_O$ can be classically represented for any Pauli string $O$.

If one measures each term from $\{\Re\langle \phi_0|O_i|\phi_0(t)\rangle\}_{i=1}^{I}$ individually with a Hadamard-test circuit, it will take a total of $\mathcal{O}(I \epsilon_1^{-2})$ measurements to achieve a uniform error of $\epsilon_1$. With classical shadow tomography, we significantly reduce this measurement sample complexity. As shown in \cref{eq:o_shadow}, we can rewrite each expectation value as the trace of a Hermitian, \emph{low-rank} observable $\Gamma_O$ over a state $\rho(t)\in \mathbb{L}(\mathcal{H}_{a} \otimes \mathcal{H})$. Moreover, we emphasize that the state preparation circuit of $\rho(t)$ only involves a single ancillary qubit, matching the overhead of a Hadamard-test circuit. As shown in \cref{fig:overview}, after preparing $\rho(t)$ with the quantum time evolution circuit, we apply a global Clifford shadow protocol on the quantum state and collect shadows $\{\hat{\sigma}_q=U_q^{\dagger}|b_q\rangle\langle b_q |U_q \}_{q=1}^{Q}$. On a classical computer, we employ the stored shadows to construct an empirical estimator for the trace,
\begin{align}
    \langle\Gamma_O\rangle_{\text{est}}=\dfrac{1}{Q}\sum_{q=1}^{Q}\Tr [ \mathcal{M}^{-1}(\hat{\sigma}_q) \Gamma_O ],
\end{align}
where $\mathcal{M}\in \mathbb{L}(\mathcal{H}_{a} \otimes \mathcal{H})$ is linear. For classical shadows obtained from global random circuits or shallow random circuits, $\mathcal{M}$ can be calculated efficiently \cite{Preskill_shadow,PhysRevLett.133.020602,Akhtar2023scalableflexible,PhysRevLett.130.230403}. It can be proven~\cite{Preskill_shadow} that when $Q=\mathcal{O}(\log(I)\ \epsilon_1^{-2})$, we can estimate all the $I$ expectation values to $\epsilon_1$ error with high probability using the global Clifford classical shadow, \emph{i.e.}, $|\langle \Gamma_{O_i}\rangle_{\text{est}}-\Tr(\rho(t)\Gamma_{O_i})|<\epsilon_1$ for all $\Gamma_{O_i}$. In addition, one can also read out the imaginary part, $\Im \langle \phi_0|O|\phi_0(t)\rangle$, by setting $\Gamma_O = i~({\rm Id}_{a} \otimes O) \ket{1,\phi_0} \bra{0,\phi_\perp} + {\rm h.c.}$.

To realize global Clifford random unitaries, one needs linear-depth quantum circuits. This could pose a serious challenge on the near-term quantum platforms due to the severe two-qubit gate errors. Fortunately, recent finding shows that shallow quantum circuits with depth $\mathcal{O}(\log L)$ can form global random unitaries on $L$ qubits~\cite{SHH24}. Moreover, experiments have demonstrated that classical shadows using log-depth quantum circuits can be made robust against various quantum errors through new theoretical advancements \cite{hu2024demonstrationrobustefficientquantum}. These emerging developments enable one to fully leverage the robust shallow shadow technique in experiments to achieve a low measurement overhead. In particular, since $\Gamma_O$ has the low-rank property, it falls within the class of observables that shallow shadows have been shown to predict reliably, even in the presence of realistic quantum noise. Without loss of generality, we will focus on the global Clifford classical shadow tomography for MODMD, assuming access to high-fidelity scrambling unitaries.

\subsubsection{Symmetry-simplified time evolution circuit}

For many-body systems, we can conveniently initialize the states $\ket{\phi_0}$ and $\ket{\phi_\perp}$ in \cref{eq:Phi_t} according to particular symmetry sectors of the Hamiltonian $H$. The presence of symmetry may \textit{entirely eliminate} the need for controlled time evolution, enabling the use of standard uncontrolled evolution for significantly more native and, thus, compact circuit implementation on hardware.

For the case of molecular Hamiltonians, we note that it is natural to prepare $\ket{\phi_0}$ and $\ket{\phi_\perp}$ with distinct fermionic occupations which, after second quantization, correspond to two computational basis states with distinct Hamming weights (under Jordan-Wigner encoding). In particular, if we initialize $\ket{\phi_{\perp}}$ to be the vacuum state $\ket{00\cdots0}$ with zero particle occupation, it will stay invariant under the time evolution: $e^{-iHt} \ket{\phi_{\perp}} = \ket{\phi_{\perp}}$. This allows us to implement uncontrolled time evolutions on the system registries alone in actual experiments, provided that the observables $O_i$ do not induce any transitions between the particle sector of interest and the vacuum state.

For general local Hamiltonians, including those arising in familiar spin systems, the requirement of controlled time evolution can also be relaxed by exploiting canonical anti-commutation relations between Pauli operators and grouped Hamiltonian terms. To illustrate, we restrict ourselves to a Hamiltonian $H$ consisting of a single group of mutually commuting Pauli terms; the same argument generalizes to many Hamiltonian terms under Trotterized dynamics. Notably, conjugating the time evolution with an anti-commuting Pauli operator $\tilde{P}$ (a chiral symmetry) effectively flips the sign of evolution time, \emph{i.e.},
\begin{align}
    \begin{bmatrix}
        {\rm Id} & 0 \\
        0 & e^{-iHt}
    \end{bmatrix} = e^{-iH\frac{t}{2}}  \begin{bmatrix}
        \tilde{P} & 0 \\
        0 & {\rm Id}
    \end{bmatrix} e^{-iH\frac{t}{2}}  \begin{bmatrix}
        \tilde{P} & 0 \\
        0 & {\rm Id}
    \end{bmatrix},
\end{align}
which constructs the controlled unitaries without directly controlling the Hamiltonian. This symmetry-based trick helps simplify circuit implementation; see~\cite{Dong2022_qetu} for details as well as concrete examples.
\bigskip

\subsection{Main algorithm}
\label{subsec:main algorithm}

With our efficient shadow implementation, we can estimate the density operator $\rho(t) $ and the associated expectation values for an arbitrary pool of operators $\{ O_i \}_{i=1}^{I}$. By doing so, we facilitate an exponential expansion of the signal subspace relative to the measurement cost for the shadow reconstruction, as estimating $I$ observables only requires $Q = \mathcal{O}(\log(I))$ samples $\{\hat{\sigma}_q\}_{q=1}^{Q}$. The collection of expectation values takes the form, 
\begin{align}
    \Vec{s}(t) = \begin{bmatrix}
    \braket{\phi_0|O_1|\phi_0(t)} \\
    \braket{\phi_0|O_2|\phi_0(t)} \\
    \vdots \\
    \braket{\phi_0|O_I|\phi_0(t)}
    \end{bmatrix} = \sum_{n=0}^{N-1} \Vec{c}_n e^{-i E_n t},
    \label{eq:multi_signals}
\end{align}
for $\Vec{c}_n \in \mathbb{C}^{I}$ with $c_{n,i} = \sum_{m=0}^{N-1} \alpha_m^{\ast} \alpha_n \braket{\psi_m | O_i | \psi_n}$. 

Given our access to real-time expectations $\vec{s}_{k} = \vec{s}(k\Dt)$ sampled at time step $\Dt$, we formulate a LS problem as the multi-dimensional variant of \cref{eq:ODMD_sys},
\begin{align}
 \underbrace{\begin{bmatrix}
        \vec{s}_1 & \vec{s}_2 & \cdots & \vec{s}_{K+1} \\
        \vec{s}_2 & \vec{s}_3 & \cdots & \vec{s}_{K+2}  \\
        \vdots & \vdots & \ddots & \vdots \\
        \vec{s}_{d} & \vec{s}_{d+1} & \cdots & \vec{s}_{K+d} \\
    \end{bmatrix}}_{\mathbf{X}' \in \mathbb{R}^{d I \times (K+1)}}  \lseq{}  & A \underbrace{\begin{bmatrix}
        \vec{s}_0 & \vec{s}_1 & \cdots & \vec{s}_K \\
        \vec{s}_1 & \vec{s}_2 & \cdots & \vec{s}_{K+1}  \\
        \vdots & \vdots & \ddots & \vdots \\
        \vec{s}_{d-1} & \vec{s}_{d} & \cdots & \vec{s}_{K+d-1} \\
    \end{bmatrix}}_{\mathbf{X} \in \mathbb{R}^{d I \times (K+1)}},
    \label{eq:MODMD_sys}
\end{align}
where $(\mathbf{X}, \mathbf{X}')$ are time-shifted data matrices containing the observable evaluations through shadows. The system matrix $A \in \mathbb{C}^{d I \times d I }$ is now block companion with $dI^2$ free parameters, \textit{i.e.}, its last $I$ rows. In the special case where the observable pool contains a single operator $O_1$, namely the identity ${\rm Id}$ on the system Hilbert space, our approach reduces to the original ODMD setting.

The system matrix $A$ captures the evolution of multi-dimensional expectations propagated by unitary dynamics $e^{-iH\Dt}$. Hence, the eigenenergies and corresponding eigenstates of the Hamiltonian $H$, as the generator of the dynamics, can be simply estimated via the eigenvalue decomposition of $A$,
\begin{align}
    A = \sum_{\ell=0}^{dI-1} \Tilde{\lambda}_{\ell} \mathbf{\Psi}_{{\rm R}, \ell} \mathbf{\Psi}_{{\rm L}, \ell},
\end{align}
where $\Tilde{\lambda}_{\ell}$ give the DMD eigenvalues while $(\mathbf{\Psi}_{{\rm R},\ell}, \mathbf{\Psi}_{{\rm L},\ell})$ denote the corresponding right and left eigenvectors. We note that $\Tilde{\lambda}_{\ell} \in \mathbb{C}$ since $A$, being block companion, is not Hermitian.

\begin{algorithm2e}[t!]
\caption{MODMD eigenenergy estimation \hspace{-2em}~\label{alg:MODMD}}

\vspace{5pt}
\KwIn{Reference state $\ket{\phi_0}$, operator pool $\{ O_i \}_{i=1}^{I}$, time step $\Delta t$, noise threshold $\tilde{\delta}$.}
\KwOut{Estimated energies $\tilde{E}_{n}$ and eigenstates $\ket{\tilde{\psi}_n}$.}
\vspace{5pt}

\nl $k \leftarrow 0$ \\
\While{${\rm not~converged}$}{
  \nl $\vec{s}_k \leftarrow e^{-i H k\Dt} \ket{\phi_0}$ \Comment*{\footnotesize{classical shadows}}
  \nl $\mathbf{X}, \mathbf{X}' \leftarrow \mathrm{Hankel}\left(\Vec{s}_0, \Vec{s}_1, \ldots, \Vec{s}_k \right)$ \Comment*{\footnotesize{\cref{eq:MODMD_sys}}}
   \nl $\mathbf{X}_{\tilde{\delta}} \leftarrow \mathbf{X}$ \Comment*{\footnotesize{least-squares regularization}}
  \nl $A \leftarrow \mathbf{X}' \mathbf{X}_{\tilde{\delta}}^{+}$ \Comment*{\footnotesize{update system matrix}}
  \nl $\tilde{E}_n  \leftarrow  - \Im \log(\tilde{\lambda}_{n}) / \Dt $ \Comment*{\footnotesize{energies}}

   \nl $\ket{\Tilde{\psi}_n}  \leftarrow  \sum_{(a,i)} z_{(a,i),n} O_i e^{-iH a \Dt} \ket{\phi_0} $ \Comment*{\footnotesize{states}}
  \nl $k \leftarrow k+1$
}
\end{algorithm2e}

We can read off our eigenpair approximations from the ordering of the phases ${\rm arg}(\Tilde{\lambda}_{\ell})$. Without loss of generality, we assume that the phases are arranged in a descending order, ${\rm arg}(\Tilde{\lambda}_0) \geq {\rm arg}(\Tilde{\lambda}_1) \geq \ldots$, such that the eigenvalue with the maximal phase, $\Tilde{\lambda}_0 := \lvert \Tilde{\lambda}_0 \rvert e^{-i \Tilde{E}_0 \Dt}  \approx e^{-i E_0 \Dt}$, encodes the DMD approximation $\Tilde{E}_0$ to the exact ground state energy $E_0$. Likewise, the eigenvalue $\Tilde{\lambda}_1$ provides an approximation to the first excited state energy $E_1$. The eigenstates, on the other hand, can be accessed from the DMD eigenvectors $\boldsymbol{\Psi}_{{\rm L},n} = [z_{0,n}, z_{1,n}, \ldots, z_{dI-1,n}]$. The left eigenvectors satisfy the eigenvalue equation,
\begin{align}
    \boldsymbol{\Psi}_{{\rm L}, n} \bX' = \Tilde{\lambda}_{n} \boldsymbol{\Psi}_{{\rm L}, n} \bX,
    \label{eq:left_eigvec}
\end{align}
where \cref{eq:left_eigvec} can be viewed as an equality relating the matrix elements of $\bX$ and $\bX'$. Such an equality restricted to the first columns of the data matrices, for example, implies
\begin{align}
   \boldsymbol{\Psi}_{{\rm L}, n} \begin{bmatrix}
       \Vec{s}_{1} \\
       \Vec{s}_{2} \\
       \vdots \\
       \Vec{s}_{d}
   \end{bmatrix} = \Tilde{\lambda}_{n} \boldsymbol{\Psi}_{{\rm L}, n} \begin{bmatrix}
       \Vec{s}_{0} \\
       \Vec{s}_{1} \\
       \vdots \\
       \Vec{s}_{d-1}
   \end{bmatrix},
\end{align}
which can be expressed in terms of the eigenvector coordinates and real-time observables,
\begin{align}
   \begin{split}
       \bra{\phi_0} & e^{iH\Dt} \left( \sum_{a=0}^{d-1}  \sum_{i=1}^{I} z_{(a,i), n}^{\ast} e^{iH a \Dt} O_i \ket{\phi_0} \right)  \\
       &= \Tilde{\lambda}_{n}^{\ast} \bra{\phi_0} \left( \sum_{a=0}^{d-1} \sum_{i=1}^{I} z_{(a,i), n}^{\ast} e^{iH a \Dt} O_i \ket{\phi_0} \right),
   \end{split}
   \label{eq:eigstate}
\end{align}
where $z_{(a,i), n} := z_{aI+i-1, n}$ are the vectorized coefficients of $\boldsymbol{\Psi}_{{\rm L}, n}$. Since $\tilde{\lambda}_{n} \approx e^{-i E_n \Dt}$, the dynamic mode above closely
follows the eigenstate oscillation driven at desired
frequency, $\braket{\phi_0|e^{iH\Dt}|\psi_n} = e^{iE_{n} \Dt} \braket{\phi_0|\psi_n}$. Hence, we can approximate $\ket{\psi_n}$ by \cref{eq:eigstate},
\begin{align}
   \ket{\Tilde{\psi}_n} = \sum_{a=0}^{d-1} \sum_{i=1}^{I} z_{(a,i), n}^{\ast} e^{iH a \Dt} O_i \ket{\phi_0},
   \label{eq:eigstate_2}
\end{align}
where $z_{(a,i),n}$ are now scaled  to give a normalized state. Any
eigenstate properties can in turn be derived in terms of the pool of operators $\{ O_i \}_{i=1}^{I}$ and time-evolved states $\{ e^{-iH a\Dt} \ket{\phi_0} \}_{a}$.

The formal solution to \cref{eq:MODMD_sys} entails computing the Moore–Penrose pseudo-inverse $\bX^{+} = (\bX^{\dagger} \bX)^{-1} \bX^{\dagger}$ of the data matrix $\bf{X}$. To ensure stability and filter out perturbative noise, we employ the following truncated singular value decomposition (SVD) of the data matrix,
\begin{align}
    \bX_{\Tilde{\delta}} = \sum_{\ell: \sigma_{\ell} > \Tilde{\delta}\sigma_{\rm max}} \sigma_{\ell} \boldsymbol{u}_{\ell} \boldsymbol{v}_{\ell}^{\dagger},
    \label{eqn:SVD_trunc}
\end{align}
where $\sigma_{\ell}$ and $(\boldsymbol{u}_{\ell}, \boldsymbol{v}_{\ell})$ are the singular values and vectors respectively. Here $\Tilde{\delta} > 0$ is a truncation threshold defined relative to the largest singular value $\sigma_{\rm max} = \max_{\ell} \sigma_{\ell}$ of $\bX$. This thresholding procedure, which removes smaller singular values associated with noise, serves to regularize the LS problem of \cref{eq:MODMD_sys}. 

In summary, our shadow-based algorithm requires as input the selected observables $\{ O_{i} \}_{i=1}^{I}$, time step $\Dt$, and singular value threshold $\Tilde{\delta}$. The algorithm is described in \cref{alg:MODMD}, which we call the \emph{multi-observable dynamic mode decomposition} (MODMD).

\subsection{Selection of hyperparameters}
\label{subsec:selection of hyperparameters}
The performance of our algorithm clearly relies on the choice of the input parameters. We first remark that the convergence of MODMD, just as any subspace method, is influenced by the choice of reference state $\ket{\phi_0}$, where the estimation error scales inversely with the squared overlap with the eigenstates of interest. Although having a larger overlap is ideal, it suffices to prepare $\ket{\phi_0}$ using simplified single-particle calculations, which generally translates to a sparse sum of product states.  Another possibility is to prepare $\ket{\phi_0}$ as a matrix product state (MPS), which is a dense sum of product states but with highly structured coefficients that allow efficient classical manipulation and storage. With this foundation, we now turn our attention to the optimization of other key hyperparameters.

\textbf{Time step}. The time step $\Dt$ determines algorithmic convergence as it sets the separation of the eigenphases $e^{-i E_n \Dt}$ over the unit circle. A larger time step is thereby advantageous for better resolving the nearby eigenphases, until an aliasing ambiguity arises when $\Delta_{0,N-1} \Dt \geq 2\pi$, where the energy gaps are defined as $\Delta_{m,n}  = E_n - E_m$ (so $\Delta_{0,N-1}$ is the spectral range). Additionally, $\Dt$ must satisfy a further compatibility condition~\cite{shen2023estimating},
\begin{align}
    \Dt \lesssim \frac{2\pi}{\displaystyle \Delta_{0,N-1} + \max_{0 \leq n < N_{\rm eig}} (\Delta_{n,n+1} - \Delta_{0,n} )},
    \label{eq:dt}
\end{align}
where $N_{\rm eig}$ counts the eigenenergies $\{ E_0, \ldots, E_{N_{\rm eig} - 1} \}$ of interest. For unambiguous and appropriately ordered (so that $\tilde{E}_n$ approximates $E_n$) estimation, we suggest bounding the spectral
range of the Hamiltonian and then linearly shifting the range
to be in $[-C\pi, C\pi]$ for some positive constant $C < 1$. In this case, the time step can be set to $1$, uniquely restricting eigenangles $E_n\Dt$ in the $2\pi$-window $(-\pi, \pi)$. Note that \cref{eq:dt} holds for the relevant energy lower and upper bounds so the exact eigenenergies do not need to be known in advance.

In implementing the time evolution on quantum device via, for example, a Trotter-Suzuki formula, the time step $\Dt$ is closely tied to the quantum resources requirements. Since we aim to extract eigenvalues of $e^{-iH\Dt}$ through the observable time-series data, approximating $e^{-iH\Dt}$ to a target accuracy $\epsilon_{\rm TS}$ sets the required circuit complexity. Access to more observables accelerates our search toward the target eigenspaces, providing the flexibility to employ a potentially smaller $\Dt$. This in turn reduces the overall circuit complexity. We defer detailed analysis of the first-order Trotter scheme to \cref{subsubsec:Trotter_noise}. 

\textbf{Observables}. The choice of observables $\{ O_i \}_{i=1}^{I}$ also plays a critical role in steering convergence. Strategically choosing observables helps construct a subspace enriched with low-energy spectral content, which can significantly influence performance of MODMD. While expanding the observable pool improves quality of spectral information, it also raises classical post-processing costs, since the LS problem in \cref{eq:MODMD_sys} grows proportionally in size with the number of observables $I$. Such a trade-off is important to keep in mind, though leveraging many observables incurs little overhead in MODMD and remains highly beneficial for optimizing quantum resource allocation.

Drawing on the signal subspace intuition from \cref{eq:multi_signals}, the observables should be spectrally `independent' in the sense that the matrix of oscillation amplitudes, 
\begin{align}
    \boldsymbol{c} = \begin{bmatrix}
        c_{0,1} & c_{0,2} & \ldots & c_{0,I} \\
        c_{1,1} & c_{1,2} & \ldots & c_{1,I} \\
        \vdots & \vdots & \ddots & \vdots \\
        c_{N-1,1} & c_{N-1,2} & \ldots & c_{N-1,I}
    \end{bmatrix},
\end{align}
maintains a full column rank of $I$. Otherwise, the multi-dimensional signals contain redundant information. As a convention, we always fix $O_1 = {\rm Id}$ corresponding to 
the ODMD algorithm.

It is worth noting that the shadow reconstruction involves strictly classical computation, thus requiring each observable $O_i$ to have a sparse representation in the Pauli basis,
\begin{align}
     O_i = \sum_{\nu =1}^{M_i} \kappa_{i,\nu} P_{i,\nu},
\end{align}
where $\{P_{i,\nu} \}_{\nu=1}^{M_i}$ is a set of $M_i = \mathcal{O}({\rm poly}(L))$ distinct $L$-qubit Pauli strings with associated weights $\{ \kappa_{i,\nu} \}_{\nu=1}^{M}$. Although it is rather convenient to select the observables $O_i$ randomly from the $4^L$ Pauli strings, the resulting real-time signals may suffer from diminished utility because of probable suppression of the target oscillation amplitudes $\lvert c_{n,i} \rvert = \lvert \braket{\psi_n|\phi_0} \rvert \lvert \braket{\phi_0|O_i|\psi_n} \rvert$. For ground state estimate, this occurs when $\lvert \braket{\phi_0|O_i|\psi_0} \rvert \approx 0$, which deteriorates the quality of the signals. As an example, a $1$-local Pauli $X$ operator changes the Hamming weight of reference state, and can hence lead to zero amplitude if the Hamiltonian preserves the total $Z$-spin. 

Alternatively, we propose the systematic generation of observable pools starting from the problem Hamiltonian, 
\begin{align}
    H = \sum_{\nu =1}^{M} \kappa_{\nu} P_{\nu},
    \label{eqn: hamiltonian_pauli_rep}
\end{align}
where the terms are ordered by the magnitude of their coefficients, $\lvert \kappa_{1} \rvert \geq ... \geq \lvert \kappa_{M} \rvert$. Such sorting induces a family of partial sums, $\sum_{\nu \in \mathcal{V}} \kappa_{\nu} P_{\nu}$, with $\mathcal{V} \subseteq \{1,\ldots,M\}$ labeling a subset of Pauli strings. Our observables can be selected from these partial sums based on importance of the $M$ Pauli weights $\{\kappa_{\nu}\}_{\nu=1}^{M}$. That is, we may consider $O_1 = \sum_{\nu=1}^{M} \kappa_{\nu} P_{\nu} = H$, $O_2 = \sum_{\nu=1}^{M-1} 
\kappa_{\nu} P_{\nu}$, etc. Let us assume that the target Hamiltonian is linearly shifted, as discussed for selection of time step, such that the low-lying energies are large in magnitude. In contrast to randomly selecting a Pauli string, the low energy amplitudes of interest, for example $\lvert c_{0,1}  \rvert = \lvert \braket{\psi_n|\phi_0} \rvert^2 |E_0|$, are effectively `magnified' relative to amplitudes associated with energies interior in the spectrum.

We remark that integer powers of the partial sums and their linear combinations can also be desired additions to the observable pool for generating high-quality real-time signals. This imposes no computational bottleneck since the observable predictions can be performed in a parallel and distributed manner on classical computers, and the variance only depends on the $1$-norm, $\max_{1 \leq i \leq I}\lVert \Vec{\kappa}_i \rVert_1$, of the Pauli weights (see \cref{subsec:error}).

\textbf{Time-delayed embedding}. The dimensions $(d,K)$ define the shape of the block Hankel matrices, $\bf X$ and $\bf X'$, assembled from the observable time-series data. To avoid potential instabilities due to rank deficiency, we require that $d \leq K$. For consistency, we fix $\frac{K}{d} = \frac{5}{2}$ throughout our manuscript, which has been observed to yield reliable empirical performance (for example relative to $\frac{K}{d} = 1$). We note that more refined $(d,K)$ choices  may be achieved from minimizing known upper bound on the conditioning number of the Hankel data matrices~\cite{Potts_2016}.

\textbf{SVD threshold}. The SVD threshold $\Tilde{\delta}$ is subject to the noise level, which can be controlled as we collect real-time data via low-rank shadow techniques. As a practical default, we set $\tilde{\delta} \approx 10 \varepsilon_{\rm noise}$, placing it roughly an order of magnitude above the statistical uncertainty $\varepsilon_{\rm noise}$ due to shot noise. Setting $\tilde{\delta}$ too high (low) risks erasing relevant eigenmode contributions (retaining excessive noise).

\textbf{Parametric validation}. To ensure the robustness of MODMD in different parametric regimes, we conduct a comprehensive sweep over key hyperparameters. Results illustrating the effect of time step, number of observables, and SVD threshold are shown in \cref{app:parameter_sweeps}. The final accuracy of our algorithm improves with larger time steps -- provided aliasing is suppressed -- and larger observable pools. While accuracy depends non-monotonically on the SVD threshold as expected, it remains rather insensitive to reasonable (over an-order-of-magnitude) variations in the exact threshold value.

\subsection{Hamiltonian properties beyond energies}
\label{subsec:Hamiltonian properties}

The MODMD framework extends beyond estimating the eigenenergies, providing access to useful Hamiltonian properties including eigenstate properties and dynamical responses. To illustrate, we first recall from \cref{subsec:main algorithm} that the MODMD eigenpairs $(\tilde{\lambda}_{\ell}, \mathbf{\Psi}_{{\rm L}, \ell})$ of system matrix $A$ can also be leveraged to construct approximations $\ket{\tilde{\psi}_n}$ to the low-lying eigenstates $\ket{\psi_n}$, which we confirm using two fidelity metrics within \cref{app:eigenstate_convg}. Such eigenstate information can be explicitly expressed and implemented as a linear combination of time evolutions on quantum hardware, and is typically inaccessible from the outputs of recent signal processing phase estimation algorithms. As a consequence, arbitrary eigenstate properties can be predicted as
\begin{align}
    f( \ket{\psi_n}) \approx f ( \ket{\tilde{\psi}_n}),
\end{align}
for any scalar-valued function $f: \mathcal{H} \rightarrow \mathbb{C}$. This predictive capability straightforwardly applies to any state property within the low-lying energy subspace.

In addition, the state shadows $\{\hat{\sigma}_q\}_{q=1}^{Q}$ stored on the classical computer can be utilized to calculate the time-dependent expectations $\braket{\phi_0(t)|O_i|\phi_0(t)}$ (note that they differ from $\braket{\phi_0|O_i|\phi_0(t)}$). Specifically, we recognize that  
\begin{align}
    \braket{O_i(t)}_{\rm est} = \frac{2}{Q}\sum_{q=1}^{Q} \Tr[\mathcal{M}^{-1}(\hat{\sigma}_q) (\ket{1_a}\bra{1_a} \otimes O_i)],
\end{align}
gives an unbiased estimate of $\braket{O_i(t)} = \braket{\phi_0(t)|O_i|\phi_0(t)}$. These additional data can be taken as an input to a separate set of MODMD calculations.

Augmenting the capability of real-time subspace methods to compactly represent the eigenstates, the general framework of dynamic mode decomposition (DMD) moreover enables the prediction of system dynamics over longer timescales~\cite{gomes2023hybridmethodquantumdynamics,Kemper2024,kaneko2024forecastinglongtimedynamicsquantum}. A multi-observable signal $\vec{s}(t)$, composed of time correlation functions, contains essential dynamical fingerprint that characterizes, for instance, how a many-body system reacts to an external perturbation in the linear-response regime~\cite{chandler1987introduction,limmer2024statistical}. Here, the response represents a dynamical property distinct from the stationary properties governed by a single eigenstate. To further study these dynamical properties, we analyze how time correlation functions are predicted within the MODMD framework. 

First, the one-point correlators $\braket{\phi_0| O_i e^{-iHt} |\phi_0}$ can be propagated forward in time in increments of $\Dt$: \cref{eq:MODMD_sys} suggests that integer powers of the system matrix $A$ can be used to (approximately) fast-forward the observables beyond the measurement window. That is, for any integer $k > K+d$, MODMD predicts the dynamics at a later time $k\Dt$ via
\begin{align}
    \braket{\phi_0| O_i e^{-iH k \Dt} |\phi_0} \stackrel{{\rm MODMD}}{\approx} \boldsymbol{e}_{i}^{\dagger} A^{k-K-d+1} \boldsymbol{x}_{K},
\end{align}
where $\boldsymbol{x}_{K} \in \mathbb{R}^{dI}$ is the last column of the data matrix $\bX$, and $\boldsymbol{e}_{i} = [0, \ldots,0,1,0,\ldots,0]^{\top} \in \mathbb{R}^{dI}$ labels the standard basis vector with the $[(d-1)I+i]$th entry equal to $1$. The system matrix $A$ functions as the forward $\Dt$-propagator for evolving observables in time, establishing $A^{-1}$ as the respective backward propagator. Such predictive abilities of MODMD are demonstrated in \cref{subsec:predict dynamics}.

Next, the two-point correlators can be approximated, with MODMD eigenenergy and eigenstate estimates, as
\begin{widetext}
    \begin{align}
    \langle \phi_0|O_i(k\Dt) O_j(l\Dt) |\phi_0 \rangle \stackrel{{\rm MODMD}}{\approx} \sum_{n=0}^{dI-1} e^{-i\tilde{E}_n (k-l)\Dt} \braket{\phi_0 | e^{i H k\Dt} O_i |\tilde{\psi}_n } \braket{\tilde{\psi}_n| O_j e^{-iH l\Dt} |\phi_0}, 
\end{align}
\end{widetext}
where $O(t) = e^{iHt}Oe^{-iHt}$ labels a time-evolved operator in the Heisenberg picture. The behavior of the  one-point correlators $\braket{\tilde{\psi}_n| O_i e^{-iH k\Dt} |\phi_0}$ over longer times can then be predicted via short-time snapshots $\braket{\tilde{\psi}_n| O_i e^{-iH a\Dt} |\phi_0}$ for $0 \leq a \leq K+d$. Notice that the inner products involve the approximate eigenstates $\ket{\tilde{\psi}_n}$ alongside $\ket{\phi_0}$. Thus by \cref{eq:eigstate_2}, measuring these snapshots in general incurs an additional cost of $\mathcal{O}(I^2 (K + d)^2)$, which makes predictions of two-point correlators more demanding. However in the case where $\ket{\phi_0} = \ket{\tilde{\psi}_{m}}$, we can fully leverage information in the signal subspace with no extra measurements as, for instance, $e^{-iH a\Dt} \ket{\tilde{\psi}_{0}} \approx e^{-i\tilde{E}_0 a\Dt} \ket{\tilde{\psi}_{0}}$. This reduces the cost back to $\mathcal{O}(\log(I)(K + d))$. Importantly, the favorable $\log(I)$ scaling is absent in conventional subspace methods where matrix elements of the forms $\braket{\phi_0|O_i e^{-iH k\Dt} O_j|\phi_0}$ or $\braket{\phi_0|e^{iH k\Dt} O_i e^{-iH\Dt} O_j e^{-iH l\Dt}|\phi_0}$ can be measured at the costs of at least $\mathcal{O}(I\log(I)(K + d))$ or $\mathcal{O}(I^2 (K + d)^2))$, respectively. While MODMD offers versatile capabilities discussed in this section, our work focuses on estimation of eigenenergies, leaving the specific explorations of other applications for future studies.

\section{Theoretical Guarantees}
\label{sec:theory}

We establish in this section fundamental connections between MODMD and modern spectral approaches, furnishing a theoretical framework that guarantees its convergence. We first explicitly show a speedup of MODMD over ODMD by having an expansive pool of observables. Next, we exploit the linearity of quantum dynamical evolution and consider MODMD as a multi-reference scheme within a suitably defined function space through Koopman operator analysis~\cite{Koopman1931HamiltonianSA,Koopman1932,BruntonKutz2015,Williams2015,BruntonKutz2016,Arabi2017}. These two analytical viewpoints reinforce each other, underpinning the reliable performance of our algorithm for excited state problems.

\subsection{Multi-observable dynamic mode decomposition}

For the least-squares (LS) problem of \cref{eq:MODMD_sys} with observables $\{ O_i \}_{i=1}^{I}$, the multivariate solution is,
\begin{align}
    A = \left[ \begin{array}{c|ccc}
    \begin{matrix}
        0 \\ \vdots \\ 0
    \end{matrix} & {\rm Id}_{d-1} \otimes {\rm Id}_{I} \\
    \hline
    -A_0 & \begin{matrix}
        -A_1 & \cdots & -A_{d-1}
    \end{matrix}
    \end{array} \right],
\end{align}
where each $A_{\ell} \in \mathbb{R}^{I \times I}$ for $ \ell = 0, 1, \cdots, d-1$ represents a submatrix, and ${\rm Id_{d-1}}$ and ${\rm Id_{I}}$ are identity operators of respective dimensions. The block companion structure of the system matrix $A \in \mathbb{C}^{dI \times dI}$ immediately follows from the block Hankel structure of the matrices $\mathbf{X}$ and $\mathbf{X}'$. The multi-observable system matrix has a characteristic polynomial, 
\begin{align}
    \mathcal{C}_{A}(z) = {\rm det}(z-A) = {\rm det}\left( \sum_{\ell=0}^{d} z^{\ell} A_{\ell}  \right),
\end{align}
where $A_d \equiv {\rm Id}_{I}$. Hence, the roots of $\mathcal{C}_{A}(z)$ correspond precisely to the $dI$ eigenvalues of the system matrix $A$. To understand the convergence of the $A$-spectrum to the actual eigenphases, we first consider two limiting cases. For the case $(d,I)=(N,1)$, we essentially recover the celebrated Prony's method~\cite{prony1795essai} as the single-observable ODMD with 
$A_{\ell} \in \mathbb{C}$, such that $\mathcal{C}_{A}(z) = \prod_{n=0}^{N-1} (z-\lambda_n)$ for $\lambda_n = e^{-i E_n \Dt}$. While for the complementary case of $(d,I) = (1,N)$, 
$\mathcal{C}_{A}(z) = {\rm det}(z-\overline{A})$ where
$\overline{A} \in \mathbb{C}^{N \times N}$ satisfies the linear homogeneous equation,
\begin{align}
    \begin{bmatrix}
     \braket{\phi_0|O_1|\phi_0(t + \Dt)} \\
     \braket{\phi_0|O_2|\phi_0(t + \Dt)} \\
     \vdots \\
    \braket{\phi_0|O_N|\phi_0(t + \Dt)} \\
    \end{bmatrix} = \overline{A}  \begin{bmatrix}
     \braket{\phi_0|O_1|\phi_0(t)} \\
     \braket{\phi_0|O_2|\phi_0(t)} \\
     \vdots \\
    \braket{\phi_0|O_N|\phi_0(t)} \\
    \end{bmatrix},
\end{align}
for any time $t \in \mathbb{R}$. Such a matrix $\overline{A}$ indeed exists because the real-time expectations $\braket{\phi_0|O_i|\phi_0(t)}$ reside within the $N$-dimensional space spanned by single-frequency signals $e^{-iE_n t}$ driven at the individual eigenfrequencies $E_n$. To investigate the general case $1 \ll dI \ll N$, we start with a simpler version of \cref{eq:MODMD_sys}, assuming the $d$ matrix blocks to be diagonal:
\begin{align}
    A_{\ell} = {\rm Diag}\begin{bmatrix}
        A_{\ell, 11} & & \\
        & A_{\ell, 22} & & \\
        & & \ddots & \\
        & & & A_{\ell, II} \\
    \end{bmatrix},
\end{align}
where $A_{\ell, ij} \equiv 0$ for $i \neq j$. In this case, the individual LS residuals associated with the observables $O_i$ are independent, and we can show that the resulting MODMD estimates are bounded by the $I$ single-observable ODMD estimates, \textit{e.g.},
\begin{align}
     \min_{1\leq i \leq I} \lvert \tilde{E}_{i,0} - E_0 \rvert \leq \lvert \tilde{E}_{0} - E_0 \rvert \leq  \max_{1\leq i \leq I} \lvert \tilde{E}_{i,0} - E_0\rvert,
     \label{eq:single-observale estimates}
\end{align}
where $\tilde{E}_0$ and $\tilde{E}_{i,0}$ indicate, respectively, the ground state energy estimate using the entire observable pool (the full system matrix $A$) and one single observable (only $i$th row $A_{\ell,ii}$ of the submatrices $A_{\ell}$). 

More interesting convergence arises when the $I$ single-observable residuals are coupled to one another, allowing reductions in the individual residuals due to the flexibility of off-diagonal elements in the submatrices $A_{\ell}$. Specific scenarios in which MODMD substantially improves upon the ODMD residuals are considered within \cref{app:multi-observable dynamic mode decomposition}. Intuitively, we expect a reduced total residual to, in turn, improve the eigenenergy estimates, where \cref{eq:single-observale estimates} holds accordingly with tighter lower and upper bound -- ideally, both approaching zero. Moreover, the reduction in total LS residual from MODMD indicates a more expressive system matrix as a proxy for the underlying dynamics, which is crucial for accurately predicting a multi-observable signal over longer times as discussed in \cref{subsec:Hamiltonian properties}.

\bigskip

\subsection{Koopman operator analysis}
We now establish convergence properties of MODMD from a functional-theoretic perspective. Our analysis will revolve around the study of the Koopman operator, a pivotal mathematical object in understanding the complexities of a dynamical system. The Koopman operator $\Ko$ probes the underlying dynamics of a system by acting on scalar-valued functions. For any function $f: \mathcal{H} \rightarrow \mathbb{C}$, its action is
\begin{align}
    \Ko[f](\ket{\phi}) = f ( e^{- i H \Delta t} \ket{\phi} ),
\end{align}
which gives a push-forward of the dynamics via the time evolution operator $e^{-iH \Dt}$. For a quantum-dynamical system, we first observe that $f_{n}(\ket{\phi}) = \braket{\psi_n|\phi}$ constitutes an eigenfunction of $\Ko$ since 
\begin{align}
    \Ko [f_{n}]  = e^{-iE_n \Dt} f_n,
\end{align}
where $e^{-i E_n \Delta t}$ is the corresponding eigenvalue. Therefore scalar functions of the form $f_{\psi,O}(\ket{\phi}) := \braket{\psi|O|\phi} = \sum_{n=0}^{N-1} \braket{\psi|O|\psi_n} \braket{\psi_n|\phi}$ lie in a $\Ko$-invariant subspace spanned by these eigenfunctions. By choosing $\ket{\psi}$ and $O$, we recast the task of identifying Hamiltonian eigenmodes as an equivalent, finite-dimensional Koopman eigenvalue problem. We use $\vec{g} = [g_1, g_2, \ldots, g_D]^{\top}$ to denote a vector of $D$ distinct scalar functions $g_i$ in the invariant function subspace, all taking the form of $f_{\psi,O}$.

We first examine the case where $g_i = (\Ko)^{i-1}[f_{\psi, {\rm Id}}]$, which gives rise to the ODMD approach when $D = d$. We seek to determine the closest approximation $\bK_{\Dt} \in \mathbb{C}^{d \times d}$ of the Koopman operator when restricted to the invariant subspace. The closest-fitting problem has a least-squares formulation,
\begin{align}
    \bK_{\Dt} = \argmin_{\bK \in \mathbb{C}^{d \times d}} \sum_{k=0}^{K} \norm{\Ko[\Vec{g}](\ket{\phi_k}) - \bK  \Vec{g} (\ket{\phi_k}) }_{2}^2,
    \label{eq:ODMD_LS}
\end{align}
where $\Ko$ acts component-wise on $\vec{g}$, and the $\ell^2$-residual is being minimized with respect to some states $\{ \ket{\phi_k} \}_{k=0}^{K}$ sampled from the Hilbert space. Formally, we can express \cref{eq:ODMD_LS} as a matrix equation
\begin{widetext}
    \begin{align}
         \underbrace{\begin{bmatrix}
            \vertbar & & \vertbar\\
            \vec{g}(e^{-iH\Dt} \ket{\phi_0}) & \cdots & \vec{g}(e^{-iH\Dt}\ket{\phi_K}) \\
            \vertbar & & \vertbar\\
        \end{bmatrix} }_{\bG' \in \mathbb{C}^{d \times (K+1)} } = \bK_{\Dt} \underbrace{\begin{bmatrix}
            \vertbar & & \vertbar\\
            \vec{g}(\ket{\phi_0}) & \cdots & \vec{g}(\ket{\phi_K}) \\
            \vertbar & & \vertbar\\
        \end{bmatrix}}_{\bG \in \mathbb{C}^{d \times (K+1)} },
    \end{align}
\end{widetext}
whose solution $\bK_{\Dt} = \bG' (\bG)^{+}$ involves computing the pseudo-inverse $\bG^{+}$. Utilizing $\bG^{+} = (\bG^{\dagger} \bG)^{-1} \bG^{\dagger}$, we can directly show that $\bK_{\Dt}$ yields an equivalent factorization $\bK_{\Dt} = \bW (\bZ)^{+}$, where $\bZ, \bW \in \bbC^{d \times d}$ are constructed by
\begin{align}
    \bZ &= \frac{1}{K+1} \sum_{k=0}^{K} \vec{g}(\ket{\phi_k}) \vec{g}^{\dagger}(\ket{\phi_k}), \label{eq:V_finite} \\
    \bW &= \frac{1}{K+1} \sum_{k=0}^{K} \vec{g}(e^{- iH \Dt}\ket{\phi_k}) \vec{g}^{\dagger}(\ket{\phi_k}).
    \label{eq:W_finite}
\end{align}
This alternative factorization is intimately related to the powerful Krylov approaches for operator diagonalization \cite{Paige1971_thesis,Saad1980_cor,horn_johnson_1985}. Suppose we sample the states $\{\ket{\phi_k}\}_{k}$ according to a probability measure $\mu$ defined over the Hilbert space. With sufficiently many samples, we have
\begin{align}
    \bZ_{ij} &\rightarrow \int_{\ket{\phi} \in \mathcal{H}} d\mu\hspace{0.03cm} g_i g_j^{\ast} = \braket{g_j, g_i }_{\mu},\\
    \bW_{ij} &\rightarrow \int_{\ket{\phi} \in \mathcal{H}} d\mu\hspace{0.03cm} \Ko[g_i] g_j^{\ast} = \braket{g_j, \Ko[ g_i]}_{\mu},
\end{align}
where the $L^2(\mu)$ inner product gives the continuum limit of Monte-Carlo averages in \cref{eq:V_finite,eq:W_finite}. Therefore eigenpairs $(\lambda_{\bK}, \vec{v}_{\bK})$ of $\bK_{\Dt}$ should satisfy the generalized eigenvalue equation,
\begin{align}
    \bW \vec{v}_{\bK} = \lambda_{\bK} \bZ \vec{v}_{\bK},
\end{align}
where $\bZ$ and $\bW$ can now be reinterpreted, respectively, as the matrix representation of the functional overlap and Koopman operator in the finite basis $\{g_1, \cdots, \Ko^{d-1}[g_1] \}$. This special nonorthogonal basis is known as the order-$d$ Krylov basis, and the associated subspace is the Krylov subspace $\mathcal{K}_{d}(g_1) \subset {\rm span}\{f_n \}_{n}$. 

Projecting the full Koopman eigenvalue problem onto this subspace of size $d \ll N$ allows an efficient retrieval of spectral information, including the extremal eigenvalues. In the ODMD algorithm, we choose $\ket{\phi_k} = e^{-i H k \Dt} \ket{\phi_0}$ for a fixed state $\ket{\phi_0}$ and take the sampling measure $\mu$ to be the empirical measure (a sum of Dirac-delta measures), 
\begin{align}
    \mu = \frac{1}{K+1} \sum_{k=0}^{K} \delta_{\ket{\phi_k}},
\end{align}
along the orbit $(\ket{\phi_0}, e^{-iH\Dt}\ket{\phi_0}, \cdots, e^{-iH K \Dt} \ket{\phi_0})$. In this case, $\lim_{K \rightarrow \infty} \lVert \Ko[f] \rVert_{L^2(\mu)} = \lim_{K \rightarrow \infty} \lVert f \rVert_{L^2(\mu)}$ for a continuous function $f$ (we expect the limit to exist~\footnote{$ \lim_{K \rightarrow \infty} \frac{1}{K+1} \sum_{k=0}^{K} f(e^{-iH k \Dt} \ket{\phi_0})$ converges uniformly to $\int_{\mathcal{H}} d\bar{\mu} f $  for all $\ket{\phi_0} \in \mathcal{H}$, provided that the limit measure $\bar{\mu}$ is uniquely invariant. For quantum systems, the action of dynamical evolution can be regarded as a collection of independent rotations up to a change of basis. If the Hamiltonian spectrum and evolution time step remain general, $\bar{\mu}$ is then the uniform measure on $\mathcal{H}$ as the unique invariant measure.} for a generic many-body Hamiltonian $H$ and time step $\Dt$), since
\begin{widetext}
    \begin{align}
    \lim_{K \rightarrow \infty} \braket{ \Ko[f], \Ko[f]}_{\mu} &= \lim_{K \rightarrow \infty} \frac{1}{K+1} \sum_{k=1}^{K+1} \abs{f( e^{-iH k \Dt} \ket{\phi_0}) }^2 = \lim_{K \rightarrow \infty} \frac{1}{K+1} \sum_{k=0}^{K} \abs{f( e^{-iH k \Dt} \ket{\phi_0}) }^2 \label{eq:isometric} = \lim_{K \rightarrow \infty} \braket{f, f}_{\mu} ,
\end{align}
\end{widetext}
where the middle equality  holds due to a vanishing difference, $\lim_{K \rightarrow \infty}\frac{1}{K+1} [ |f(e^{-iH(K+1)\Dt} \ket{\phi_0})|^2 - |f(\ket{\phi_0})|^2 ] = 0$. In other words, the Koopman operator $\Ko$, when restricted to the invariant subspace, is isometric and hence normal. By the spectral theorem, the $\Ko$-eigenfunctions  $\{ f_n \}_{n=0}^{N-1}$ are orthogonal to each other. Under this condition, exponentially rapid convergence of standard Krylov approach has been thoroughly analyzed~\cite{saadbook}, which aligns with our ODMD observations.

Now we extend the result above to eigenfunctions that are nonorthogonal, a  regime of practical interest for finite $K$. In such general cases, the functional Krylov approach still converges exponentially, but its convergence depends explicitly on the conditioning of the eigenfunctions under the $L^2(\mu)$ inner product. In particular, we define
\begin{align}
    \mathcal{B}_{mn} = \frac{\braket{f_m,f_n}_{\mu}}{\lVert f_m \rVert_{L^2(\mu)} \lVert f_n \rVert_{L^2(\mu)} },
    \label{eq:ortho_Krylov}
\end{align}
where $\{ f_n \}_{n=0}^{N-1}$ are mutually orthogonal if and only if $\mathcal{B}$ is the identity. Departures from orthogonality are captured by the condition number ${\rm cond}(\mathcal{B}) = \lVert \mathcal{B} \rVert_2 \lVert \mathcal{B}^{-1} \rVert_2$~\cite{Saad_GMRES,Liesen2004}. In practice, $K + 1 < N$ always holds. Since the empirical $L^2(\mu)$ inner product only resolves functions through their values on the sampled orbit, the Gram matrix $\mathcal{B}$ has rank of at most $K + 1$ and is thus rank deficient. In this regime, The relevant notion of conditioning is the one defined on the identifiable subspace, given by the intrinsic condition number ${\rm cond}(\mathcal{B}) \mapsto \lVert \mathcal{B} \rVert_2 \lVert \mathcal{B}^{+} \rVert_2$. It reduces to the usual condition number when $\mathcal{B}$ is full rank, and measures how stably a finite-$K$ signal distinguishes resolvable Koopman channels.

In the single observable setting, we invoke the following theorem for the ground state energy error.
\medskip

\noindent \textbf{Theorem 1}. Let $\Tilde{E}_{0}(d)$ be the approximate ground state energy extracted in the $d$-dimensional function subspace spanned by $\{g_1, \Ko[g_1], \cdots, \Ko^{d-1}[g_1] \}$. For $d \geq 1$, there exists time step $\Delta t$ so that the error $\delta E_{0}(d) = \Tilde{E}_{0}(d) - E_{0}$ is bounded by
\begin{align}
    \delta E_{0}(d) \leq \frac{ {\rm cond}(\mathcal{B})^{1/2} \abs{ \sin[(E_{N-1} - E_{0})\Dt]} }{\Tilde{\epsilon}_{0 \rightarrow 1}^{2(d-1)} \Dt } \tan^2{\Xi} ,
    \label{ineq:loeigval_bound}
\end{align}
where $\cos^2{\Xi} = \lvert \langle f_0, g_1 \rangle_{\mu} \rvert^2 /(  \lVert f_0 \rVert_{L^2(\mu)}^2  \lVert g_1 \rVert_{L^2(\mu)}^2 )$ specifies the squared overlap between the reference function $g_1$ and the true ground state $\Ko$-eigenfunction, while $\Tilde{\epsilon}_{0 \rightarrow 1} = 1 + 3(E_1 - E_0)\Dt/2\pi$ $\in [1,2]$ characterizes the normalized spectral gap of the Hamiltonian $H$.

\noindent \textit{Proof.} The proof is provided in \cref{subsec:theorem_1}.
\bigskip

Note that ${\rm cond}(\mathcal{B}) \rightarrow 1$ as $K \rightarrow \infty$ when eigenenergies are non-degenerate. By construction, $\mathcal{B}_{nn} \equiv 1$ and
\begin{align}
    |\mathcal{B}_{mn}| &= \frac{1}{K+1} \left \lvert \sum_{k=0}^{K} e^{i(E_n - E_m)k\Dt}  \right \rvert , \nonumber \\
    &= \frac{\lvert \sin{( \frac{K+1}{2} \theta_{mn})} \rvert}{(K+1) \lvert \sin{(\frac{1}{2}\theta_{mn})} \rvert} \rightarrow 0,
\end{align}
for $\theta_{mn} = (E_n - E_m) \Dt \neq 0$. Consequently, $\mathcal{B}$ converges asymptotically to the identity operator and ${\rm cond}(\mathcal{B}) \to 1$. To build intuition at $K + 1 < N$, we turn to the harmonic oscillator where $E_{n} - E_{m} = (n-m) \Delta E$, yielding a clean toy model with a semi-dense spectrum. As an illustrative choice, setting $\Dt  = \frac{2\pi}{N\Delta E}$ makes $\mathcal{B}$ circulant, allowing its eigenvalues $\{0, \frac{N}{K+1} \}$ to be computed exactly. In this case $\lVert \mathcal{B} \rVert_2 \lVert \mathcal{B}^{+}\rVert_2 \equiv 1$.

In contrast, setting $\Dt <\frac{2\pi}{N\Delta E}$ leads to a less favorable conditioning: $\mathcal{B}$ loses its circulant symmetry and is only Toeplitz, causing its nonzero eigenvalues to disperse. As $\Dt$ keeps shrinking, the matrix becomes closer to rank-1. In this case, the conditioning deteriorates, demonstrating how near-resonances or overly fine time steps can impair mode separability.

On the other hand, if the ground state energy admits a degeneracy, we can replace $\mathcal{B}$ by an effective Gram matrix constructed over distinct energy subspaces, ensuring that the conditioning factor remains finite. Degeneracy does not affect energy convergence, but it precludes resolution of individual eigenstates. Further details are provided in \cref{app:conditioning_degenerate}.


Building on the single-observable Krylov idea (ODMD) above, the MODMD approach can thus be viewed as an enriched extension, where we allow for an extra degree of freedom in selecting multiple functions
\begin{align}
    \vec{g} = \begin{bmatrix}
        f_{\phi_0, {\rm Id}} \\
        f_{\phi_0, {O_2}} \\
        \vdots \\
        f_{\phi_0, {O_I}} 
    \end{bmatrix} \implies \vec{g}(e^{-iHt}\ket{\phi_0}) = \vec{s}(t).
\end{align}
By leveraging the classical shadows, each quantum circuit $e^{-iHk\Dt}$ originally capable of computing a single overlap can now simultaneously compute the Koopman action on a vector $\vec{g}$ of scalar functions. Each function corresponds to a unique choice of observable. This key algorithmic improvement enables a \textit{block} Krylov scheme, significantly accelerating the energy convergence. The rate of convergence in the MODMD setting is described by the theorem below.
\bigskip

\noindent \textbf{Theorem 2}. Let $\Tilde{E}_{n}(d)$ be the approximate $n$th eigenenergy extracted in the $dI$-dimensional function subspace ${\rm span} \{\vec{g}, \Ko[\vec{g}], \cdots, \Ko^{d-1}[\vec{g}] \}$, and $\delta E_{n}(d) = \Tilde{E}_{n}(d) - E_{n}$ be the approximation error. Consider the diagonal error matrix
\begin{align}
    \bDel_I(d) = \begin{bmatrix}
        \delta E_{0}(d) & & \\ 
         & & \ddots & \\
         & & & \delta E_{I-1}(d) \\
    \end{bmatrix},
\end{align}
which contains approximations to the lowest $I$ energies. For $d \geq 1$, there exists a time step $\Delta t$ so that the spectral approximation $\bDel_I(d)$ is bounded by
\begin{align}
    \lVert \bDel_{I}(d) \rVert_2 \leq \frac{ {\rm cond}(\mathcal{B})^{1/2} \abs{\sin[(E_{N-1} - E_{0})\Dt]} }{\tilde{\epsilon}_{I-1 \mapsto I}^{2(d-1)} \Dt } \lVert \tan^2 \Theta \rVert_2,
    \label{ineq:loeigval_bound2}
\end{align}
for the operator norm $\lVert \cdot \rVert_2$. Here $\Theta$ denotes the canonical angle between the two subspaces ${\rm span}\{ f_n : 0 \leq n \leq I-1 \}$ and ${\rm span}\{ g_i: 1 \leq i \leq I \}$, which generalizes the squared overlap in Theorem 1 (see \cref{subsec:canonical_angles,subsec:theorem_2}). In the denominator, $\tilde{\epsilon}_{I-1 \mapsto I} \in [1, 2]$ depends on the $I$th spectral gap $(E_{I} - E_{I-1})$ of the Hamiltonian $H$.

\noindent \textit{Proof.} The proof is provided in \cref{subsec:theorem_2}.
\bigskip

Despite the formal similarity between the bounds from Theorem 1 and Theorem 2, we highlight that convergence in multi-observable setting offers a more favorable scaling for excited state calculations. The ODMD bound for the $I$th lowest eigenenergy, as in standard subspace methods employing the reference state $\ket{\phi_0}$, includes an additional multiplicative factor of $\Omega\big(2^{I-1} \tilde{\epsilon}_{I-1 \mapsto I}^{2I-2}\big)$~\cite{shen2022real}. Notably, this prefactor grows exponentially as we approach the higher excited states, counteracting the single-observable error decay from Theorem 1 unless $d > I$. While it is natural to extend the standard subspace methods using $I$ reference states $\ket{\phi_i} \propto O_i\ket{\phi_0}$, the quantum cost of measuring the expectations $\braket{\phi_i|e^{-iHt}|\phi_j}$ is at least $\mathcal{O}(I\log(I))$. This is because each state $\ket{\phi_i}$ must be time-evolved, making it much more expensive than MODMD.

Beyond the sharper asymptotic error bound, the multi-observable structure of MODMD introduces qualitative advantages that single-observable ODMD cannot deliver. First, single-observable methods can recover at most one representative in a degenerate or near-degenerate energy cluster, which often results in stalled convergence due to the collapse of spectral gap between closely spaced levels. By contrast, the block Krylov nature of MODMD allows multiple eigenmodes to be resolved simultaneously within a single invariant subspace, thus revealing richer spectral information~\cite{Li2015}. Second, block subspace approaches are known to accelerate convergence for the excited states by approximating invariant subspaces rather than individual eigenstates~\cite{saadbook}, a property that MODMD inherits in the low-noise regimes. Finally, the use of many non-collinear observables improves the conditioning of the LS problem, effectively attenuating the regression errors and reducing the nonasymptotic measurement overhead for any desired precision. Since classical shadows let us acquire multiple observables at virtually the same quantum circuit depth, with scrambling unitaries adding a precision-independent depth increment, such gains come at no cost to maximal runtime. Together, these advantages show that MODMD offers not only quantitative error improvements but also qualitatively more expressive, rapid, and stable spectral resolution.


\subsection{Error analysis}
\label{subsec:error}

To account for noisy quantum hardware, we present in this subsection a preliminary error analysis. For purpose of basic premises, we examine error components of three distinct types, (1) the statistical uncertainty arising from the randomizing shadow protocol, (2) the deterministic error due to approximate compilation of the time evolution, and (3) the device-level error that originate in noise on near-term hardware. We show that the first two error types remain completely independent of the problem size across a practical range of observable selections. Specifically, the circuit approximation error grows polynomially with the maximal evolution time when we consider Trotterized evolution as viable proxy for exact dynamics. For device noise, we focus on the depolarization channel -- an elementary error model applicable across many quantum platforms -- and discuss how its rate can be quantified by the MODMD algorithm.

\subsubsection{Statistical noise}
\label{subsubsec:shot_noise}
First, we note that the prediction error associated with randomized measurements in the classical shadow techniques admits, in our case, a variance independent of the system size $L$. 
The classical shadows can predict observables, $\gamma_{i}(t) = \Tr [ \rho(t) \Gamma_{O_i} ]$, of the form \cref{eq:o_shadow}, where we recall that $\rk(\Gamma_{O_i}) \leq 2$ even when $\rk(O_i) =  2^L = N$. The low-rank property implies \cite{Preskill_shadow},
\begin{align}
    \vV[\gamma_i] &\leq \norm{ \GOi - \frac{\Tr[\GOi]}{N}({\rm Id}_a \otimes {\rm Id}) }_{\rm sh}^2, \\
    &\stackrel{\mathcal{U} = {\rm Cl}(N)}{=} 3\Tr[\GOi^2 ],
    \label{eq:Cliff_var}
\end{align}
where $\norm{\cdot}_{\rm sh}$ is the shadow norm conditional on the measurement primitive and the trace $\Tr[\GOi^2]$ on the RHS of \cref{eq:Cliff_var} can be unfolded by the defining relation $\GOi = {\rm Id}_{a} \otimes O_i \ket{1,\phi_0} \bra{0,\phi_\perp} + \ket{0,\phi_\perp} \bra{1,\phi_0} {\rm Id}_{a} \otimes O_i$. Hence, the prediction variance for a Pauli string, $P = \bigotimes_{\ell = 1}^{L} \sigma_{\ell}$ with $\sigma_{\ell} \in \{{\rm Id}_{\ell}, X_{\ell}, Y_{\ell}, Z_{\ell} \}$, can be uniformly bounded through the Cauchy–Schwarz inequality, $\Tr[\Gamma_P^2] \leq 2\norm{P}_2^2 \leq 2$. This immediately implies
\begin{align}
    O_i = \sum_{\nu=1}^{M_i} \kappa_{i, \nu} P_{\nu} \implies \vV[\gamma_i] \leq \mathcal{O}(\hspace{0.05cm} \norm{\vec{\kappa}_i}_1^2),
\end{align}
for a general Hermitian operator $O_i = O_i^{\dagger}$. Observe that the variance remains rather insensitive to the operator locality or the system size. Accordingly, each data matrix element in the MODMD setting incurs an additive error of $\epsilon_1 = \epsilon_{\rm noise}$ if we sample $Q = \mathcal{O}( \log(I) \displaystyle \max_{1 \leq i\leq I}\norm{\vec{\kappa}_i}_1^2 \epsilon_{\rm noise}^{-2} )$ quantum measurements.

\subsubsection{Trotter error}
\label{subsubsec:Trotter_noise}
A second error source pertains to inexact implementation of the unitary evolution $e^{-iH t}$, which perturbs both the eigenfrequencies $e^{ -i E_n t}$ and the time-evolved states $\rho(t)$. For near-term implementation, we assume query access to an approximate compilation of the evolution, for example, through Trotter–Suzuki factorization \cite{suzuki,Hatano2005} or linear combination of unitaries (LCU) \cite{lcu_2012}. We consider specifically the Trotter scheme for illustration and comment that similar analysis should hold for other schemes. For $H = \sum_{\nu=1}^{M} H_{\nu} = \sum_{\nu =1}^{M} \kappa_{H,\nu} P_{\nu}$, a first-order Trotter formula gives,
\begin{align}
    \epsilon_1 = \left \lVert e^{-iH t} - \Bigg( \prod_{\nu=1}^{M} e^{-i H_{\nu} t/r} \Bigg)^r \right \rVert_2 = \mathcal{O} \left( \frac{ \norm{\vec{\kappa}_H}_{1}^2 t^2}{r} \right), 
    \label{eq:Trotter}
\end{align}
where the Trotterized Hamiltonian simulation can be performed efficiently on the quantum computer if $M = \mathcal{O}({\rm polylog}(N))$. Equivalently, for
\begin{align}
    r = \mathcal{O}( \lVert \vec{\kappa}_H \rVert_{1}^2 t^2 \epsilon_{1}^{-1})
\end{align}
Trotterized blocks with a time discretization $\Dt^{\rm Trotter} = t/r$, we incur a Hamiltonian simulation error $\mathcal{O}(\epsilon_1)$ in the operator norm. For a $\mathsf{p}$th-order Trotter scheme, the error for simulating $e^{-iH t}$ is $\mathcal{O}( ( \lVert \vec{\kappa}_H \rVert_{1} t)^{\mathsf{p} + 1} r^{-\mathsf{p}})$~\cite{Berry2007}. We will work with this general 
$\mathsf{p}$th-order setting, as tracking the resulting error contributions is straightforward.



For any maximal runtime $t_{\rm max} = k_{\rm max} \Dt$, $e^{-iHt_{\rm max}}$ can be Trotterized with depth $\mathcal{O}(M( \lVert \vec{\kappa}_H \rVert_{1} t_{\rm max} )^{1+1/\mathsf{p}} \epsilon_1^{-1/\mathsf{p}})$, where we omit the $\mathcal{O}(5^{\mathsf{p}/2 - 1})$ dependence due to Suzuki recursion, which builds a $\mathsf{p}$th-order product formula from five copies of the $(\mathsf{p-2})$th-order formula. This factor is constant for a fixed $\mathsf{p}$. As preserving the Hankel structure of data matrices $\mathbf{X}$ and $\mathbf{X}'$ is essential, we pick the same Trotter step to implement $e^{-iHt_k}$ for $t_k \leq t_{\rm max}$; changing the Trotter step size across time steps $k$ would break the Hankel structure.

Let $\gamma_i(t) := \Re \braket{\phi_0|O_i e^{-i H t} |\phi_0}$. To achieve a sampling uncertainty $\epsilon_{\rm noise}$ when estimating the expectation values $\gamma_i(t_k)$ under Trotterized evolution up to $t_{\rm max}$, the number of operator exponentials required is
\begin{align}
    \tilde{\mathcal{O}} \left( \frac{\log(I) k_{\rm max} M ( \lVert \vec{\kappa}_H \rVert_{1} t_{\rm max} )^{1+1/\mathsf{p}}}{\epsilon_1^{1/\mathsf{p}} \epsilon_{\rm noise}^{2} } \right),
    \label{eq:total_op_exp}
\end{align}
because $\sum_{k=1}^{k_{\rm max}} t_k/ \Dt^{\rm Trotter} = \mathcal{O}(k_{\rm max} r_{\rm max})$ while at each time step $k$, we simulate $e^{-iH t_k}$ with $\tilde{\mathcal{O}}( \log(I)\epsilon_{\rm noise}^{-2})$ shots to collect the classical shadow dataset. To ensure that the solution to LS system in \cref{eq:MODMD_sys} is spectrally $\epsilon_1$-close to its Trotterized version, we take $\epsilon_{1} \rightarrow \mathcal{O}( \epsilon_1 I^{-1/2} k_{\rm max}^{-1})$ and $\epsilon_{\rm noise} \rightarrow \mathcal{O}( \epsilon_1 I^{-1/2} k_{\rm max}^{-1})$, using the fact that the spectral norm of the data matrix deviation is at most $\sqrt{dI(K+1)}$ times the entry-wise error. Assigning $\epsilon_1 = \mathcal{O}(\epsilon)$ to be the algorithmic error then yields an \emph{upper bound} on the total gate complexity for processing a multi-observable signal of length $k_{\rm max}$~\footnote{For a given problem, the scrambling unitaries required in classical shadow tomography have finite depth, as low as logarithmic in the system size~\cite{hu2024demonstrationrobustefficientquantum}, for each independent circuit run. So the gate count associated with the classical shadow component is $\mathcal{O}(t_{\rm tot} Q) = \mathcal{O}(\log(I) k_{\rm max}^4 \epsilon_{1}^{-2}) $ and is therefore subleading.},
\begin{align}
     N_{\rm gate, \rm tot}^{\rm MODMD}  = \tilde{O} \left( \frac{ \rchi  \log(I) I^{1+1/2\mathsf{p}} k_{\rm max}^{4+2/\mathsf{p}} M  \lVert \vec{\kappa}_H \rVert_{1}^{1+1/\mathsf{p}}  }{\epsilon^{2+1/\mathsf{p}}} \right),
\end{align}
with a factor $\rchi = \lVert \mathbf{X}^{+} \rVert_2 + \lVert \mathbf{X}^{+} \rVert^2_2 \lVert \mathbf{X}' \rVert_2$ arising from the matrix perturbation bound for pseudoinverse (recall that the LS solution is formally derived via pseudoinverse)~\cite{Wedin1973,Stewart1990}. In particular, $\rchi$ captures the influence of $k_{\rm max}$, $I$, and the observables $\{O_i\}_{i=1}^{I}$ through the ideal data matrices.


 This scaling invites a heuristic comparison with that of quantum phase estimation (QPE)~\cite{Kit97,Zwierz2010,Giovannetti2011,Nielsen_Chuang_2010}, where the time evolution is likewise Trotterized with a $\mathsf{p}$th-order scheme. Here we generalize the derivation in \cite{klymko2022real} that only considers the second-order scheme. The Trotterized QPE implements time steps $t_k^{\rm QPE, Trotter} = 2^{k-1} \Dt^{\rm QPE, Trotter}$. Note that the total energy error resulting from a Trotter step of $\Dt^{\rm QPE, Trotter}$ is roughly~\cite{klymko2022real},
\begin{align}
   \epsilon \Dt^{\rm QPE, Trotter} \approx \pi \sqrt{2^{-2k_{\rm max}^{\rm QPE}-2} + \epsilon_1^2},
\end{align}
which reflects the interplay between phase estimation and Trotterization. Setting both error sources to $2^{-k_{\rm max}^{\rm QPE}-1} = \epsilon_1 = \epsilon \Dt^{\rm QPE, Trotter}/(2\sqrt{2}\pi)$ locks the final error at $\epsilon$. The phase error imposes $k_{\rm max}^{\rm QPE} \leq \log_2(\epsilon t/(2\sqrt{2} \Dt^{\rm QPE, Trotter}))$ while the Trotterization error imposes
\begin{align}
    \epsilon_1 = ( \lVert \vec{\kappa}_H \rVert_{1} \Dt^{\rm QPE, Trotter}  )^{\mathsf{p} + 1} = \frac{\epsilon \Dt^{\rm QPE, Trotter}}{2\sqrt{2}\pi},
\end{align}
which yields $\Dt^{\rm Trotter, QPE} = \lVert \vec{\kappa}_H \rVert_{1}^{-(\mathsf{p}+1)/\mathsf{p}} (\epsilon/(2\sqrt{2}\pi))^{1/\mathsf{p}}$ and therefore
\begin{align}
    k_{\rm max}^{\rm QPE} \leq \frac{\mathsf{p} + 1}{\mathsf{p}} \log_2 \left(\frac{2\sqrt{2} \pi \lVert \vec{\kappa}_H \rVert_{1}  }{\epsilon} \right).
\end{align}
Accordingly, the total gate complexity for QPE is
\begin{align}
    N_{\rm gate, \rm tot}^{\rm QPE} = \tilde{O} \left( \sum_{n \in \mathcal{N}_{\rm eig} } \frac{1}{\displaystyle \lvert \alpha_n \rvert^2 } \frac{M (2\sqrt{2}\pi \lVert \vec{\kappa}_{H} \rVert_{1})^{1+1/\mathsf{p}} }{\epsilon^{1+1/\mathsf{p}}} \right),
    \label{eq:QPE_depth}
\end{align}
where we recall that $\mathcal{N}_{\rm eig}$ is the set of target eigenindices (in our case, those labeling the low-lying eigenstates) and $\lvert \alpha_n \rvert^2 = \lvert \braket{\psi_n|\phi_0} \rvert^2$. The summation over $\mathcal{N}_{\rm eig}$ accounts for the circuit repetitions necessary to land on each target eigenstate.


In regimes where simulation of time evolution achieves sufficient fidelity, as anticipated in the early fault-tolerant era, MODMD remains robust against constant shot noise $\epsilon_{\rm noise} = \mathcal{O}(1)$, assuming access to well-prepared reference state with dominant overlap on the target eigenstates. In particular, a recent theoretical analysis~\cite{ding2024espritalgorithmhighnoise} elucidates the notable noisy super-resolution property: MODMD under uniform time steps, $t_k = k \Delta t$, can achieve an algorithmic precision $\epsilon$ by evolving up to $t_{\rm max} = \tilde{\mathcal{O}}(\epsilon^{-2/3})$ simulation time, enabling a shallower circuit than the sharp $\Omega(\epsilon^{-1})$ lower bound for QPE. This improved scaling arises when the reference state is sufficiently supported on the target eigenstates. However, the MODMD total runtime, $t_{\rm tot} = \tilde{\mathcal{O}}(\epsilon^{-4/3})$, gets asymptotically worse than the Heisenberg limit. Alternatively, another analysis~\cite{Li2023} under adaptive time step, $t_k = \prod_{0 \leq k'<k} m_{k'}$  with $m_0 = 1$ and $m_{k'>1} \geq 2$ for $k' > 1$, indicates that MODMD can similarly saturate the Heisenberg limit in total runtime. 

In near-term regimes where Trotterization overhead is explicit, we underscore that the associated circuit depth, determined by a $\mathsf{p}$-th order scheme in use, is an especially practical consideration beyond the total gate complexity.
For MODMD, this depth is $\tilde{\mathcal{O}}( \lVert \vec{\kappa}_H \rVert_{1}^{1 + 1/\mathsf{p}} t_{\rm max}^{1 + 1/\mathsf{p}}{\epsilon^{-1/\mathsf{p}}} )  = \tilde{\mathcal{O}} (\epsilon^{-2/3 - 5/3\mathsf{p}} )$ for the maximal runtime $t_{\rm max} = \tilde{\mathcal{O}}(\epsilon^{-2/3})$, \emph{i.e.}, when the initial state has dominant overlap with the set of target eigenstates. For QPE, the Trotterized depth scales as $\Omega(\lVert \vec{\kappa}_H \rVert_{1}^{1+1/\mathsf{p}} \epsilon^{-1-1/\mathsf{p}}) = \Omega(\epsilon^{-1-1/\mathsf{p}})$ by \cref{eq:QPE_depth}. This suggests that, as circuit depth still poses limitations on current devices, higher-order Trotterization for $\mathsf{p} \geq 3$ can yield a smaller relative depth overhead for MODMD. This, albeit, comes at the expense of larger total runtime due to more circuit runs.

\subsubsection{Depolarizing noise channel}
For pre-fault-tolerant implementation, hardware noise need to be considered since device errors can significantly degrade simulation coherence and, thus, the performance of MODMD. Perhaps the simplest and most pervasively modeled device error is the global depolarization channel, $\mathcal{D}_{t}: \rho(0) \mapsto \tilde{\rho}(t)$. This phenomenological model serves a coarse approximation to decoherence effects and captures uniform, unstructured noise.

Under this channel, the ancilla-system composite state evolves toward a maximally mixed state,
\begin{align}
    \tilde{\rho}(t) 
    = e^{- r_{\rm D} \lvert t \rvert}\rho(t) + \frac{1-e^{- r_{\rm D} \lvert t \rvert}}{2N} {\rm Id}_{a} \otimes {\rm Id},
\end{align}
where $r_{\rm D} > 0$ parametrizes the rate of depolarization. We note that the exponential decay degrades the intensity of time-evolved signal where $\Tr[\tilde{\rho}(t) \Gamma_O] = e^{- r_{\rm D} \lvert t \rvert} \Tr[\rho(t)\Gamma_O]$ for any operator $O$. Therefore, the exponential damping is encoded in our multi-observable signal as the real part of the calculated MODMD phases $\tilde{\lambda}_n$ (in contrast to the imaginary part encoding eigenenergy estimates). That is, $r_{\rm D} \approx -\Re \log(\tilde{\lambda_n})/\Dt$. It is worth stating that Fourier-based algorithms~\cite{ding2024quantummultipleeigenvaluegaussian,chan2023algorithmic}, under such damping in the time domain, are prone to corresponding spectral broadening in the energy domain. This can obscure spectral features of interest unless additional deconvolution techniques are applied.

 We further argue that MODMD also performs robustly under specific Pauli noise channels, given a few additional assumptions. We illustrate with the Pauli channel $\mathcal{D}_{t}^{\rm P}: \rho(t) \mapsto \tilde{\rho}(t)$, 
\begin{align}
    \tilde{\rho}(t) = g_{\rm P} \rho(t) + (1 -  g_{\rm P}) P_{\rm noise} \rho(t) P_{\rm noise},
\end{align}
where $g_{\rm P} \in [0, 1]$ controls the weight of the ideal evolution relative to the Pauli error, and $P_{\rm noise}$ acts as a single fixed Pauli string for simplicity. We assume, in particular, that the ancilla qubit does not undergo any bit-flip errors. We 
factor $P_{\rm noise}$ into ancilla and system components: $ P_{\rm noise}^{a} \otimes P_{\rm noise}^{\rm sys}$. Then for any Pauli string $O$,
\begin{align}
    \begin{split}
        \Tr[ P_{\rm noise} & \rho(t) P_{\rm noise} \Gamma_O] = \\
         &\pm \Re \braket{\phi_{\perp}|P_{\rm noise}^{\rm sys}| \phi_{\perp}}
        \braket{\phi_0(t)| O P_{\rm noise}^{\rm sys}|\phi_0 },
    \end{split}
\end{align}
where a sign ambiguity arises here from both the ancillary contribution of $\braket{1|P_{\rm noise}^{a}|1}$ and system contribution due to the (anti)commutation relation between $P_{\rm noise}^{\rm sys}$ and $O$. Observe that if $\ket{\phi_{\perp}}$ is a computational basis state, $P_{\rm noise}$ acts nontrivially only if it is composed entirely of $Z$ gates. Moreover, if $\ket{\phi_{0}}$ is a computational basis state, then
\begin{align}
    \tilde{\gamma}_i(t) &= \Tr[\tilde{\rho}(t) \Gamma_O] = [g_{\rm P} \pm (1 - g_{\rm P})] \gamma_i(t),
\end{align}
where the RHS is non-vanishing for $g_{\rm P} \neq \frac{1}{2}$. In practice, as knowledge about the locality of the system error $P_{\rm noise}^{\rm sys}$ gets learned, one can prepare a more interesting reference state, such as sparse superposition of computational basis states or low-bond-dimension MPS, that is unaffected by the error with high probability. Quantifying robustness under physically realistic Pauli noise models, for example channels induced by continuous-time Lindblad dynamics, is an exciting direction beyond the scope of this work.

Finally, we consider a general channel $\mathcal{D}_{t}^{\omega}: \rho(0) \mapsto \tilde{\rho}(t)$ subsuming the depolarizing and Pauli noise,
\begin{align}
    \tilde{\rho}(t) = g_t \rho(t) + (1-g_t) \omega(t),
\end{align}
where $g_t$ again defines the survival probability of the ideal time-evolved state, and $\omega(t)$ is a noisy background state. Notice that $\Tr[\omega(t) \Gamma_O] \equiv 0$ for a general operator $O$ when $\omega(t)$ is block-diagonal in the ancilla basis. The deviation between exact observables, \emph{e.g.}, $\gamma_i(t)$ from \cref{subsubsec:Trotter_noise}, and their counterparts, $\tilde{\gamma}_i(t)$, depends on specific models of the survival probability $g_t$. However, a universal upper bound on this deviation can be derived using the matrix H\"{o}lder inequality~\cite{bhatia97},
\begin{align}
    \abs{\gamma_i(t) - \tilde{\gamma}_{i}(t)} \leq (1-g_t) \left \lVert \vec{\kappa}_i \right \rVert_1 \left \lVert \rho(t) - \omega(t) \right \rVert_{\ast 1}
\end{align} 
where $\lVert \cdot \rVert_{\ast1}$ denotes the Schatten 1-norm and we also rely on the fact that $\lVert \Gamma_{O_i} \rVert_2 \leq \lVert O_i\ket{\phi_0} \rVert_2 \leq \lVert \vec{\kappa}_i \rVert_1$.


\bigskip


\section{Applications}
\label{sec:applications}

In this section, we detail numerical studies conducted on representative many-body systems from condensed-matter physics and quantum chemistry to demonstrate the efficacy of the MODMD framework. Our numerical calculations precisely follow prescriptions in \cref{sec:MODMD framework}.

\begin{figure*} 
    \centering
    \begin{subfigure}[t]{0.49\textwidth}
        \centering
        \includegraphics[scale = .55]{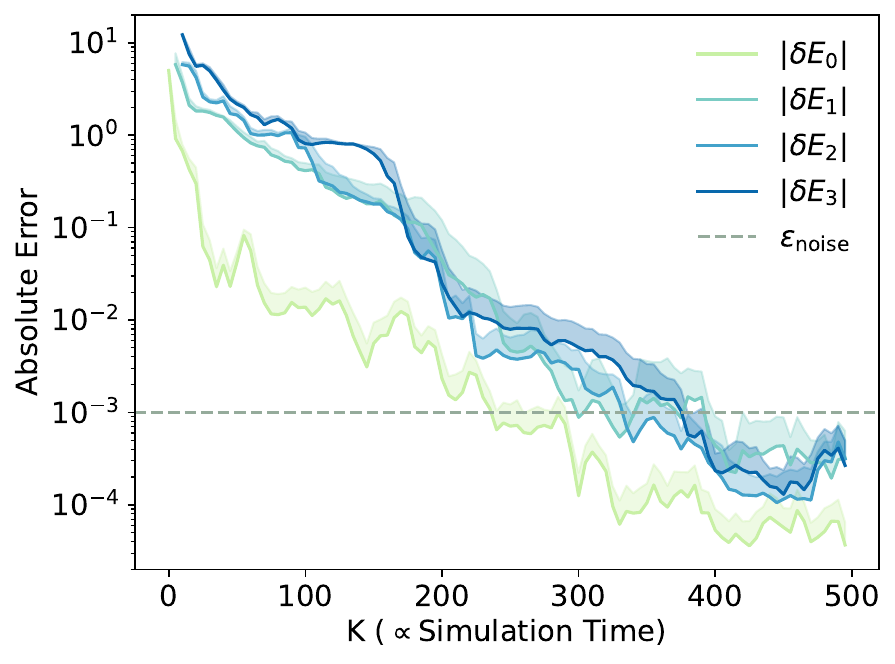}
        \caption{MODMD algorithm}
        \label{fig: TFIM K Convergence MODMD}
    \end{subfigure}%
    ~ 
    \begin{subfigure}[t]{0.49\textwidth}
        \centering
        \includegraphics[scale = .55]{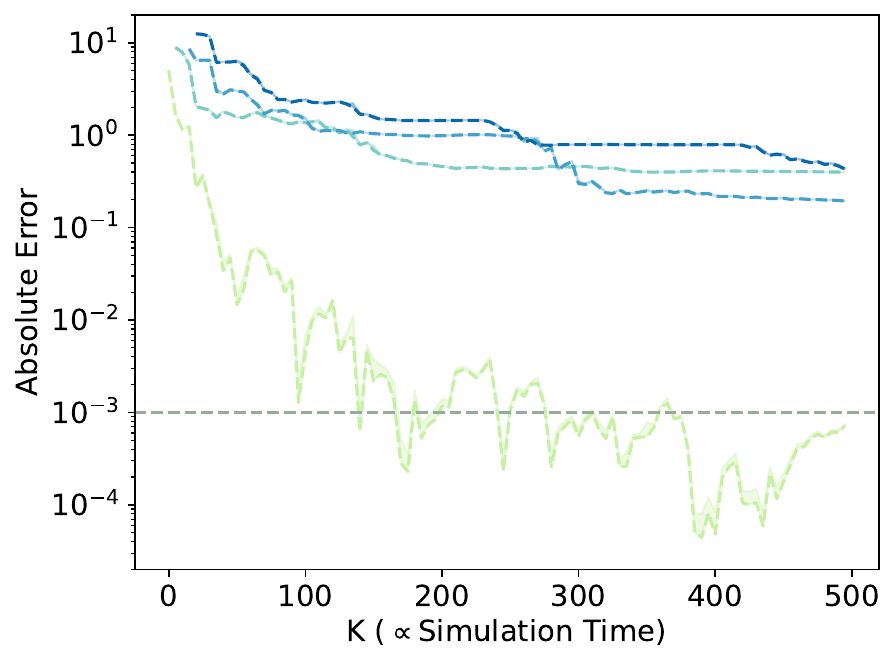}
        \caption{ODMD algorithm}
        \label{fig: TFIM K Convergence ODMD}
    \end{subfigure}
    \caption{Convergence of eigenenergies for the transverse-field Ising model (TFIM). To obtain eigenenergy estimates $\tilde{E}_n$, we fix the (M)ODMD parameters $\frac{K}{d}=\frac{5}{2}$ and $\Tilde{\delta}=10^{-2}$ for constructing and thresholding the pair of data matrices $\mathbf{X},\mathbf{X}' \in \mathbb{R}^{d I \times (K+1)}$. Gaussian $\mathcal N(0,\varepsilon_{\rm noise}^2)$ noise with $\varepsilon_{\rm noise} = 10^{-3}$ is added independently to the real or/and imaginary parts of the matrix elements. The absolute errors, $\lvert \tilde{E}_n - E_n \rvert$, in the first four eigenenergies of the TFIM Hamiltonian are shown with respect to $K$ proportional to the non-dimensional maximal simulation time. The reference state $\ket{\phi_0}$ is an equal superposition of six computational basis states (see \cref{app:reference state}) and we employ a time step of $\Dt \approx 0.08$. Shading above the solid/dashed lines shows the standard deviation across trials for each quantity. For TFIM, the model Hamiltonian is of size $32768 \times 32768$ and the largest linear system in the corresponding MODMD LS problem has size $1400 \times 501$. \textbf{Left}. Absolute errors from the multi-observable (MODMD) algorithm with $I = 7$ distinct observables, where the convergence results are averaged over 20 trials, each involving a Gaussian noise realization and a selection of $I-1$ random observables. \textbf{Right}. Absolute errors from the single-observable (ODMD) algorithm where the convergence results are averaged over 20 trials, each involving a Gaussian noise realization. }
    \label{fig: TFIM K Convergence}
\end{figure*}

\subsection{Condensed-matter physics}
\label{subsec:TFIM}

We examine the convergence of the ground and excited state eigenenergies of the 1D transverse field Ising model (TFIM) with a total of $L=15$ spins and open boundary conditions. The system Hamiltonian is given by
\begin{align}
    H_{\text{TFIM}}=- J\sum_{i=1}^{L-1}  Z_i Z_{i+1} - h \sum_{i=1}^{L} X_i,
    \label{eq:H_TFIM}
\end{align}
for coupling constant $J$ and external field strength $h$.

We first demonstrate the performance of MODMD for TFIM parameters fixed at $J=h=1$, examining how the algorithm behaves with varying number of time steps, or equivalently the maximal simulation time. As discussed in \cref{sec:methods,sec:theory}, increasing the dimensions $d$ and $K$ of the data matrix $\mathbf{X} \in \mathbb{R}^{d I \times (K+1)}$ facilitates convergence of the MODMD eigenvalues $\tilde{\lambda}_{n} = \lvert \tilde{\lambda}_{n} \rvert e^{-i\tilde{E}_{n}\Dt}$ (note that these eigenvalues are not necessarily confined to the unit circle in complex plane) to the eigenphases $\lambda_n = e^{-i E_n \Dt}$ of the Hamiltonian. \cref{fig: TFIM K Convergence} illustrates this convergence for the first $N_{\rm eig} = 4$ eigenenergy estimates. Specifically, we report the absolute error $\lvert \delta E_{n} \rvert = \lvert \tilde{E}_{n} - E_{n} \rvert$ as a function of the dimension $K$, with the ratio $\frac{K}{d} = \frac{5}{2}$ held constant throughout subsequent calculations as this ratio provides near-optimal performance for ODMD~\cite{shen2023estimating}. 

\begin{figure} [t]
\centering
\includegraphics[scale = .55]{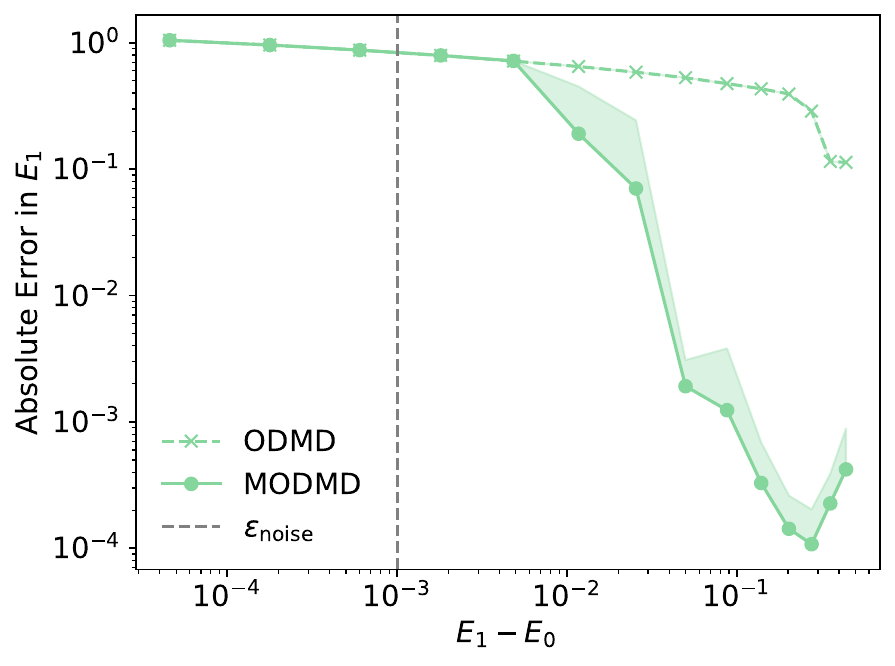}
\caption{Convergence of the first excited state energy of the TFIM. Absolute error in the first excited state energy is shown as a function of the spectral gap between the ground and the first excited state for fixed $K=500$. The vertical dotted line marks the noise level $\varepsilon_{\rm noise} = 10^{-3}$. Convergence results are averaged over 20 trials each involving a Gaussian noise realization and, in the MODMD case, also a selection of $I-1$ random observables. The shading above the solid/dashed lines shows the standard deviation across trials for each quantity.}
\label{fig:TFIM Gap}
\end{figure}

\begin{figure*}[t]
    \centering
    \begin{subfigure}[t]{0.49\textwidth}
        \centering
        \includegraphics[scale = .55]{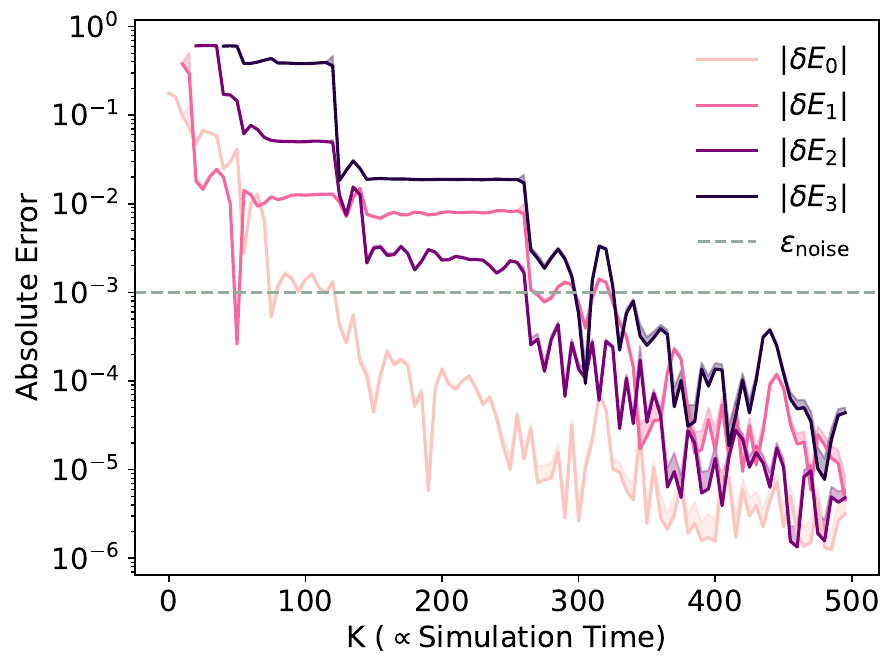}
        \caption{MODMD algorithm}
        \label{fig: LiH K Convergence MODMD}
    \end{subfigure}%
    ~ 
    \begin{subfigure}[t]{0.49\textwidth}
        \centering
        \includegraphics[scale = .55]{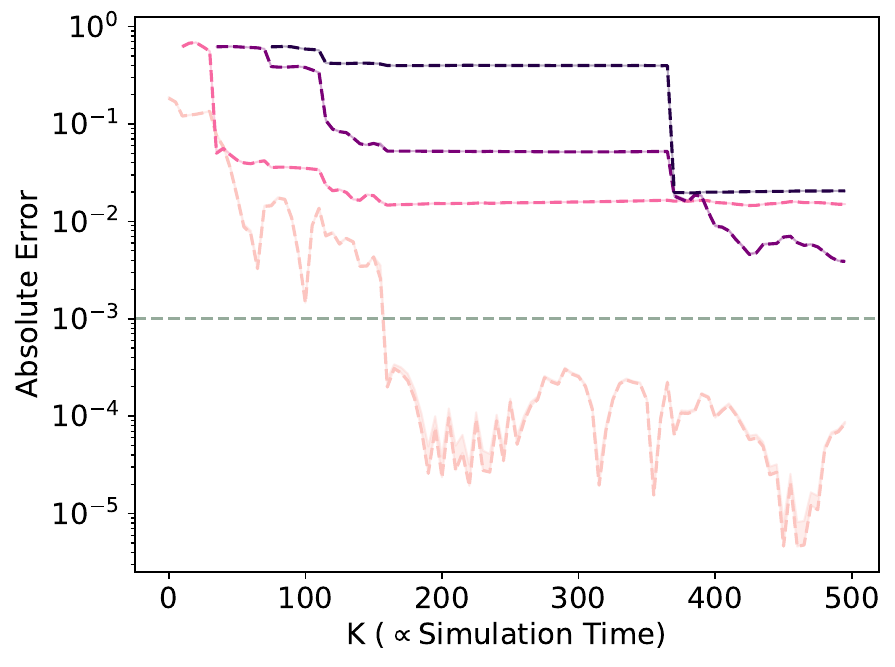}
        \caption{ODMD algorithm}
        \label{fig: LiH K Convergence ODMD}
    \end{subfigure}
    \caption{Convergence of eigenenergies for the lithium hydride (LiH) molecule. To obtain eigenenergy estimates $\tilde{E}_n$, we fix the (M)ODMD parameters $\frac{K}{d}=\frac{5}{2}$ and $\Tilde{\delta}=10^{-2}$ for constructing and thresholding the pair of data matrices $\mathbf{X},\mathbf{X}' \in \mathbb{R}^{d I \times (K+1)}$. Gaussian $\mathcal N(0,\varepsilon_{\rm noise}^2)$ noise with $\varepsilon_{\rm noise} = 10^{-3}$ is added independently to the real or/and imaginary parts of the matrix elements. The reference state $\ket{\phi_0}$ is a superposition of six Slater determinants (see \cref{app:reference state}) and we employ a time step of $\Delta t\approx0.39$. The absolute errors, $\lvert \tilde{E}_n - E_n \rvert$, in the first four eigenenergies of the LiH Hamiltonian are shown with respect to $K$ proportional to the non-dimensional maximal simulation time. Convergence results are averaged over 20 trials of Gaussian noise realizations with shading  above the solid/dashed lines showing the standard deviation across trials for each quantity. For LiH, the model Hamiltonian is of size $1024 \times 1024$ and the largest linear system in the corresponding MODMD LS problem has size $1400 \times 501$. \textbf{Left}. Absolute errors from the multi-observable (MODMD) algorithm with $I = 7$ distinct observables. \textbf{Right}. Absolute errors from the single-observable (ODMD) algorithm.}
    \label{fig: LiH K Convergence}
\end{figure*}

The left panel of \cref{fig: TFIM K Convergence} shows convergence in the multi-observable setting where $I=7$. We assign $O_1=\text{Id}$ as per the convention given in \cref{subsec:selection of hyperparameters}, and sample $\{O_i\}_{i=2}^I$ randomly from the set of 1-local Pauli strings. The initial state $\ket{\phi_0}$ is an equal superposition of six computational basis states. The use of simple random Pauli observables proves effective here since $(1)$ the TFIM does not preserve the Hamming weight (local bit-flip $X$ does not annihilate $\braket{\phi_0|O_i|\phi_0(t)}$) and $(2)$ our reference is a superposition (so
local phase-flip $Z$ does not add a trivial $\pm 1$ overall phase). The reference $\ket{\phi_0}$ contains relatively small but sufficient overlap with the first few Hamiltonian eigenstates, where the squared overlap sums to $\sum_{n=0}^{N_{\rm eig}-1} \lvert \braket{\psi_n|\phi_0} \rvert ^2 \approx 10^{-1}$. This allows for the algorithm to generate an adequate signal without requiring substantial similarity between the reference state and target eigenstates. To stimulate the shadow-induced errors, we additionally introduce Gaussian $\mathcal N(0,\varepsilon_{\rm noise}^2)$ noise with $\varepsilon_{\rm noise}=10^{-3}$ to the multi-observable signal. The absolute energy errors shown are averaged over a total of 20 realizations of both the Gaussian noise and 1-local Pauli observables. 
For comparison, the right panel displays the single-observable convergence from the standard ODMD algorithm as our benchmark. We observe considerably faster convergence in the excited state energies in MODMD for cases where ODMD nearly stagnates. We highlight that the quantum cost, or total number of shots required, is at most $2(K+d)\log(I)\varepsilon_{\rm noise}^{-2}$, only a factor of $2\log(I) < 4$ more than that of ODMD in this case.

The convergence of MODMD naturally divides into two regimes, each defined by a distinct error scaling. The noise level $\varepsilon_{\rm noise}$ essentially determines the crossover between the two convergence regimes. When $\lvert \delta E_{n} \rvert > \varepsilon_{\rm noise}$, we observe an exponentially decaying error typical of the classical subspace methods~\cite{Paige1971_thesis,saadbook,horn_johnson_1985}. Conversely, in the regime where $\lvert \delta E_{n} \rvert < \varepsilon_{\rm noise}$, increasing simulation time leads to slower, algebraic error decay with precision ultimately limited by Heisenberg scaling. This crossover between the exponential and algebraic error decay is shown explicitly within \cref{app:crossover of convergence}. In practice, the onset of the algebraic error behavior can be considerably delayed, for example, via increased and noise-mitigated sampling of classical shadows. Similar analysis for the Heisenberg model, serving as another condensed-matter system, can be found in \cref{sec: Additional Quantum Systems}.


Despite its simplicity, the TFIM is an instructive toy model as it undergoes a quantum phase transition, where the spectral gap between the first two eigenstates can be systematically tuned by varying the $\frac{h}{J}$ ratio. In the thermodynamic limit $L \rightarrow \infty$, the gap closes at $\frac{h}{J} = 1$ and increases monotonically with $\frac{h}{J}$. To investigate the gap dependence near a phase transition, we demonstrate the difference in performance between the single-observable (ODMD) and multi-observable (MODMD) algorithms in the presence of near-degenerate target energies. In \cref{fig:TFIM Gap}, we focus on comparing the error $\lvert \delta E_1 \rvert$ in the first excited energy against the gap $E_1-E_0$. The convergence results indicate that, in our multi-observable approach, the first excited state energy $E_{1}$ can be accurately estimated when the noise level is slightly
smaller than the spectral gap. In contrast, ODMD requires a visibly higher gap-to-noise ratio to achieve a comparable accuracy. This highlights a significant improvement of MODMD in distinguishing near-degenerate eigenstates.

\subsection{Quantum chemistry}

Electronic structure calculation is a fundamental problem in quantum chemistry. Here we evaluate performance of the MODMD algorithm for molecular Hamiltonians, whose second quantization can be efficiently implemented on the quantum computer. A minimal STO-3G basis set is used to construct the finite-dimensional Hamiltonian, which is then transformed into qubit form via the parity fermion-to-qubit mapping~\cite{Seeley_2012}.



Moreover, we construct observables based on the Pauli representation of the Hamiltonian as per \cref{eqn: hamiltonian_pauli_rep}, and deterministically select a subset of $I=7$ Pauli operators $P_{\nu}$ (where $O_1=\text{Id}$) with medium-magnitude weights $\kappa_{\nu}$ to maximize spectral independence among observables. The reference state $\ket{\phi_0}$ is prepared to be a superposition of six Slater determinants. These effective single-particle states map simply to computational basis states derived from a mean-field calculation. This choice ensures both a nonzero overlap with the target eigenstates and viability for experimental preparation.

We illustrate MODMD for the lithium hydride (LiH) molecule, where exact eigenenergies can be calculated by full diagonalization of the electronic Hamiltonian. \cref{fig: LiH K Convergence} shows the convergence of the first $N_{\rm eig} =4$ eigenenergy estimates for LiH at its equilibrium bond length of $1.59${\AA}. Similar to the TFIM results from \cref{fig: TFIM K Convergence}, \cref{fig: LiH K Convergence} exhibits a characteristic exponential-to-algebraic crossover in the error decay, with the transition occurring near noise level $\varepsilon_{\rm noise}$ (see \cref{app:crossover of convergence}). The plateaus observed for the higher energy levels at intermediate values of $K$ can be attributed to the anomalous convergence to different but higher-lying eigenenergies, which we verified numerically. As we approach the interior of the Hamiltonian spectrum, MODMD can resolve the higher excited state energies for sufficiently large $K$, whereas
ODMD stagnates. Results of ODMD and MODMD on additional chemical systems, $\text{BeH}_2$ and $\text{N}_2$, can be found in \cref{sec: Additional Quantum Systems}.

\begin{figure} [t]
    \centering
    \includegraphics[scale=.55]{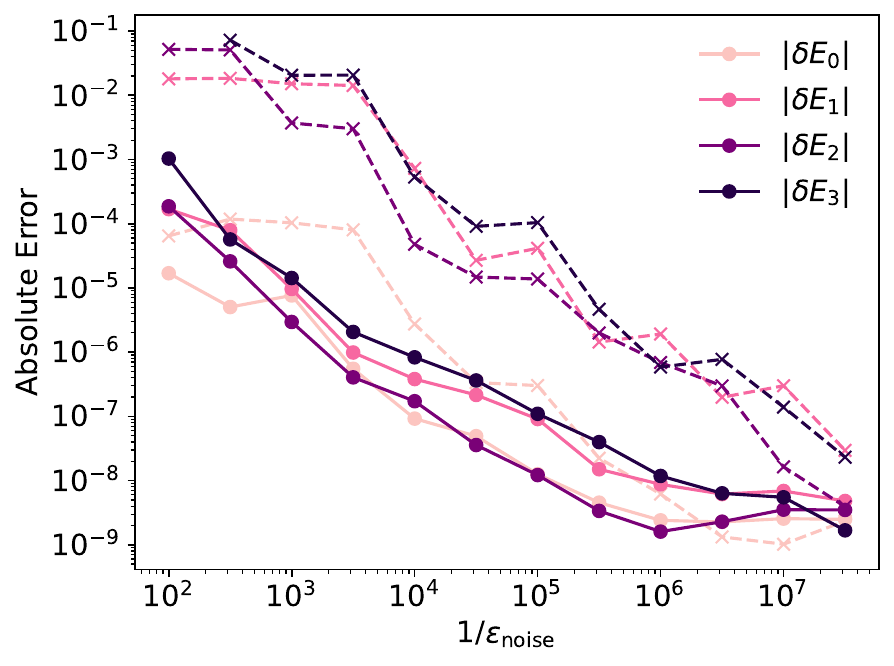}
    \caption{Convergence of eigenergies for LiH under varying noise level. Absolute errors in the four lowest eigenenergies of the LiH Hamiltonian are plotted against the noise level $\varepsilon_{\rm noise}$ for fixed $K=500$ and $\Tilde{\delta} = 10 \varepsilon_{\rm noise}$. All convergence results reflect an average over 20 trials of Gaussian noise realizations. The solid lines correspond to the multi-observable (MODMD) algorithm with $I = 7$ distinct observables and the dashed lines to the single-observable (ODMD) algorithm. }
    \label{fig: LiH Noise}
\end{figure}

In \cref{fig: LiH Noise} we assess the noise robustness of the MODMD algorithm in our LiH example. Adopting the same hyperparameters as used in \cref{fig: LiH K Convergence}, we now fix $K=500$ and vary the noise level $\varepsilon_{\rm noise}$. Importantly, we maintain the SVD threshold $\Tilde{\delta} = 10 \varepsilon_{\rm noise}$ at a consistently larger value. We observe a power law scaling of the absolute error with respect to the noise level, suggesting that a conservative truncation strategy can be used to protect the actual signal from noise across a wide range of noise magnitudes. By comparison, the ODMD algorithm yields larger errors for any given $\varepsilon_{\rm noise}$. Furthermore, reducing noise only significantly improves the ODMD performance when $\varepsilon_{\rm noise}$ reaches relatively small values. In other words, ODMD requires a more aggressive truncation $\tilde{\delta}$, which may cause a serious loss of the signal information and thereby slower convergence.

\subsection{Comparison with alternative approaches}
\subsubsection{QMEGS}
\label{subsec:QMEGS}
To evaluate the practical performance of MODMD, we compare its convergence with alternative state-of-the-art real-time methods. We first benchmark against Quantum Multiple Eigenvalue filtered Search (QMEGS)~\cite{ding2024quantummultipleeigenvaluegaussian} as a leading signal processing algorithm. QMEGS estimates multiple Hamiltonian eigenenergies by reconstructing a Gaussian-filtered signal,
\begin{align}
    &\mathcal{G}(E,O) = \left \lvert \frac{1}{K} \sum_{k=0}^{K-1} \hat{s}_k e^{i E t_k} \right \rvert, \nonumber \\ 
    &\hat{s}_k \in \{ \pm 1 \pm i\}, \quad \mathbb{E}(\hat{s}_k) = \braket{\phi_0 | Oe^{-i H t_k} | \phi_0},
    \label{eq:qmegs}
\end{align}
where the times $\{t_{k}\}_k$ are sampled i.i.d. from a truncated Gaussian distribution. Peaks of $\mathcal{G}(E,O)$ correspond to eigenvalues of $H$, reflecting a Gaussian-broadened version of the underlying eigenstate overlap $\lvert \braket{\psi_n|\phi_0}\rvert^2$. QMEGS retrieves the dominant energies upon locating the peaks using a classical search-and-block optimization strategy, with quantum data collected from single-shot Hadamard-test measurements for the observable $O$. Here we take the observable to be unitary (\textit{e.g.},  $O = {\rm Id}$ or Pauli strings), consistent with the Hadamard-test setting.

Notably, the QMEGS algorithm differs from MODMD in its theoretical guarantee, which depends on the choice of a suitable reference state, although the algorithm itself can be initialized with arbitrary reference state. QMEGS reaches Heisenberg-limited scaling, under the assumption that the reference state satisfies the \textit{sufficiently dominant condition}, namely that there is a set of eigenindices $\mathcal N_{\rm eig}$ so that
\begin{align}
    \min_{n \in \mathcal N_{\rm eig}}|\langle \psi_n | \phi_0\rangle|^2 > \sum_{n \notin \mathcal N_{\rm eig}} |\langle \psi_n |\phi_0 \rangle|^2.
    \label{eq:dominance_condition}
\end{align}
Additionally, while MODMD can resolve the lowest-lying eigenenergies, QMEGS most naturally identifies energies with the largest eigenstate overlaps. To facilitate a direct comparison between the methods, we use an initial state $\ket{\varphi_0}$ for which the first $N_{\rm eig} = 4$ eigenstates are weighted preferentially but \cref{eq:dominance_condition} is not fulfilled. Such an initial state is constructed in \cref{app:reference state}.

Specifically, we perform similar simulations to those in \cref{subsec:TFIM}, analyzing the error in $\lvert \mathcal{N}_{\rm eig} \rvert$ eigenenergies of the TFIM Hamiltonian as a function of simulation time. To maintain consistency, the hyperparameters of the two algorithms are adjusted so that the total number of shots remains the same for each method. The size of QMEGS dataset $K_{\rm QMEGS}$, given MODMD parameters $K_{\rm MODMD}$, $d_{\rm MODMD}$, $I$, as well as shot noise level $\varepsilon_{\rm noise}$, is calibrated as $K_{\rm QMEGS} = (K_{\rm MODMD} + d_{\rm MODMD})\log (I)\varepsilon_{\rm noise}^{-2}$. For any maximal MODMD evolution time $t_{\max}=(K+d)\Dt$, we choose the Gaussian width and truncation in QMEGS to match $t_{\max}$. As in \cref{subsec:TFIM}, the observables $O_i$ in each MODMD trial consist of $O_1=\text{Id}$ and six additional candidates $O_{i \geq 2}$ drawn from the $1$-local Pauli strings. We account for the fact that QMEGS uses a single observable by running the algorithm to reconstruct $\mathcal{G}(E,O_i)$ for each $O_i$ in the MODMD observable set. Focusing on low-lying eigenenergies and noting that additional observables can only improve or leave unchanged their visibility, we take the minimum error $\displaystyle|\delta E_n^{\rm{QMEGS}}|=\min_{1\leq i\leq I}|\delta E_n^{\rm{QMEGS}}(O_i)|$ across observables. For demonstration, we focus on a 12-qubit system with a simulated noise level of $\varepsilon_{\rm noise}=10^{-5/2}$ and MODMD threshold $\tilde\delta=10^{-3/2}$.

\begin{figure} [t]
    \centering
    \includegraphics[scale=.55]{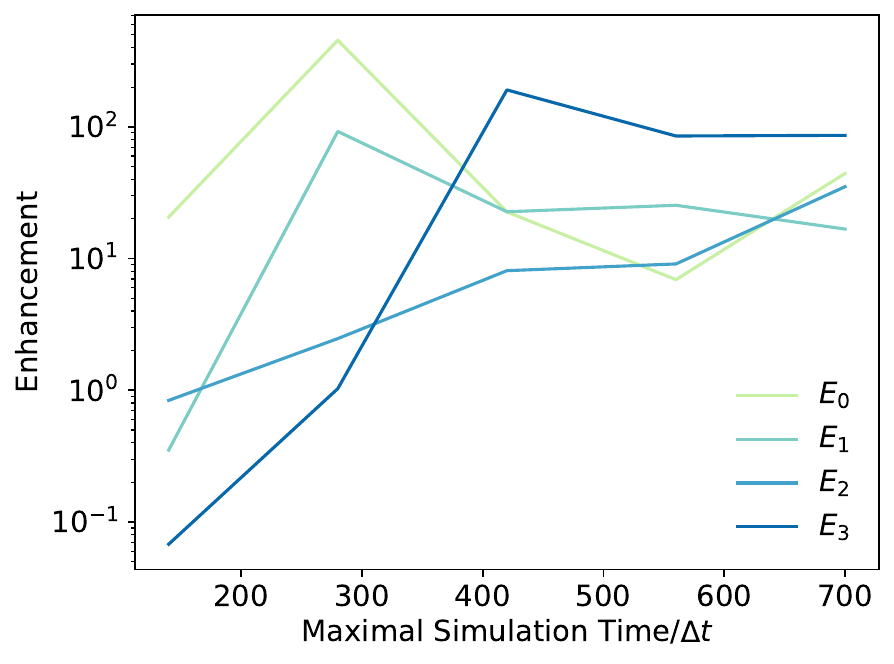}
    \caption{Performance comparison of MODMD and QMEGS on the transverse-field Ising model (TFIM). Relative performance enhancement, $\left \lvert \frac{\delta E_n^{\rm QMEGS}}{\delta E_{n}^{\rm MODMD}} \right \rvert$, is computed from absolute errors in the first four eigenenergies (obtained by QMEGS and MODMD with $I =7$ distinct observables), and is shown with respect to $K_{\rm MODMD}$ proportional to the non-dimensional maximal simulation time. MODMD and QMEGS errors are averaged over 5 trials: each employing a common observable set with $O_1 = {\rm Id}$ and $I-1$ observables randomly selected from a pool of Pauli strings. We search for 20 peaks in QMEGS, while the four lowest-lying eigenstates have the 1st, 7th, 10th, and 5th highest initial state overlap respectively. Note that the absolute errors for both algorithms do not reach the same order of magnitude as $\epsilon_{\rm{noise}}$ until the larger values of maximal simulation time in this plot.}
    \label{fig: QMEGS Comparison}
\end{figure}

The relative enhancement of MODMD over QMEGS, $\left \lvert \frac{\delta E_n^{\rm QMEGS}}{\delta E_{n}^{\rm MODMD}} \right \rvert$, as a function of the non-dimensional maximal simulation time is shown in \cref{fig: QMEGS Comparison}. Beyond small simulation time, MODMD consistently outperforms QMEGS in the convergence of all four lowest-lying eigenenergies. The QMEGS hyperparameters $\alpha_{\rm QMEGS}$ and $q_{\rm QMEGS}$~\cite{ding2024quantummultipleeigenvaluegaussian} are tuned to optimize performance, with details provided in \cref{app: QMEGS sweep}.

\subsubsection{VQPE and UVQPE}
\label{subsec:(U)VQPE}
As a further comparison, we examine the performance of MODMD relative to VQPE~\cite{klymko2022real,shen2022real}, a leading real-time subspace method. We benchmark both approaches using the TFIM Hamiltonian under a similar setting described in \cref{subsec:TFIM}. Recall our previous initial state $\ket{\phi_0}$ was chosen so that $\sum_{n=0}^{3} \lvert \braket{\psi_n|\phi_0} \rvert ^2 \approx 10^{-1}$, thereby situating the system well beyond the \emph{sufficiently dominant} regime,  
where signal processing methods are effective. To pursue our comparative analysis under general instance, here we consider a less structured (yet classically tractable) initial state $\ket{\varphi_0}$ specified in \cref{app:reference state}. This state does not preferentially overlap with low-lying eigenstates, allowing us to more rigorously assess the performance of MODMD and VQPE in the absence of spectral bias.

 VQPE solves the subspace eigenvalue problem, 
\begin{align}
    \mathbf{H} \mathbf{\Psi} = \tilde{E}\mathbf{S} \mathbf{\Psi},
\end{align}
where the Hamiltonian and overlap matrix elements, 
\begin{align}
    \mathbf{H}_{jk}= \braket{\phi_j|H|\phi_k}, \quad \mathbf{S}_{jk}= \braket{\phi_j|\phi_k},
\end{align}
are obtained by projecting the full many-body eigenvalue problem onto the subspace spanned by the nonorthogonal real-time basis $\{ \ket{\phi_k} = e^{-iHk\Dt}\ket{\phi_0} \}_{k}$. Notice that these matrix elements correspond to two types of observables, $O = H$ and $O = {\rm Id}$. Their real and imaginary parts are measured separately in VQPE using the Hadamard test, which requires sampling at least four distinct circuits per time step (even assuming access to block encoding of the problem Hamiltonian $H$). Importantly, VQPE remains effective  despite imperfections in time evolution~\cite{Kirby2024analysisofquantum}, such as those introduced by Trotterization,  since the algorithm ultimately projects the exact Hamiltonian onto subspace spanned by some prepared states -- regardless of whether they are exactly or approximately time-evolved.

Unitary VQPE (UVQPE), as a direct algorithmic twin, solves an isospectral subspace eigenvalue problem, $\mathbf{U} \mathbf{\Psi} = e^{- i \tilde{E} \Dt} \mathbf{S} \mathbf{\Psi}$, where $\mathbf{U}_{jk} = \braket{\phi_j|e^{-iH\Dt}|\phi_k} \equiv \mathbf{U}_{jk+1}$ define the matrix elements of projected time-evolution operator. UVQPE mitigates resource overhead by circumventing the need to measure the Hamiltonian matrix elements; on the other hand, like MODMD, it still requires sufficiently accurate time evolution to extract spectral information. We view UVQPE as effective variant of VQPE, bridging and encompassing state-of-the-art subspace approaches.

\begin{figure} [t]
    \centering
    \includegraphics[scale=.55]{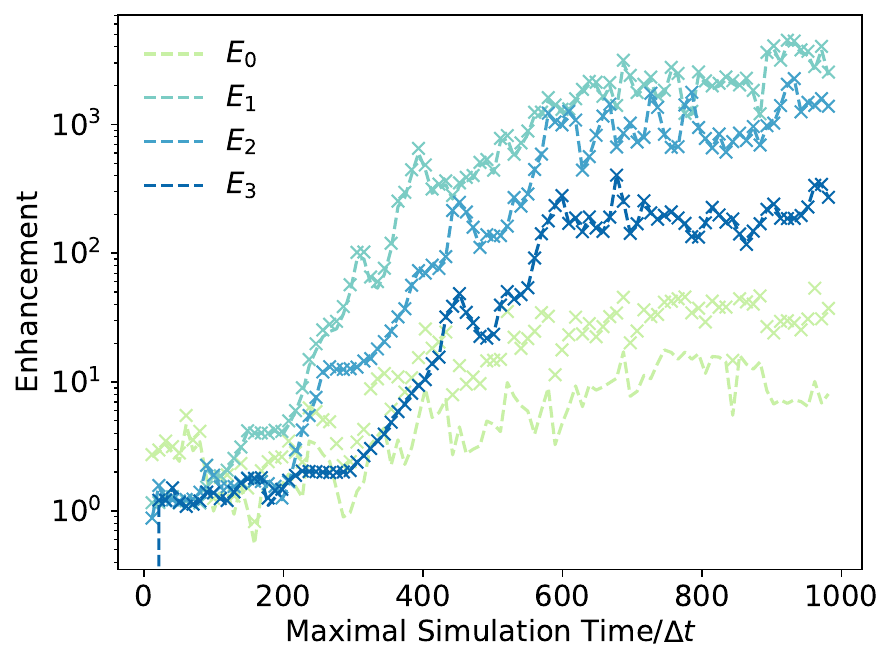}
    \caption{Performance comparison of MODMD and (U)VQPE on the transverse-field Ising model (TFIM). Relative performance enhancement, $\left \lvert \frac{\delta E_n^{\rm (U)VQPE}}{\delta E_{n}^{\rm MODMD}} \right \rvert$, is computed from absolute errors in the first four eigenenergies (obtained by (U)VQPE and MODMD with $I = 7$ distinct observables), and is shown with respect to $K$ proportional to the non-dimensional maximal simulation time. MODMD errors are averaged over 20 trials, each involving a Gaussian noise realization and a set of $I-1$ observables randomly selected from a pool of candidate operators. Cross markers represent comparison with UVQPE while dash lines represent comparison with VQPE.}
    \label{fig: VQPE Comparison}
\end{figure}

In \cref{fig: VQPE Comparison}, we compare (U)VQPE and MODMD for the first $N_{\rm eig}=4$ eigenenergy estimates. In particular, we show the relative enhancement, $\left \lvert \frac{\delta E_n^{\rm (U)VQPE}}{\delta E_{n}^{\rm MODMD}} \right \rvert$, as a function of the maximal simulation time. It is worth noting that a $d \times (K+1)$ linear system in MODMD corresponds to a generalized eigenvalue problem of size $(d+K+1) \times (d+K+1)$ in (U)VQPE.

When the simulation is stopped at $K=500$, MODMD consistently achieves lower energy errors than (U)VQPE, often by one to three orders of magnitude. The advantage of MODMD becomes evident as the simulation extends, where (U)VQPE struggles to resolve eigenstates reliably. Leveraging a richer observable signal with essentially no increase in the measurement overhead per time step (note that $\log(I) \sim 4$ for $I \sim 10^2$, matching a minimal factor required by VQPE), MODMD delivers accurate energy estimates with significantly reduced quantum costs.

\subsection{Prediction of longer-time dynamics}
\label{subsec:predict dynamics}
To demonstrate the utility of the MODMD framework beyond eigenenergy estimation, we present additional numerical simulations, where the system matrix $A$ is constructed from a set of observable measurements and then used to forecast future dynamics. 

Following \cref{eq:multi_signals} and \cref{eq:MODMD_sys}, we measure the multi-observable expectations $\vec s_k=\vec s(k\Delta t)$ for $k=0,1,\ldots, k^{\ast}$ and arrange these real-time snapshots to construct $\mathbf{X}$ and $\mathbf{X}'$. Here, $k^{\ast}$ is a hyperparameter controlling how many snapshots are collected -- and therefore the extent of past information exploited for prediction. After constructing the system matrix $A$, we propagate the dynamics through the MODMD prescription $\vec s_{k+1}\approx A \vec s_k$, as is outlined in \cref{subsec:Hamiltonian properties}. This allows us to predict $\vec s_k$ for future steps $k= k^*+1, k^*+2, \ldots$ by computing $A^{k-k^*}\mathbf{X}'$. 

In \cref{fig: Dynamics Prediction}, we visualize real part of the $i$th observable component, \emph{i.e.}, $\Re \langle\phi_0|O_i|\phi_0(k\Delta t)\rangle$, up to 200 time steps beyond $k^{\ast}$. For transverse-field Ising and lithium hydride Hamiltonians, we observe that the predictive accuracies improve with $k^*$ in both cases. This aligns with our basic intuition that longer observation windows grant ability to predict dynamics further into the future. 

 Though we only predict the dynamics for the particular sets of observables used in \cref{sec:applications}, one could perform extrapolation of other quantities of interest depending on the system and dynamics in question. In support of such broader applicability, we also investigate the approximate eigenstates extracted by MODMD in Appendix~\ref{app:eigenstate_convg}, where our method exhibits reliable performance -- suggesting its competence in capturing a wide range of low-energy properties. 

\begin{figure*} [h!]
    \centering
    \begin{subfigure}{\textwidth}
        \centering
        \includegraphics[width = \linewidth]{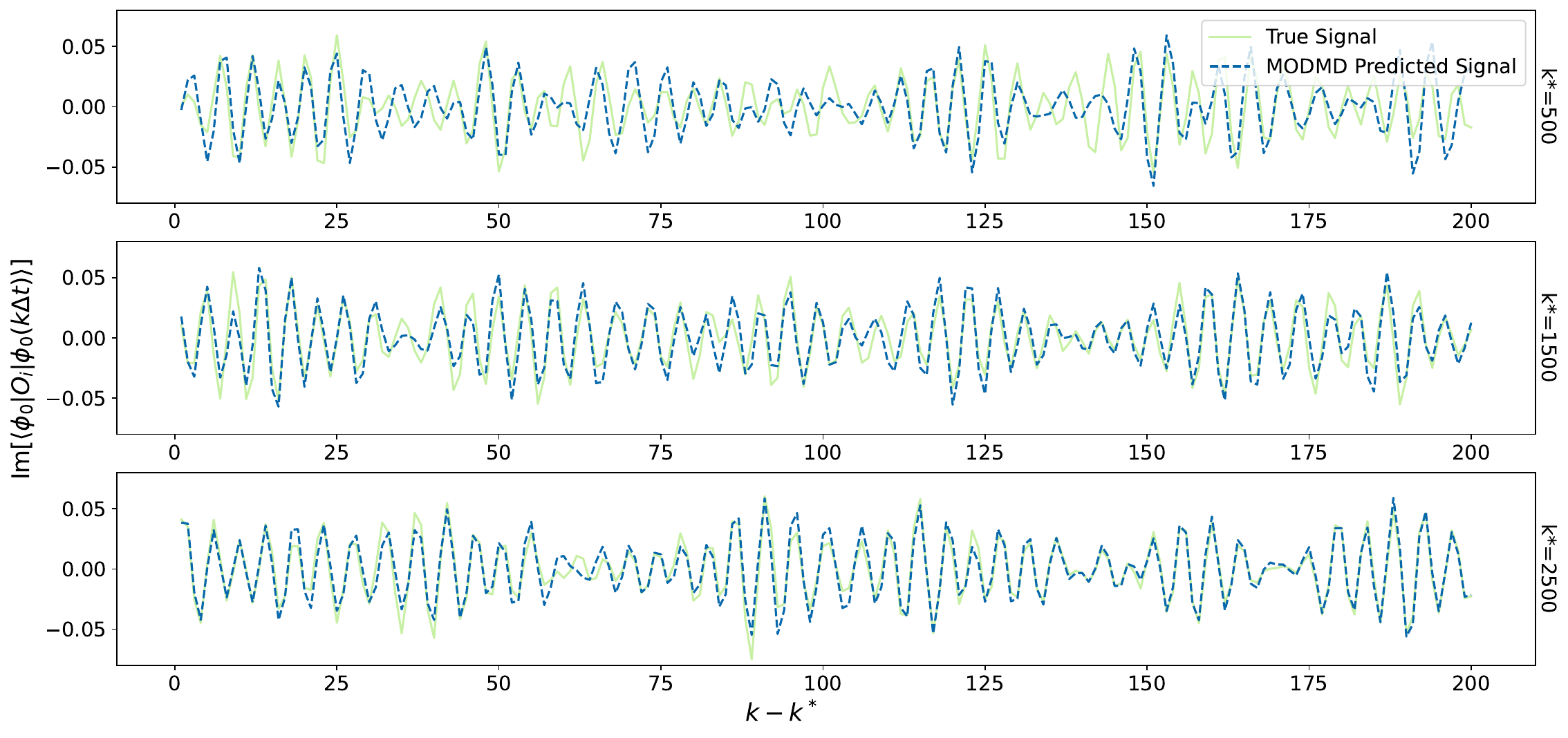}
        \caption{Dynamics of TFIM}
        \label{fig: TFIM Dynamics Prediction}
    \end{subfigure}%
\end{figure*}
\begin{figure*} [h!]
\ContinuedFloat
    \begin{subfigure}{\textwidth}
        \centering
        \includegraphics[width = \linewidth]{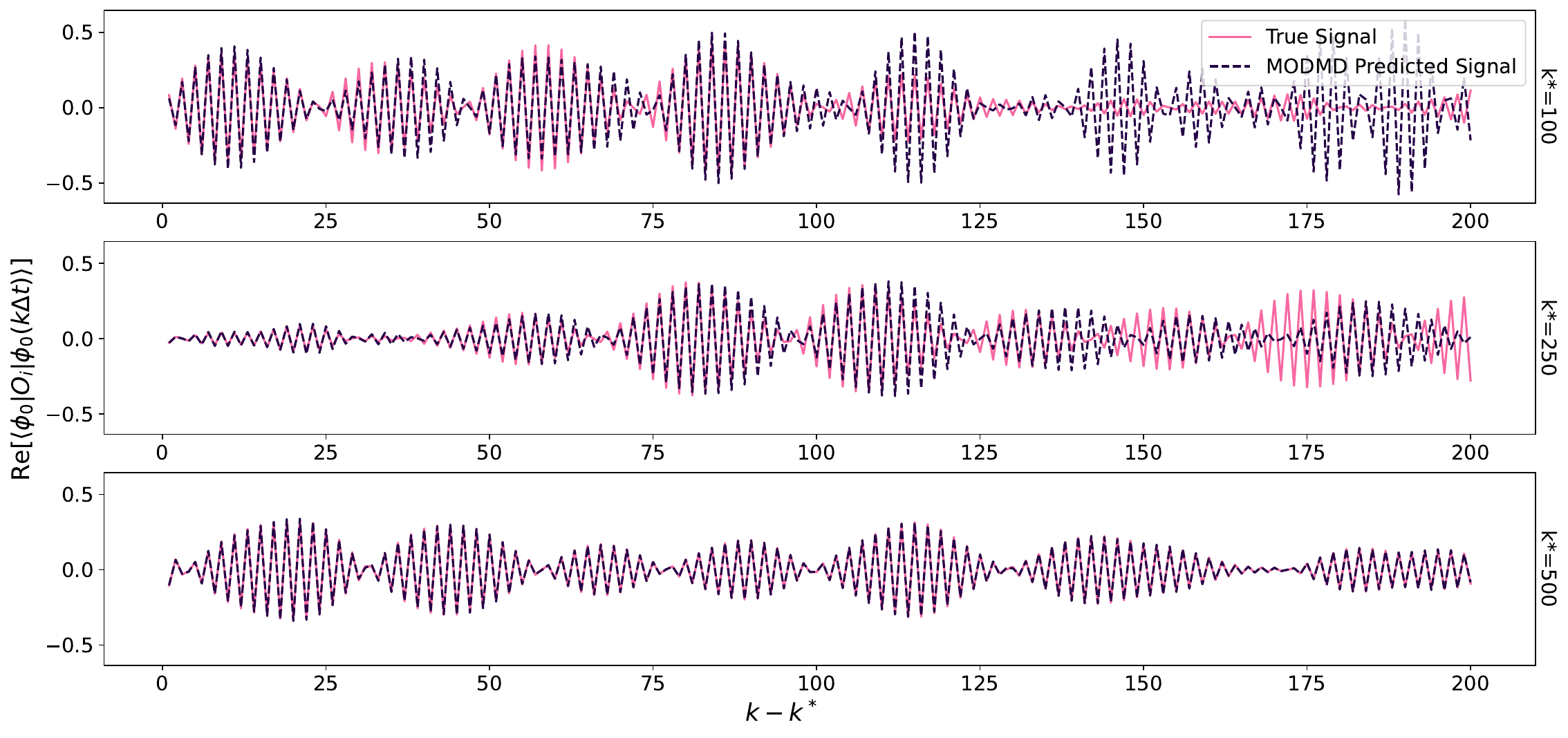}
        \caption{Dynamics of LiH}
        \label{fig: LiH Dynamics Prediction}
    \end{subfigure}
    \caption{Predicted observable dynamics from MODMD. For different values of $k^*$ (as specified on the right side of each panel), we construct the system matrix $A$ from the multi-observable snapshots $\vec s_k=\vec s(k\Delta t)$ for $k=0,1,\ldots, k^{\ast}$. We then predict $\vec s_k$ for $k=k^*+1,k^*+2,\dots,k^*+200$ by computing $\vec s_k=A^{k-k^*}\mathbf{X'}$ and plot the $i$th element of predicted $\vec s_k$, \emph{i.e.}, $\Re \langle\phi_0|O_i|\phi_0(k\Delta t)\rangle$. The figure displays the results for $i=5$, though similar convergence behavior is observed across all observables. The MODMD hyperparameters $\frac{K}{d}=\frac{5}{2}$ and $\Tilde{\delta}=10^{-2}$ are fixed for constructing and thresholding the pair of data matrices $\mathbf{X},\mathbf{X}' \in \mathbb{R}^{d I \times (K+1)}$. Gaussian $\mathcal N(0,\varepsilon_{\rm noise}^2)$ noise with $\varepsilon_{\rm noise} = 10^{-3}$ is added independently to the matrix elements. The reference states, time steps, and observable sets are the same as those used in \cref{sec:applications}. The
    results are averaged over 20 trials, each involving a Gaussian noise realization. \textbf{Top}. Dynamics predictions for the transverse-field Ising model (TFIM) Hamiltonian. \textbf{Bottom}. Dynamics predictions for the lithium hydride (LiH) Hamiltonian. }
    \label{fig: Dynamics Prediction}
\end{figure*}

\bigskip

\section{Conclusion}
In this work, we developed a hybrid quantum-classical measurement-driven framework for effectively extracting information about the low-energy eigenspaces of quantum many-body systems. Our novel MODMD approach leverages real-time evolution on quantum hardware and classically unravels multi-dimensional signals, composed of real-time observables, from a limited number of randomized measurements. The simultaneous prediction of many observables leads to accurate estimates of eigenenergies and shallower circuits with shorter evolution time. We explored the theoretical underpinnings of MODMD, which exponentially suppresses spectral error in the low-noise regime. We numerically demonstrated its rapid convergence in the presence of perturbative noise
using examples from condensed matter physics and quantum chemistry.

Compared to state-of-the-art real-time approaches, we highlight the unique strengths of our method in addition to its reliable convergence and noise resilience. To our best knowledge, MODMD is among the most resource-efficient for generating real-time signals. This is because (1) we evolve a single reference state for a duration shorter than required by single-observable approaches, where the reduction in the simulation time becomes more substantial as the number of observables included in the signal subspace increases, and (2) the reference state does not have to possess large overlaps with the low energy eigenstates of interest. Furthermore, our classical post-processing consists of solving a simple least-squares problem followed by a standard eigenvalue problem, which is ansatz-free and thus circumvents an exponential growth in optimization costs, whether quantum or classical, associated with the number of desired eigenenergies. 

In fact, our MODMD framework is capable of retrieving ground and excited state properties beyond eigenenergy levels, demonstrating an extensive and timely application of the low-rank shadow. This versatility suggests that similar ideas may extend to dynamical properties as well. Non-equilibrium behavior is particularly intriguing in this context, and exploring such directions can further extend the scope of our multi-observable framework.


\section*{Acknowledgements}
This work was funded by the U.S. Department of Energy (DOE) under Contract No.~DE-AC02-05CH11231, through the Office of Science, Office of Advanced Scientific Computing Research (ASCR) Exploratory Research for Extreme-Scale Science (YS, KK, DC, RVB).
This research used resources of the National Energy Research Scientific Computing Center (NERSC), a U.S. Department of Energy Office of Science User Facility located at Lawrence Berkeley National Laboratory, operated under Contract No. DE-AC02-05CH11231. HYH would like to acknowledge the support from the Harvard Quantum Initiative.
SFY and HYH would like to acknowledge funding from the DOE through the QUACQ center (DE-SC0025572).

\section*{Data availability}
Our numerical calculations that support the findings of this article are openly available through the GitHub repository \faGithub\href{https://github.com/alexbuzalicse/modmd}{MODMD}~\cite{modmd_github}.

\clearpage
\onecolumngrid

\appendix
\section*{Appendix}


\section{Shadow under locally scrambling dynamics}

Here we adopt a dynamical perspective on the efficient generation of the classical shadow, which introduces random scrambling via real-time evolution under a disordered local Hamiltonian $H_{s}(t)$. For concreteness, we focus on quantum spin glass, in particular the disordered transverse field Ising model (TFIM) with Hamiltonian,
\begin{align}
    H_s(\tau) = - \sum_{\langle i,j \rangle}^{L} J_{ij}(\tau) X_i X_{j} - h(\tau) \sum_{i=1}^{L} Z_i,
    \label{eq:dTFIM}
\end{align}
where $J_{ij}$ and $h$ set the spin-spin coupling and external field strength respectively. We assume for simplicity that $J_{ij} = \sum_{k} J_{ij,k} \mathds{1}_{[\tau_k, \tau_{k+1})}(\tau) $ and $h = \sum_{k} h_k \mathds{1}_{[\tau_k, \tau_{k+1})}(\tau)$ remain piecewise constant in time, with $\mathds{1}_{[\tau_k, \tau_{k+1}]}$ being an indicator function of the $k$th-step interval. To incorporate scrambling disorder, we employ quenched random interactions, \textit{e.g.}, $J_{ij,k} \sim \mathcal{N}(0, \sigma_J^2)$ and $h_k \sim \mathcal{N}(0, \sigma_h^2)$. Thus under the time evolution $U_s = \prod_{k= K_s}^{0} e^{-iH_s(\tau_k) (\tau_{k+1}- \tau_k) }$, local quantum information gets scrambled and the resulting entanglement produced by the disorder dynamics facilitates the recovery of $\rho_\Phi(t)$. We remark that the number of scrambling time steps $K_s$ serves the role analogous to the circuit depth when considering Clifford gates \cite{Nakata_spinglass}. For any integer $T \geq 0$, $\lim_{K_s \rightarrow \infty} \mathcal{U}(\{ \tau_k \}_{0 \leq k \leq K_s}) $ is a $T$-design, whereby its first $T$ statistical moments align with those of the Haar measure. To simulate long-time dynamics, we draw upon our recent tool of algebraic circuit compression \cite{pra_cd2022,siam_cd2022}, particularly suited to a disordered TFIM Hamiltonian $H_s(\tau)$ in \cref{eq:dTFIM}, to keep the effective depth of the time evolution $U_s$ shallow. Remarkably, the depth post compression should exhibit no dependence on the maximal runtime $\tau_{K_s}$, therefore allowing a significantly more efficient exploration of the Haar limit.

\section{Observable dynamic mode decomposition}
\label{app:observable dynamic mode decomposition}

The standard dynamic mode decomposition (DMD), originally developed in the field of numerical fluid dynamics, is a measurement-driven approximation for the temporal progression of a classical dynamical system \cite{Mezi2004ComparisonOS,Mezi2005SpectralPO,Rowley2009SpectralAO,schmid_2010,Mezi2013AnalysisOF}. Specifically, DMD samples the system snapshots at regular time intervals $\Delta t$ and uses them to construct an efficient representation of the full dynamical trajectory. For simplicity, we consider a system whose $N$-dimensional state manifold is $\mathbb{C}^N$.
The optimal linear approximation for the discretized time step $k \mapsto k+1$ is expressed as the least-squares (LS) relation,
\begin{align}
    \boldsymbol{\phi}_{k+1} \lseq{} A_k  \boldsymbol{\phi}_k ,
\end{align}
where $\boldsymbol{\phi}_k \in \mathbb{C}^{N}$ specifies the system state at time $k\Dt$ and $A_k \in \mathbb{C}^{N \times N}$ is the system matrix, \textit{i.e.}, the linear operator that minimizes the residual $\Vert \boldsymbol{\phi}_{k+1} - A_k \boldsymbol{\phi}_k \Vert_2$ to yield the LS relation above. Similarly, the optimal linear approximation for a sequence of successive snapshots $k = 0, 1, \ldots, K+1$ can be determined by the solution,
\begin{align}
     \begin{bmatrix}
        \vertbar & \vertbar & & \vertbar\\
        \boldsymbol{\phi}_{1} & \boldsymbol{\phi}_{2} & \cdots & \boldsymbol{\phi}_{K+1} \\
        \vertbar & \vertbar & & \vertbar\\
    \end{bmatrix} \lseq{} A \begin{bmatrix}
        \vertbar & \vertbar & & \vertbar\\
        \boldsymbol{\phi}_{0} & \boldsymbol{\phi}_{1} & \cdots & \boldsymbol{\phi}_{K} \\
        \vertbar & \vertbar & & \vertbar\\
    \end{bmatrix},
    \label{eq:dmd_lls}
\end{align}
where the system matrix $A$ minimizes the sum of squared residuals over the length-$K$ sequence. The linear flow described by \cref{eq:dmd_lls} naturally generates approximate dynamics $\boldsymbol{\phi}_{\rm DMD}(t) = A^{\frac{t}{\Delta t}} \boldsymbol{\phi}_0$ governed by eigenmodes of $A$. DMD-based approaches can be remarkably effective despite their formal simplicity, since they are rooted in the general Koopman operator theory developed to describe the behavior of general (non)linear dynamical systems \cite{Koopman1931HamiltonianSA,Koopman1932,BruntonKutz2015,BruntonKutz2016,Arabi2017}.

The standard DMD approach described above for classical dynamics cannot be immediately translated to quantum dynamics. The DMD approximation of the system evolution would require complete knowledge of the system state, as specified by an $N$-dimensional complex vector at each time step.  However, we do not have direct access to the full many-body quantum state. Instead, we can only access the state of a quantum system via measurement sampling of observables~\footnote{While ``observable'' is typically used in quantum mechanics to refer specifically to Hermitian operators (with real expectation value), here we use a broader definition, encompassing also complex scalar quantities that can be computed from measurements on a quantum computer.}. To address this challenge, we employ a technique motivated by Takens' embedding theorem \cite{takens1971,takens1981,Gutierrez:2021wik} to obtain an effective state vector consisting of an operator measured at a sequence of successive times.  We reformulate the linear model underpinning DMD in terms of these observable-vectors to approximate the system dynamics.

Takens' embedding theorem \cite{takens1971,takens1981} establishes a connection between the manifold of states, which an observer cannot directly access, and time-delayed measurements of an observable.
In particular, the theorem asserts, under generous conditions, that a state on an $N$-dimensional (sub)manifold can be completely determined using a sequence of at most $d_{\star} \leq 2N + 1$ time-delayed observables. This correspondence reads
\begin{align}
    \boldsymbol{\phi}(t) \leftrightarrow
     \boldsymbol{o}_{t,d_\star} = \begin{bmatrix}
   o(t) \\
    o(t+\Delta\tau) \\
     \vdots \\
    o(t+(d_*-1)\Delta\tau)
    \end{bmatrix},
    \label{eq:Takens}
\end{align}
where $\Delta\tau > 0$ is the time delay, $\boldsymbol{\phi}(t)$ is the system state, $o(t) = o[\boldsymbol{\phi}(t)]$ is the measured observable, and $\boldsymbol{o}_{t}$ is the $d_{\star}$-dimensional ``observable trajectory'' containing the dynamical information.
The RHS of \cref{eq:Takens} is known as a $d_{\star}$-dimensional delayed embedding of the observable.
Takens' theorem relates the evolution of microscopic degrees of freedom to the evolution history of macroscopic observables, providing a concrete probe into the dynamical properties of the system without direct access to the full states.
Here we adopt the term Takens' embedding technique to refer to the method of applying time delays on the system observables, motivated by the rigorous results of Takens' embedding theorem.

In anticipation of efficiently leveraging near-term quantum resources, we choose the time delay in Takens' embedding technique to equal the DMD time interval, \textit{i.e.}, $\Delta\tau = \Delta t$.
Given this choice, we then measure the system along time steps $\{t_k = k\Delta t\}_{k=0}^{K+1}$ and acquire the sequence of observable trajectories $\{ \boldsymbol{o}_{t_k,d}\}_{k=0}^{K+1}$, each of some length $d \leq d_{\star}$,
\begin{align}
\boldsymbol{o}_{t_k,d} = \begin{bmatrix}
o(t_k) \\
o(t_{k+1}) \\
\vdots \\
o(t_{k+d-1})\\
\end{bmatrix}, \qquad 0 \leq k \leq K+1.
\end{align}
By construction, the first $(d-1)$ entries of $\boldsymbol{o}_{t_k,d}$ are identical to the last $(d-1)$ entries of $\boldsymbol{o}_{t_{k-1},d}$. Consequently, the matrix assembled by arranging successive trajectories $ \boldsymbol{o}_{t_k}$ as columns
\begin{align}
\mathbf{X}_{k_1:k_2} = \begin{bmatrix} \boldsymbol{o}_{t_{k_1},d} & \boldsymbol{o}_{t_{k_1+1},d} & \cdots & \boldsymbol{o}_{t_{k_2},d} \end{bmatrix},
\end{align}
has a Hankel form, \textit{i.e.}, the matrix elements on each anti-diagonal are equal. In the embedding space, we can identify the closest linear flow,
\begin{align}
\mathbf{X}_{1:K+1} \lseq{} A \mathbf{X}_{0:K} \implies A = \mathbf{X}_{1:K+1} (\mathbf{X}_{0:K})^{+}, \label{eq:LS_ODMD}
\end{align}
where $+$ denotes the Moore–Penrose pseudo-inverse. 
Here the system matrix $A$ assumes a companion structure with just $d$ free parameters. The approximation to the system dynamics is then stored in the $d$ parameters inferred from measurements of $K+d+1$ delayed observables. We hence name our least-squares embedding in the observable space the \textit{observable dynamic mode decomposition} (ODMD).

\section{Multi-observable dynamic mode decomposition}
\label{app:multi-observable dynamic mode decomposition}
Recall that the multi-observable system matrix has a characteristic polynomial,
\begin{align}
    \mathcal{C}_{A}(z) = {\rm det}(z-A) = {\rm det}\left( \sum_{\ell=0}^{d} z^{\ell} A_{\ell}  \right),
\end{align}
where $A_d \equiv {\rm Id}_{I}$. We examine the simpler version of \cref{eq:MODMD_sys} in which we assume the $d$ matrix blocks,
\begin{align}
    A_{\ell} = {\rm Diag}\begin{bmatrix}
        A_{\ell, 11} & & \\
        & A_{\ell, 22} & & \\
        & & \ddots & \\
        & & & A_{\ell, II} \\
    \end{bmatrix},
\end{align}
to be diagonal with $A_{\ell, ij} \equiv 0$ for $i \neq j$. In this case, the multivariate residual is
\begin{align}
     \overrightarrow{{\rm Res}}_{k}(A) = \sum_{\ell=0}^{d} A_{\ell} \vec{s}_{\ell + k} &=  \sum_{\ell=0}^d A_{\ell} \sum_{n=0}^{N-1} \vec{c}_n \lambda_n^{\ell + k} = \sum_{n=0}^{N-1} \lambda_n^k \sum_{\ell=0}^{d} \lambda_{n}^{\ell} A_{\ell} \vec{c}_{n} = \sum_{n=0}^{N-1} \lambda_n^k \sum_{i=1}^{I} c_{n,i} \left( \sum_{\ell=0}^{d} \lambda_{n}^{\ell} A_{\ell, ii} \right) \vec{e}_i,   \label{eq:diag_res} \\
    &= \sum_{n=0}^{N-1} \lambda_n^k \sum_{i=1}^{I} c_{n,i} \mathcal{C}_{A|i}(\lambda_n) \vec{e}_i =   \sum_{i=1}^{I} \left( \sum_{n=0}^{N-1} c_{n,i} \lambda_n^k \mathcal{C}_{A|i}(\lambda_n) \right) \vec{e}_i, \label{eq:diag_char}
\end{align}
where \cref{eq:diag_res} directly follows from the submatrices $A_{\ell}$ being diagonal and $\vec{e}_i$ denotes the canonical basis vector of $\mathbb{C}^{I}$. Moreover, we define in \cref{eq:diag_char} the $I$ single-observable characteristic polynomials,
\begin{align}
    \mathcal{C}_{A|i}(z) = \sum_{\ell=0}^{d} z^{\ell} A_{\ell,ii} = \prod_{\ell=0}^{d-1} \left(z - \Tilde{\lambda}_{\ell|i}\right),
\end{align}
each with $d$ roots $\Tilde{\lambda}_{\ell|i}(A_{0,ii},A_{1,ii},\cdots, A_{d-1,ii})$. Clearly, \cref{eq:diag_char} indicates that the multivariate residual admits $I$ decoupled components corresponding to the different observables. Minimizing the total residual, $\lVert \overrightarrow{{\rm Res}}_{k}(A) \rVert_2$, is hence equivalent to minimizing each of the $I$ observable residuals, $\big \lvert \sum_{n=0}^{N-1} c_{n,i} \lambda_n^k \mathcal{C}_{A|i}(\lambda_n) \big \rvert$. This is consistent with the fact that the matrix determinant ${\rm det}(z-A$) factorizes into independent contributions,
\begin{align}
    \mathcal{C}_{A}(z) = \prod_{i=1}^{I} \left( \sum_{\ell=0}^{d} z^{\ell} A_{\ell,ii} \right) = \prod_{i=1}^{I} \mathcal{C}_{A|i}(z),
\end{align}
such that the eigenvalues of $A$ also factorize into clusters based on single-observable residuals. The resulting MODMD estimates are bounded by the $I$ single-observable ODMD estimates, \textit{e.g.},
\begin{align}
     \min_{1\leq i \leq I} \lvert \tilde{E}_{i,0} - E_0 \rvert \leq \lvert \tilde{E}_{0} - E_0 \rvert \leq  \max_{1\leq i \leq I} \lvert \tilde{E}_{i,0} - E_0\rvert,
\end{align}
where $\tilde{E}_0$ and $\tilde{E}_{i,0}$ designate, respectively, the ground state energy estimate using the entire observable pool (the full system matrix $A$) and a single observable (one row of the system matrix $A_{\ell,ii}$).

More interesting convergence arises when the $I$ observable residuals are coupled and give a total residual,
\begin{align}
    \overrightarrow{{\rm Res}}_{k}(A) 
    &= \sum_{n=0}^{N-1} \lambda_n^k \sum_{i,j=1}^{I} c_{n,j} \left( \sum_{\ell=0}^{d} \lambda_{n}^{\ell} A_{\ell, ij} \right) \vec{e}_i, \\
    &= \sum_{i=1}^{I} \left(\sum_{n=0}^{N-1}  c_{n,i} \lambda_n^k \mathcal{C}_{A|i} (\lambda_n) + \sum_{n=0}^{N-1}\lambda_n^k \sum_{\ell=0}^{d-1}  \lambda_n^{\ell} \sum_{j\neq i}^{I} A_{\ell,ij}c_{n,j}\right) \vec{e}_i, 
\end{align}
where we can lower the residual by utilizing the flexibility of off-diagonal elements in the submatrices $A_{\ell}$. For example, the total residual may vanish completely when the multi-observable signal is $d(I-1)$-sparse in the eigenfrequency basis, where, for $1\leq i\leq I$, the coefficients $c_{n,i}$ are supported on at most $d(I-1)$ eigenindices $\mathcal{N}_{i} = \{0 \leq n \leq N-1 : c_{n,i} \neq 0 \}$. That is, $\max_{1 \leq i\leq I} \lvert \mathcal{N}_{i} \rvert \leq d(I-1)$. In this case, a vanishing residual is possible, provided that $(a)$ $\lvert \cup_{i=1}^{I} \mathcal{N}_i \rvert \leq d(I-1)$ and $(b)$ the matrix elements $\{ A_{\ell,ij} \}_{\ell, j \neq i}$ can be set appropriately such that $\forall 1 \leq i \leq I$,
\begin{align}
    \begin{bmatrix}
       \beta_{0,i1}(1) & \ldots & \beta_{\ell,ij\neq i}(1) & \ldots & \beta_{d-1,iI}(1) \\
       \beta_{0,i1}(2) & \ldots & \beta_{\ell,ij\neq i}(2) & \ldots & \beta_{d-1,iI}(2) \\
       \vdots & & \ddots & & \vdots \\
       \beta_{0,i1}(\lvert \mathcal{N}_i \rvert) & \ldots &  \beta_{\ell,ij\neq i}(\lvert \mathcal{N}_i \rvert) & \ldots & \beta_{d-1,iI}(\lvert \mathcal{N}_i \rvert) \\
       \beta_{0,11}(1) & \ldots &  \beta_{\ell,1j\neq i}(1) & \ldots & \beta_{d-1,1I}(1) \\
       \vdots & & & \ddots & \vdots \\
       \beta_{0,I1}(\lvert \mathcal{N}_I \rvert) & \ldots &  \beta_{\ell,Ij\neq i}(\lvert \mathcal{N}_I \rvert) & \ldots & \beta_{d-1,II}(\lvert \mathcal{N}_I \rvert) \\
    \end{bmatrix} \begin{bmatrix}
        A_{0,i1} \\
        A_{0,i2} \\
        \vdots \\
        A_{\ell,ij} \\
        \\
        \vdots \\
        A_{d-1,iI} 
    \end{bmatrix} = - \begin{bmatrix}
      \beta_{0,ii}(1) \mathcal{C}_{A|i}\big(\lambda_{n_{1}[i]}\big) \\
      \beta_{0,ii}(2) \mathcal{C}_{A|i}\big(\lambda_{n_{2}[i]}\big) \\
      \vdots \\
      \beta_{0,ii}(\lvert \mathcal{N}_i \rvert )\mathcal{C}_{A|i}\big(\lambda_{n_{\lvert \mathcal{N}_i \rvert}[i]}\big) \\
      0 \\
      \vdots \\
      0
    \end{bmatrix},
    \label{eq:I-1 sparse condition}
\end{align}
where $n_{\xi}[i]$ numerates support eigenindices in the set $\mathcal{N}_i$ for observable $O_i$, and $\beta_{\ell,ij}(\xi) = \lambda_{n_\xi[i]}^{\ell} c_{n_{\xi}[i],j} $. Given $\{ A_{\ell,ii} \}_{\ell}$, a formal solution $\{ A_{\ell,ij} \}_{\ell, j \neq i}$ to \cref{eq:I-1 sparse condition} exists if the matrix on the LHS has a full rank. This directly implies that the total residual may vanish completely for a $dI$-sparse multi-observable signal in the eigenfrequency basis (as expected), because $\{ A_{\ell,ii}\}_{\ell}$ can always be chosen so that the single-observable characteristic polynomial $\mathcal{C}_{A|i}(z)$ encompasses the roots $\{\lambda_{n_{\xi}[i]}\}_{\xi}$ corresponding to at most $d$ members of $\mathcal{N}_{i}$. Without additional degrees of freedom from the off-diagonal matrix elements of $A$, the residual only vanishes for a $d$-sparse signal. 

We now relax the sparsity assumption on the signal to explore the general case. Observe that
\begin{align}
    \min_{A} \sum_{k=0}^{K} \left \lVert \overrightarrow{{\rm Res}}_{k}(A) \right \rVert^2_2 &= \min_{A} \sum_{k=0}^{K} \sum_{i=1}^{I} \left\lvert \sum_{n=0}^{N-1}  c_{n,i} \lambda_n^k \mathcal{C}_{A|i} (\lambda_n) +   \sum_{j\neq i}^{I} \sum_{n=0}^{N-1} c_{n,j} \lambda_n^k \mathcal{C}_{A|ij} (\lambda_n) \right\rvert^2, \\
    & \leq \min_{\left\{A_{\ell,ii}: 0 \leq \ell \leq d-1, 1 \leq i \leq I \right\}} \sum_{k=0}^{K} \sum_{i=1}^{I} \left\lvert \sum_{n \notin \mathcal{N}_{i}^{\star} }  c_{n,i} \lambda_n^k \mathcal{C}_{A|i} (\lambda_n) + \sum_{j \neq i}^{I} \sum_{n \notin \mathcal{N}_{i}^{\star}} c_{n,j} \lambda_n^k \bar{\mathcal{C}}_{A|ij} (\lambda_n) \right\rvert^2,
    \label{app:multi-residual bound}
\end{align}
where $\mathcal{C}_{A|ij}(z) = \sum_{\ell=0}^{d} z^{\ell} A_{\ell,ij} \equiv \sum_{\ell=0}^{d-1} z^{\ell} A_{\ell,ij}$ for $j \neq i$ defines a degree-$(d-1)$ polynomial via the off-diagonal matrix elements $\{ A_{\ell,ij} \}_{\ell}$, $\mathcal{N}_{i}^{\star} \subset [N]$ is an eigenindex subset of size $\lvert \mathcal{N}_{i}^{\star} \rvert = d(I-1)$, and $\bar{\mathcal{C}}_{A|ij}(z) = \sum_{\ell=0}^{d-1} z^{\ell} A_{\ell,ij}(\mathcal{N}_{i}^{\star}; \{ A_{\ell,ii}\}_{\ell}) $ fixes $\mathcal{C}_{A|ij}(z)$ by evaluating specific coefficients $\{  A_{\ell,ij} \}_{\ell}$ such that,
\begin{align}
     \sum_{j \neq i}^{I} c_{n,j} \bar{\mathcal{C}}_{A|ij} (\lambda_n) = -  c_{n,i} \mathcal{C}_{A|i} (\lambda_n), \qquad \forall n \in \mathcal{N}_i^{\star}.
     \label{app:matching polynomial}
\end{align}
Resembling \cref{eq:I-1 sparse condition}, we reserve the notation $n_{\xi}^{\star}[i]$ for the support eigenindices in $\mathcal{N}_i^{\star}$. \cref{app:matching polynomial} then holds only if the following coefficient matrix has full rank, \emph{i.e.},
\begin{align}
   {\rm rank}[ \beta_{\mathcal{N}_i^{\star}}] = {\rm rank}\begin{bmatrix}
       \beta_{0,i1}^{\star}(1) & \ldots & \beta_{\ell,ij\neq i}^{\star}(1) & \ldots & \beta_{d-1,iI}^{\star}(1) \\
       \beta_{0,i1}^{\star}(2) & \ldots & \beta_{\ell,ij\neq i}^{\star}(2) & \ldots & \beta_{d-1,iI}^{\star}(2) \\
       \vdots & & \ddots & & \vdots \\
       \beta_{0,i1}^{\star}(\lvert \mathcal{N}_i^{\star} \rvert) & \ldots &  \beta_{\ell,ij\neq i}^{\star}(\lvert \mathcal{N}_i^{\star} \rvert) & \ldots & \beta_{d-1,iI}^{\star}(\lvert \mathcal{N}_i^{\star} \rvert) \\
    \end{bmatrix} = d(I-1),
\end{align}
for $\beta_{\ell,ij}^{\star}(\xi) = \lambda_{n_\xi^{\star}[i]}^{\ell} c_{n_{\xi}^{\star}[i],j} $. An arbitrary index set $\mathcal{N}_{i}^{\star}$ can be assigned as long as the full-rank condition is met; ideally, we aim for $\mathcal{N}_i^{\star} = \mathcal{N}_i^{\star}(\{ A_{\ell,ii}\}_{\ell})$ to satisfy the optimal property,
\begin{align}
    \left\lvert \sum_{j=1}^{I} \sum_{n \notin \mathcal{N}_{i}^{\star} }  c_{n,j} \lambda_n^k \bar{\mathcal{C}}_{A|ij} (\lambda_n) \right\rvert  = \min_{\mathcal{N} \subset [N]: \lvert \mathcal{N} \rvert = d(I-1)}  \left\lvert \sum_{j=1}^{I} \sum_{n \notin \mathcal{N}}  c_{n,j} \lambda_n^k \bar{\mathcal{C}}_{A|ij} (\lambda_n) \right\rvert,
\end{align}
which minimizes the effective residual  parametrized by the $d$ diagonal matrix elements $\{ A_{\ell,ii}\}_{\ell}$, with $\bar{\mathcal{C}}_{A|ii}(z) = \mathcal{C}_{A|i}(z)$. By the triangle inequality, we can bound the RHS of \cref{app:multi-residual bound} as,
\begin{align}
    \left\lvert \sum_{n \notin \mathcal{N}_{i}^{\star} }  c_{n,i} \lambda_n^k \mathcal{C}_{A|i} (\lambda_n) + \sum_{j \neq i}^{I} \sum_{n \notin \mathcal{N}_{i}^{\star}} c_{n,j} \lambda_n^k \bar{\mathcal{C}}_{A|ij} (\lambda_n) \right\rvert \leq \sum_{j=1}^{I} \sum_{n \notin \mathcal{N}_{i}^{\star} } \left\lvert c_{n,j} \bar{\mathcal{C}}_{A|ij}(\lambda_n) \right\rvert  \leq \bar{c}_{i} \sum_{j=1}^{I} \bar{C}_{ij},
\end{align}
where $\bar{c}_{i} = \max_{1\leq j\leq I} \sum_{n \notin \mathcal{N}_{i}^{\star} } \left\lvert c_{n,j} \right\rvert$ and $\bar{C}_{ij} = \max_{n \notin \mathcal{N}_{i}^{\star}} \lvert \bar{\mathcal{C}}_{A|ij} (\lambda_n) \rvert$. With suitable choices of the reference state $\ket{\phi_0}$ and operator pool $\{ O_i\}_{i=1}^{I}$, we can strategically adjust the coefficients $\{ c_{n,i} \}_{n,i}$, therefore controlling the constant $\bar{c}_{i}$ and the conditioning of the matrix $\beta_{\mathcal{N}_i^{\star}}$ (or equivalently $\lVert \beta_{\mathcal{N}_i^{\star}}^{-1} \rVert_2$). Accordingly, for given index sets $\{ \mathcal{N}_{i}^{\star} \}_{i}$, we have
\begin{align}
   \begin{split}
       \min_{\left\{A_{\ell,ii}: 0 \leq \ell \leq d-1, 1 \leq i \leq I \right\}} & \sum_{k=0}^{K} \sum_{i=1}^{I} \left( \bar{c}_{i} \sum_{j=1}^{I} \bar{C}_{ij} \right )^2  \\
   &\leq (K+1) \sum_{i=1}^{I} \bar{c}_{i}^2 \min_{\left\{A_{\ell,ii}: 0 \leq \ell \leq d-1 \right\}} \left( \max_{n \notin \mathcal{N}_{i}^{\star}} \big\lvert \mathcal{C}_{A|i} (\lambda_n) \big\rvert  + \sum_{\ell=0}^{d-1}\sum_{j\neq i}^{I} \left\lvert A_{\ell,ij}(\mathcal{N}_{i}^{\star}; \{ A_{\ell,ii}\}_{\ell}) \right\rvert \right)^2,
   \end{split}
   \label{app:multi-residual bound2}
\end{align}
where the residual from the off-diagonal matrix elements can be bounded by,
\begin{align}
   \sum_{\ell=0}^{d-1}\sum_{j\neq i}^{I} \left\lvert A_{\ell,ij}(\mathcal{N}_{i}^{\star}; \{ A_{\ell,ii}\}_{\ell}) \right\rvert 
   \leq d(I-1) \max_{n \in \mathcal{N}_i^{\star}} \lvert c_{n,i} \rvert \lVert \beta_{\mathcal{N}_i^{\star}}^{-1} \rVert_2 \max_{n \in \mathcal{N}_{i}^{\star}} \big\lvert \mathcal{C}_{A|i} (\lambda_n) \big\rvert,
\end{align}
using an identity analogous to \cref{eq:I-1 sparse condition}. To further bound the RHS of \cref{app:multi-residual bound2}, we attempt to tightly approximate the minimizer over the $d$ diagonal matrix elements $\{ A_{\ell,ii} \}_{\ell}$, or alternatively over the set of degree-$d$ monic polynomials defined along the circular arc $\mathcal{A}_{\theta} = \{ z \in \mathbb{C}: \abs{z}=1,{\rm -\theta \leq arg}(z) \leq \theta\}$. In particular let us consider the complex-valued Chebyshev polynomials,
\begin{align}
    \hat{\mathcal{C}}(z)
    = \argmin_{\mathcal{C}\hspace{0.02cm}\text{monic}:\hspace{0.02cm}{\rm deg}(\mathcal{C}) = d} \sup_{z \in \mathcal{A}_{\theta}} \abs{\mathcal{C}(z)},
\end{align}
whose parametric representations are explicitly constructible by tools such as Jacobi’s elliptic and theta functions~\cite{akhiezer1928functionen}. For notational convenience, we denote $\lVert \mathcal{C} \rVert_{\theta} = \sup_{z \in \mathcal{A}_{\theta}} \abs{\mathcal{C}(z)}$. Similar to the real-valued Chebyshev polynomials over the interval $[-1,1]$, $\hat{\mathcal{C}}(z)$ retains the minimal-norm property on $\mathcal{A}_{\theta}$, where $\lVert \hat{\mathcal{C}} \rVert_{\theta} \sim 2 \sin^{d} (\theta/2) \cos^2 (\theta/4)$ asymptotically for large $d$~\cite{schiefermayr2019chebyshev}. Setting $\theta_{H}(\Dt) = (E_{N-1} - E_0)\Dt/2$, the total residual decays at least exponentially:
\begin{align}
    \min_{A} \sum_{k=0}^{K} \left \lVert \overrightarrow{{\rm Res}}_{k}(A) \right \rVert^2_2 \leq (K+1) \lVert \hat{\mathcal{C}} \rVert_{\theta_{H}(\Dt)}^2 \sum_{i=1}^{I} \bar{c}_{i}^2 \left[ 1 + d(I-1) \max_{n \in \mathcal{N}_i^{\star}} \lvert c_{n,i} \rvert \lVert \beta_{\mathcal{N}_i^{\star}}^{-1} \rVert_2 \right]^2,
\end{align}
where the prefactor improves significantly, compared to the single-observable case derived in~\cite{shen2023estimating}, when
\begin{align}
   \max_{1 \leq j \leq I} \sum_{n \notin \mathcal{N}_{i}^{\star} } \left\lvert c_{n,j} \right\rvert \ll \frac{1}{\displaystyle 1 + d(I-1) \max_{n \in \mathcal{N}_i^{\star}} \lvert c_{n,i} \rvert \lVert \beta_{\mathcal{N}_i^{\star}}^{-1} \rVert_2}, \qquad \forall 1\leq i\leq I.
\end{align}

\section{Proof of theorems}
\subsection{Theorem 1}
\label{subsec:theorem_1}

\noindent \textbf{Theorem 1}. Let $\Tilde{E}_{0}(d)$ be the approximate ground state energy extracted from the $d$-dimensional function subspace spanned by $\{g_1, \Ko[g_1], \cdots, \Ko^{d-1}[g_1] \}$. For $d \geq 1$, there exists time step $\Delta t$ such that the error $\delta E_{g}(d) = \Tilde{E}_{0}(d) - E_{0}$ is bounded by
\begin{align}
    \delta E_{0}(d) \leq \frac{ {\rm cond}(\mathcal{B})^{1/2} \abs{ \sin[(E_{N-1} - E_{0})\Dt]} }{\Tilde{\epsilon}_{0 \rightarrow 1}^{2(d-1)} \Dt } \tan^2{\Xi},
\end{align}
where ${\rm cond}(\mathcal{B}) = \lVert \mathcal{B}^{-1} \rVert \lVert \mathcal{B} \rVert $ denotes the condition number of the $N \times N$ eigenfunction Gram matrix $\mathcal{B}$ with entries $\mathcal{B}_{mn} = \braket{f_m,f_n}_{\mu}/(\lVert f_m \rVert_{L^2(\mu)}  \lVert f_n \rVert_{L^2(\mu)})$, $\cos^2{\Xi} = \lvert \langle f_0, g_1 \rangle_{\mu} \rvert^2 /(  \lVert f_0 \rVert_{L^2(\mu)}^2  \lVert g_1 \rVert_{L^2(\mu)}^2 )$ is the squared overlap between the reference function $g_1$ and the true ground state $\Ko$-eigenfunction, while $\Tilde{\epsilon}_{0 \rightarrow 1} = 1 + 3(E_1 - E_0)\Dt/2\pi$ $\in [1,2]$ characterizes the normalized spectral gap of the Hamiltonian $H$.

\noindent[\textit{Proof}.] Let us define,
\begin{align}
    \theta(f) = \arg \frac{\langle f, \Ko f \rangle_{\mu}}{\langle f, f \rangle_{\mu}} \in [-\pi,\pi], \label{eq:angle_expect}
\end{align}
which returns the expected dynamical phase factor associated with a function $f$ under Koopman evolution, where $f$ belongs to the invariance subspace ${\rm span}\{f_n\}_{n=0}^{N-1}$. Given a suitable symmetrizing spectral shift such that $\norm{H} \Dt \leq \pi$, we note that $\theta(f_0) = \max_{f} \theta(f) = - E_0 \Dt$. This variational principle implies,
\begin{align}
        \Tilde{E}_0(d) &= -\frac{1}{\Dt} \max_{ f \in {\rm Kr}_{d} }  \theta(f) = -\frac{1}{\Dt}  \max_{p \in \mathcal{P}_{d-1}} \theta( p(\Ko) g_1 ),
        \label{eq:Keigval}
\end{align}
for which we use ${\rm Kr}_{d}$ to denote our $d$-dimensional function Krylov subspace ${\rm span} \{g_1, \cdots, \Ko^{d-1}[g_1] \}$ and $\mathcal{P}_{d-1}$ the set of degree-$(d-1)$ polynomials over $\mathbb{C}$. 

Hereafter, we assume that the eigenfunctions $\{f_n\}_{n=0}^{N-1}$ are orthogonal, \textit{i.e.}, \cref{eq:ortho_Krylov} simplifies to a diagonal relation where $\mathcal{B}_{mn} = 0$ if $m \neq n$. Drawing on the convergence analysis for nonnormal operators in~\cite{Liesen2004}, we establish that the error bound in the cases of nonorthogonal eigenfunctions is amplified by at most a multiplicative factor ${\rm cond}(\mathcal{B})^{1/2} \geq 1$. This is because $\mathcal{B}$ admits the representation $\mathcal{B} = \mathcal{T}^{\dagger}\mathcal{T}$ where $\mathcal{T}$ collects the eigenfunctions written in standard function basis, so that ${\rm cond}(\mathcal{B}) = {\rm cond}(\mathcal{T})^2$. In the orthogonal case, we can expand $g_1$ in the eigenfunctions,
\begin{align}
    \frac{g_1}{\lVert g_1 \rVert_{L^2(\mu)}^2} = \sum_{n=0}^{N-1} \frac{z_n f_n}{\lVert f_n \rVert_{L^2(\mu)}},
\end{align}
with $\sum_{n=0}^{N-1} \abs{z_n}^2 = 1$. Thus \cref{eq:Keigval} in the eigenbasis reads,
\begin{align}
     \Tilde{E}_0(d) &= -\frac{1}{\Dt} \max_{p \in \mathcal{P}_{d-1}} \arg \frac{ \langle p(\Ko) g_1,  \Ko p(\Ko) g_1 \rangle_{\mu} }{\langle p(\Ko) g_1, p(\Ko) g_1 \rangle_{\mu}},\\
     &= -\frac{1}{\Dt} \max_{p \in \mathcal{P}_{d-1}} \arg \frac{ \sum_{n=0}^{N-1} \lambda_n \abs{z_n p(\lambda_n)}^2 }{\sum_{n=0}^{N-1} \abs{z_n p(\lambda_n)}^2 },
\end{align}
where $\lambda_n = e^{- iE_n \Dt}$ label the eigenvalues of the Koopman operator. We note that
\begin{align}
    \max_{p \in \mathcal{P}_{d-1}} \arg \frac{ \sum_{n=0}^{N-1} \lambda_n \abs{z_n p(\lambda_n)}^2 }{\sum_{n=0}^{N-1} \abs{z_n p(\lambda_n)}^2 } \geq  \max_{p \in \mathcal{P}_{d-1} } \arg \frac{ \lambda_{0} \abs{z_{0} p(\lambda_{0})}^2 + \lambda_{N-1} \sum_{n=1}^{N-1} \abs{z_n p(\lambda_n)}^2 }{ \sum_{n=0}^{N-1} \abs{z_n p(\lambda_n)}^2 },
\end{align}
by convexity of the circular sector with angle $[-E_{N-1}\Dt, -E_0\Dt]$ in the complex plane. Then
\begin{align}
    \Tilde{E}_0(d) &\leq -\frac{1}{\Dt} \max_{p \in \mathcal{P}_{d-1} } \arg \left[ \lambda_0 + (\lambda_{N-1} - \lambda_0 ) \frac{ \sum_{n=1}^{N-1} \abs{z_{n} p(\lambda_{n})}^2 }{ \sum_{n=0}^{N-1} \abs{z_n p(\lambda_n)}^2 } \right], \\
    &= -\frac{1}{\Dt} \arg \left[ \lambda_0 + (\lambda_{N-1} - \lambda_0 ) \min_{p \in \mathcal{P}_d} \frac{ \sum_{n=1}^{N-1} \abs{z_{n} p(\lambda_{n})}^2 }{ \sum_{n=0}^{N-1} \abs{z_n p(\lambda_n)}^2 } \right],
    \label{eq:Krylov_frac}
\end{align}
where we observe $\arg (\lambda_{N-1} - \lambda_0) = \frac{1}{2}[\arg\lambda_{N-1} + \arg(-\lambda_0)] \leq \arg \lambda_0$ assuming $E_0 \leq 0$ WLOG. To proceed, we seek a family of polynomials defined over the unit circle $\mathbb{S}^1 = \{ z \in \mathbb{C}: |z| = 1 \}$ to bound the fraction on the RHS of \cref{eq:Krylov_frac}. Now let us fix some $q \in (0,1]$ and consider the handy choice of complex-valued Rogers-Szeg\H{o} polynomials~\cite{RSpoly,orthogonalpoly}, 
\begin{eqnarray}
      W_d(z|q) = \begin{cases}
      1         & d = 0 \\ 
      z + 1     & d = 1 \\
      (1+z) W_{d-1}(z|q) \\
      \hspace{.6 cm} - (1-q^{j-1}) z W_{d-2}(z|q) & d \geq 2
      \end{cases},
      \label{eq:R-S}
\end{eqnarray}
over the circle $\mathbb{S}^{1,q} = \{z: |z| = q^{-1/2} \}$. For simplicity, we rewrite $z = -q^{-1/2}\exp{(-i\vartheta)}$ where $\vartheta \in [-\pi, \pi])$ denotes an angular phase. Here a prefactor of $-1$ is included to periodically translate the polynomials so that $W_d(\vartheta|q)$ adapts the symmetry $W_d(-\vartheta)=W_d(\vartheta)^{\ast}$ (we also omit a conditional dependence of $W_d$ on $q$ for notational clarity). Such family of polynomials shares the key properties that $(i)$ $|W_d(\vartheta)|$ remains bounded below unity over some angular window $\mathcal{W} = [-\Omega, \Omega]$ $\subset [-\pi, \pi]$ and $(ii)$ $|W_d(\vartheta)|$ grows rapidly outside $\mathcal{W}$. 

Note that the constant $q$ controls the width of our truncated angular window $\mathcal{W}$. In the limit of $q \rightarrow 1$, one can verify that these polynomials converge to, 
\begin{align}
    W_d(\vartheta) \rightarrow \sum_{k=0}^{d} \binom{d}{k} \exp{\left[ -ik(\vartheta+\pi) \right]},
\end{align}
which simply gives the sum of evenly spaced points along $\mathbb{S}^1$ weighted by binomial coefficients. As a consequence, $\sup_{\vartheta} |W_d(\vartheta)| \approx2^d$ for $q \approx 1$. To bound the fraction in \cref{eq:Krylov_frac} tightly, we look for a suitable linear transformation $\mathcal{L}: \mathbb{S}^1 \rightarrow \mathbb{S}^1$ acting on the $N$ individual eigenphases, $ \lambda_{n} = e^{-i\vartheta_n(\Dt)}$, such that $\mathcal{L}$ nudges excited state angles $\vartheta_{n \geq 1}$ all inside the truncated window $\mathcal{W}$ while keeping the ground state angle $\vartheta_0$ outside. It is safe to assume $\vartheta_{N-1} - \vartheta_{1} \leq 2\Omega$ and $ 2\Omega \leq \vartheta_{N-1} - \vartheta_0 \leq \pi + \Omega$ by adapting a suitable time step size $\Delta t$, \textit{e.g.},
\begin{align}
    \Delta t = \sup_{\tau} \Big\{ \tau \in \mathbb{R}^{+}: \vartheta_{N-1}(\tau) - \vartheta_{1}(\tau) \leq 2\Omega, \vartheta_{1}(\tau) - \vartheta_{0}(\tau) \leq \Omega^{\rm c} \Big\},
    \label{eq:thm1_Dt_choice}
\end{align}
with $\Omega^{\rm c} = \pi - \Omega$. Therefore a natural choice of $\mathcal{L}$ is the phase multiplication or angular translation, $\mathcal{L}: \vartheta \mapsto \vartheta - \Omega + \vartheta_{1}$, which circularly moves $\{\lambda_{n}\}_{n=0}^{N-1}$ so that $ \mathcal{L}(\vartheta_{0}) \leq - \Omega = \mathcal{L}(\vartheta_{1}) \leq - \mathcal{L}(\vartheta_{N-1})$ as desired. With $\mathcal{L}$ chosen above, we establish a variational upper bound on the ground state energy error by substituting the trial polynomials $p = W_{d-1} \circ \mathcal{L}$ into \cref{eq:Krylov_frac},
\begin{align}
    \Tilde{E}_0(d) &\leq -\frac{1}{\Dt}  \arg\left[ \lambda_0 + (\lambda_{N-1} - \lambda_0) \frac{ \sum_{n=1}^{Q-1} \abs{z_n W_{d-1}(\mathcal{L}(\lambda_{n})) }^2 }{ \sum_{n=0}^{N-1} \abs{z_{n} W_{d-1} (\mathcal{L}(\lambda_{n}))}^2} \right],\\
    &\leq -\frac{1}{\Dt}  \arg\left[ \lambda_0 + (\lambda_{N-1} - \lambda_0) \frac{ \sum_{n=1}^{Q-1} \abs{z_n W_{d-1}(\Omega) }^2 }{ \abs{z_{0} W_{d-1} (\mathcal{L}(\lambda_{0}))}^2} \right],\\
    &= -\frac{1}{\Dt}  \arg \left[ \lambda_0 + (\lambda_{N-1} - \lambda_0) \frac{\sin^2{\Xi}}{\abs{W_{d-1} (\mathcal{L}(\lambda_{0}))}^2 \cos^2{\Xi}  } \right],\label{eq:bound1}
\end{align}
where in arriving at \cref{eq:bound1} we have utilized the property $(i)$ of $W_{d-1}(\vartheta)$ and defined the angle $\Xi$ by $\cos^2{\Xi} = |z_0|^2 = \lvert \langle f_0 , g_1 \rangle_{\mu} \rvert^2/(  \lVert f_0 \rVert_{L^2(\mu)}^2  \lVert g_1 \rVert_{L^2(\mu)}^2 )$ which specifies the $L^2$-projection of our reference function $g_1$ onto the ground state eigenfunction. For the limiting case $q = 1$, it is rather straightforward to show that $\Omega = \pi/3$ and,
\begin{align}
    \abs{W_d ( \mathcal{L}(\lambda_0) )}^{1/d} = \sqrt{2 - 2\cos{\left( \vartheta_{1} - \vartheta_{0} + \Omega \right)}} \geq  1 + C \epsilon_{0 \rightarrow 1},
     \label{eq:Linbound}
\end{align}
where $\epsilon_{0 \rightarrow 1} = (\vartheta_{1} - \vartheta_{0})/\Omega^{\rm c} = (E_1 - E_0) \Dt /\Omega^{\rm c}$ gives the normalized spectral gap times the time step and $C$ is a constant for which \cref{eq:Linbound} holds with $\epsilon_{0 \rightarrow 1} \in [0, 1]$. For example, $C = 1$ can be justified by concavity of the LHS of the inequality above with respect to $(\vartheta_{1}-\vartheta_{0})$. Hence we can further bound \cref{eq:bound1} using \cref{eq:Linbound},
\begin{align}
    \Tilde{E}_0(d) &\leq  -\frac{1}{\Dt}  \arg \left[ \lambda_0 \left( 1 - \Tilde{\epsilon}_{0 \rightarrow 1}^{-2(d-1)} \tan^2{\Xi} \right) + \lambda_{N-1} \Tilde{\epsilon}_{0 \rightarrow 1}^{-2(d-1)} \tan^2{\Xi} \right], \label{eq:bound2} \\
    &\leq \frac{1}{\Dt} \arctan \left[ \frac{ (1 - \zeta) \sin\vartheta_0 + \zeta \sin\vartheta_{N-1}  }{ (1-\zeta)\cos\vartheta_0 + \zeta \cos\vartheta_{N-1}    }  \right],\\
    &= E_0 + \frac{ \sin(\vartheta_{N-1} -\vartheta_{0}) }{\Dt} \zeta + \mathcal{O}(\zeta^2),
\end{align}
where we have defined $\Tilde{\epsilon}_{0 \rightarrow 1} = 1+ C \epsilon_{0 \rightarrow 1} > 1$ and $\zeta =  \Tilde{\epsilon}_{0 \rightarrow 1}^{-2(d-1)} \tan^2{\Xi}$. The last equality can be derived from a Taylor expansion up to leading order in $\zeta$. Accounting for the eigenfunction conditioning ${\rm cond}^{1/2}(\mathcal{B})$, we have thereby proved the statement as claimed. $\square$

\subsection{Aside: canonical angles}
\label{subsec:canonical_angles}
To set up the proof of Theorem 2, we first introduce the notion of subspace overlap for subsequent discussions.~\cite{Li2015} Suppose that we have two subspaces $\mathcal{X}$ and $\mathcal{Y}$ of $\mathbb{C}^N$ with ${\rm dim} \mathcal{X} \leq {\rm dim} \mathcal{Y}$. Let $B_{\mathcal{X}} \in \mathbb{C}^{N \times {\rm dim} \mathcal{X}}$ and $B_{\mathcal{Y}} \in \mathbb{C}^{N \times {\rm dim} \mathcal{Y}}$ be the orthornormal basis matrices of $\mathcal{X}$ and $\mathcal{Y}$ ($B_{\mathcal{X}}$ and $B_{\mathcal{Y}}$ are thus not unique). The canonical angles between the subspaces are defined as
\begin{align}
    \theta_{\ell}[\mathcal{X}, \mathcal{Y}] = \arccos{\sigma_\ell} \in [0, \frac{\pi}{2}], \quad 1 \leq \ell \leq {\rm dim} \mathcal{X},
\end{align}
where $\sigma_1 \leq \sigma_2 \leq \ldots \leq \sigma_{{\rm dim} \mathcal{X}}$  denote the singular values of $B_{\mathcal{X}}^{\dagger}B_{\mathcal{Y}}$. For convenience, we use a diagonal matrix
\begin{align}
    \Theta[\mathcal{X},\mathcal{Y}] = \begin{bmatrix}
        \theta_1 && & \\
        &\theta_2 && \\
        &&\ddots & \\
        &&& \theta_{{\rm dim} \mathcal{X}}
    \end{bmatrix}
\end{align}
to record the set of ${\rm dim} \mathcal{X}$ canonical angles. Notice when ${\rm dim} \mathcal{X}=1$, the canonical angle is determined by the familiar $\ell^2$-inner product on the Hilbert space. Here we present elementary and known results~\cite{Stewart1990} that are helpful for deriving the block convergence bound.

\noindent \textbf{Result C.2.1.} For $\mathcal{Y}_0 \subseteq \mathcal{Y}$ with ${\rm dim} \mathcal{Y}_0 = {\rm dim} \mathcal{X}$, $\theta_{\ell}[\mathcal{X}, \mathcal{Y}] \leq \theta_{\ell}[\mathcal{X}, \mathcal{Y}_0]$ and the equality can be saturated.

\noindent \textbf{Result C.2.2.} Suppose ${\rm dim} \mathcal{X} = {\rm dim} \mathcal{Y}$ and $\theta_{\ell}[\mathcal{X},\mathcal{Y}] \neq \frac{\pi}{2}$.

\noindent (a) For $\mathcal{Y}_0 \subseteq \mathcal{Y}$, there exists unique $\mathcal{X}_0 \subseteq \mathcal{X}$ with ${\rm dim} \mathcal{X}_0 = {\rm dim} \mathcal{Y}_0$ such that $\Pi_{\mathcal{Y}} \mathcal{X}_0 = \mathcal{Y}_0$ for orthogonal projection $\Pi_{\mathcal{Y}}$ onto $\mathcal{Y}$. Moreover,
\begin{align}
    \theta_{\ell}[\mathcal{X}_0,\mathcal{Y}_0] \leq \theta_{\ell}[\mathcal{X},\mathcal{Y}] ,
\end{align}
for $1 \leq \ell \leq {\rm dim} \mathcal{Y}_0$ and hence
\begin{align}
   \lVert \sin \Theta[\mathcal{X}_0,\mathcal{Y}_0] \rVert \leq \lVert \sin \Theta[\mathcal{X},\mathcal{Y}] \rVert
\end{align}
for any unitarily invariant norm $\lVert \cdot \rVert$.

\noindent (b) For any orthonormal vectors $[y_1, \ldots, y_{Q}]$ in $\mathcal{Y}$, there exists linearly independent vectors $[x_1, \ldots, x_{Q}]$ in $\mathcal{X}$ so that $\Pi_{\mathcal{Y}}x_{\ell} = y_{\ell}$ and $(a)$ holds for $\mathcal{X}_0 = {\rm span} \{ x_1, \ldots, x_{Q} \}$ and $\mathcal{Y}_0 = {\rm span}\{ y_1, \ldots, y_{Q} \}$ with ${\rm dim} \mathcal{X}_0 = {\rm dim} \mathcal{Y}_0 = Q$.

\subsection{Theorem 2}
\label{subsec:theorem_2}

\noindent \textbf{Theorem 2}. Let $\Tilde{E}_{n}(d)$ be the approximate $n$th eigenenergy extracted from the $dI$-dimensional function subspace ${\rm span} \{\vec{g}, \Ko[\vec{g}], \cdots, \Ko^{d-1}[\vec{g}] \}$, and $\delta E_{n}(d) = \Tilde{E}_{n}(d) - E_{n}$ be the approximation error. Consider the error matrix,
\begin{align}
    \bDel_I(d) = {\rm diag}\begin{bmatrix}
        \delta E_{0}(d) & & \\ 
         & & \ddots & \\
         & & & \delta E_{I-1}(d) \\
    \end{bmatrix},
\end{align}
which contains approximations to the lowest $I$ energies. For $d \geq 1$, there exists time step $\Delta t$ such that the spectral approximation $\bDel_I(d)$ is bounded by
\begin{align}
    \lVert \bDel_{I}(d) \rVert_2 \leq \frac{{\rm cond}(\mathcal{B})^{1/2} \abs{\sin[(E_{N-1} - E_{0})\Dt]} }{\tilde{\epsilon}_{I-1 \mapsto I}^{2(d-1)} \Dt } \lVert \tan^2 \Theta \rVert_2,
\end{align}
for the operator norm $\lVert \cdot \rVert_2$. Here $\Theta$ denotes the canonical angle between the two subspaces ${\rm span}\{ f_n : 0 \leq n \leq I-1 \}$ and ${\rm span}\{ g_i: 1 \leq i \leq I \}$, which generalizes the squared overlap in Theorem 1 (see \cref{subsec:canonical_angles,subsec:theorem_2}). In the denominator, $\tilde{\epsilon}_{I-1 \mapsto I} \in [1, 2]$ depends on the $I$th spectral gap $(E_{I} - E_{I-1})$ of the Hamiltonian $H$.

\noindent[\textit{Proof}.] We begin by treating the case of mutually orthogonal eigenfunctions.

Here, it suffices to establish the more general result for the error matrix $\bDel = {\rm diag}[\delta E_{n_1}(d), \delta E_{n_1+1}(d), \ldots, \delta E_{n_2}(d)]$ for $n_2 - n_1 + 1 =I$. From here on, we will use the operator norm as a specific example of a unitarily invariant norm by setting $\lVert \cdot \rVert = \lVert \cdot \rVert_2$. Our key contribution lies in extending the proof of the theorem, building on existing results for block Krylov methods~\cite{Li2015}, to the real-time setting. Applying result C.2.2(b), we know there exists $B_I = [x_1, \ldots, x_I]$ with ${\rm span} B_I = {\rm span}\{ g_1, \ldots, g_I \}$ so that $\Pi_{{\rm invariant~subspace} } B_I = [f_{n_1}, \ldots, f_{n_2}]$. WLOG we assume that $B_I \mapsto B_I (B_I^{\dagger} B_{I})^{-1/2}$ has orthonormal columns, \textit{i.e.}, $\braket{x_{i}, x_{j}}_{\mu} = \delta_{ij}$ for $1\leq i,j\leq I$. For a degree-$(d-1)$ polynomial $p$, we define 
\begin{align}
    R_I = p(\Ko)B_I = \sum_{n=0}^{N-1} f_n p(\lambda_n) f_n^{\dagger} B_I,
\end{align}
where the dual $f_n^{\dagger}$ are defined with respect to the functional inner product $\braket{\cdot, \cdot}_{\mu}$. We will further write the equality above as
\begin{align}
    R_I = F p(\Lambda) F^{\dagger} B_I = \left[ F_{a} p(\Lambda_{a}) F^{\dagger}_{a} + F_{b} p(\Lambda_{b})  F^{\dagger}_{b} +  F_{c} p(\Lambda_{c}) F^{\dagger}_{c} \right] B_I,
\end{align}
where $F = [F_{a}~F_{b}~F_{c}]$ and $\Lambda = \Lambda_{a} \oplus \Lambda_{b} \oplus \Lambda_{c}$ contain the $N$ Koopman eigenfunctions and eigenvalues respectively in a block form. Here we use three subscripts $a$, $b$, and $c$ to label the partition of Hamiltonian spectrum into three disjoint blocks with energies below $E_{n_1}$, from $E_{n_1}$ to $E_{n_2}$, and above $E_{n_2}$. In particular, the middle block $b$ is represented by $F_{b} = [f_{n_1}, \ldots, f_{n_2}]$ and $\Lambda_b = {\rm diag}[\lambda_{n_1}, \ldots, \lambda_{n_2}]$.

By our construction, $F_b^{\dagger} B_I$ is nonsingular so
\begin{align}
   R_I (F_b^{\dagger} R_I)^{-1} = R_I \left[ p(\Lambda_b) F_b^{\dagger} B_I \right]^{-1} =  F_{a} p(\Lambda_{a}) F^{\dagger}_{a} B_I \left[ p(\Lambda_b) F_b^{\dagger} B_I \right]^{-1} + F_b + F_{c} p(\Lambda_{c}) F^{\dagger}_{c} B_I \left[ p(\Lambda_b) F_b^{\dagger} B_I \right]^{-1},
   \label{eq:R_FdaggerRinv}
\end{align}
whenever $p(\Lambda_b)$ is nonsingular by suitable choice of the polynomial $p$. Since ${\rm span}R_I$ is clearly a subspace of the Krylov subspace ${\rm Kr}_{d,I} := {\rm span}\{\vec{g}, \cdots, \Ko^{d-1}[\vec{g}] \}$, result C.2.1 implies 
\begin{align}
    \left \lVert \tan \Theta \left[{\rm span}F_b, {\rm Kr}_{d,I} \right] \right \rVert &\leq \Big \lVert \tan \Theta [{\rm span}F_b, {\rm span}R_I ] \Big \rVert, \\
    &= \left \lVert \sin{\Theta[{\rm span}F_b, {\rm span}R_I]} \left( \cos{\Theta[{\rm span}F_b, {\rm span}R_I]} \right)^{-1}  \right \rVert, \\
    &= \left \lVert [F_a~F_c]^{\dagger} R_I ( F_b^{\dagger} R_I )^{-1}  \right \rVert = \left \lVert \begin{bmatrix}
        p(\Lambda_a) & 0\\
        0 & p(\Lambda_c)
    \end{bmatrix} \begin{bmatrix}
        F_a^{\dagger} B_I ( F_b^{\dagger} B_I )^{-1} \\
        F_c^{\dagger} B_I ( F_b^{\dagger} B_I )^{-1}
    \end{bmatrix} p(\Lambda_b)^{-1} \right \rVert, \label{eq:Fdagger_B} \\
    &\leq  \left \lVert \begin{bmatrix}
        p(\Lambda_a) & 0\\
        0 & p(\Lambda_c)
    \end{bmatrix} \right \rVert \left \lVert p(\Lambda_b)^{-1} \right \rVert \left \lVert \begin{bmatrix}
        F_a^{\dagger} B_I ( F_b^{\dagger} B_I )^{-1} \\
        F_c^{\dagger} B_I ( F_b^{\dagger} B_I )^{-1}
    \end{bmatrix} \right \rVert, \\
    &\leq \frac{\displaystyle \max_{n<n_1 \lor n>n_2} \lvert p(\lambda_n) \rvert}{\displaystyle \min_{n_1 \leq n \leq n_2} \lvert p(\lambda_n) \rvert}  \left \lVert \tan \Theta[{\rm span}F_b, {\rm span}B_I ] \right \rVert,
    \label{ineq:eigvec_tan_convg}
\end{align}
where \cref{eq:Fdagger_B} results from the substitution of \cref{eq:R_FdaggerRinv}. 

The inequality above characterizes the convergence of the Krylov subspace (towards to the target eigenstates), and our task is to choose a polynomial $p$ that can tightly bound the fraction in \cref{ineq:eigvec_tan_convg}. As in \cref{subsec:theorem_1}, we exploit the Rogers-Szeg\H{o} polynomials $W_{d-1}(z|q)$ and work in the limit $q = 1$ to derive relevant results. For $n_2 = N-1$, let us consider the polynomial
\begin{align}
    p(z) = \frac{W_{d-1}(e^{-i \varphi}z)}{W_{d-1}(e^{-i \varphi}\lambda_{n_1})}, 
\end{align}
where $e^{-i\varphi}$ is a constant phase offset shifting the angles $\vartheta_{n < n_1}$ (recall that $\vartheta_n = E_n \Dt$) inside the window $\mathcal{W} = [-\Omega, \Omega]$ while keeping the angles $\vartheta_{n_1 \leq n \leq n_2}$ outside $\mathcal{W}$. In this case, observe that
\begin{align}
    \min_{n_{1} \leq n \leq n_{2}} \lvert p(\lambda_n) \rvert &= \lvert p(\lambda_{n_1}) \rvert = 1, \\
    \max_{n<n_1 \lor n>n_2} \lvert p(\lambda_n) \rvert &= \max_{n < n_{1}} \lvert p(\lambda_n) \rvert \leq \frac{1}{\left \lvert {W_{d-1}(e^{-i \varphi}\lambda_{n_{1}}) } \right \rvert}.
\end{align}
By employing suitable
time step size $\Dt$, \textit{e.g.}, that from \cref{eq:thm1_Dt_choice}, we assume $\vartheta_{n_{1} - 1} -  \vartheta_0 \leq 2\Omega$ and $2\Omega \leq \vartheta_{N-1} -\vartheta_0 \leq \pi+\Omega$. Specifically, $\varphi = \Omega - \vartheta_{n_{1} - 1}$ circularly shifts the eigenphases $\{ \lambda_n \}_{n=0}^{N-1}$. This ensures that $\abs{\varphi + \vartheta_0} \leq \varphi + \vartheta_{n_{1}-1} = \Omega \leq \varphi + \vartheta_{n_1}$ as intended. From Theorem 1, we recall $\Omega = \pi/3$ and $ \lvert W_{d-1}(e^{-i\varphi} \lambda_{n_1} ) \rvert^{1/(d-1)} \geq 1+ C \epsilon_{n_{1} -1 \mapsto n_{1} }$
where $\epsilon_{n_{1}-1 \mapsto n_{1}} = 3(E_{n_1} - E_{n_{1}-1}) \Dt/2\pi$ encodes the normalized $n_1$th energy gap and $C$ is some appropriate constant that can be set to $1$ (c.f. \cref{eq:Linbound}). This immediately implies an exponential convergence to the eigenspaces,
\begin{align}
    \left \lVert \tan \Theta \left[{\rm span}F_b, {\rm Kr}_{d,I} \right] \right \rVert \leq \frac{1}{\tilde{\epsilon}_{n_{1} -1 \mapsto n_{1}}^{(d-1)}  }  \left \lVert \tan \Theta\left[ {\rm span}F_b, {\rm span}\{g_1, \ldots, g_I \}  \right] \right \rVert,
\end{align}
for the base case $n_{2} = N-1$. When $n_{2} < N - 1$, we consider the polynomial $ p = p_1 \cdot p_2$ with
\begin{align}
     p_1(z) = \frac{W_{d-N+n_{2}}(e^{-i \varphi}z)}{W_{d-N+n_{2}}(e^{-i \varphi}\lambda_{n_1})},
     \label{eq:thm2_p1}
\end{align}
and 
\begin{align}
    p_2(z) = \prod_{n=n_{2}+1}^{N-1} \frac{z - \lambda_n}{\lambda_{n_{2}} - \lambda_n},
    \label{eq:thm2_p2}
\end{align}
where $p$ is factorized into lower order polynomials $p_1$ of degree $(d - N + n_2)$ and $p_2$ of degree $(N - n_{2} - 1)$. Observe that by design $p(\lambda_n) = 0$ for $n \geq n_2 + 1$ and thus
\begin{align}
    \min_{n_{1} \leq n \leq n_{2}} \lvert p(\lambda_n) \rvert &\geq \min_{n_{1} \leq n \leq n_{2}} \lvert p_1(\lambda_{n}) \rvert \min_{n_{1} \leq n \leq n_{2}} \left \lvert p_2(\lambda_n) \right \rvert = \min_{n_{1} \leq n \leq n_{2}} \prod_{m=n_{2}+1}^{N-1} \left \lvert \frac{\lambda_n - \lambda_m}{\lambda_{n_{2}} - \lambda_m} \right \rvert, \\
    \max_{n<n_1 \lor n>n_2} \lvert p(\lambda_n) \rvert &\leq \max_{n < n_{1}} \lvert p_1(\lambda_n) \rvert \max_{n < n_{1}} \lvert p_2(\lambda_n) \rvert \leq \frac{1}{\left \lvert {W_{d-N+n_{2}}(e^{-i \varphi}\lambda_{n_{1}}) } \right \rvert} \max_{n < n_{1}} \prod_{m=n_{2}+1}^{N-1} \left \lvert \frac{\lambda_n - \lambda_m}{\lambda_{n_{2}} - \lambda_m} \right \rvert,
\end{align}
which directly follows from applying our base case result to the
eigensector $\{ \lambda_n \}_{n=0}^{n_2}$. Note that our argument requires $d \geq N - n_2$: a symmetric bound can be readily established using the exact same argument with the roles of $n_1$ and $n_2$ exchanged, \textit{i.e.}, for $d \geq n_1 + 1$, upon a spectral flip $H \mapsto -H$.

Now we proceed to analyze the convergence of the Krylov energy approximation. For simplicity, we focus again on approximating from below Hamiltonian eigenvalues $E_{n} \lesssim E_{N-1}$ near the right edge of the spectrum. The argument for approximating from above eigenvalues near the left edge of the spectrum can be easily adapted with a spectral flip. We observe that up to the trivial spectral rotation $\Ko \mapsto \lambda_{n_2}^{-1} \Ko$
leaving the functional Krylov subspace invariant, we can assume WLOG $E_{n_{2} - 1} \leq 0 < E_{n_{2} + 1}$ for the remainder of the proof. For $V_{0} := R_I (F_b^{\dagger}R_I)^{-1}$, we first construct $V_{-} = V_{0}(V_{0}^{\dagger}V_{0})^{-1/2}$, which contains orthonormal columns. This allows us to introduce eigenvalues, $\hat{\lambda}_{n} = \lvert \hat{\lambda}_{n} \rvert e^{-i\hat{E}_n \Dt}$ for $n_1 \leq n \leq n_2$, of the ${\rm span}V_{0}$-projected Koopman operator $V^{\dagger}_{-} \Ko V_{-}$, with the ordering ${\rm arg}(\hat{\lambda}_{n_1}) \geq \ldots \geq {\rm arg}(\hat{\lambda}_{n_2})$. These eigenvalues depend explicitly on choices of the degree-$(d-1)$ polynomial $p$ (as $R_I$ does). Since ${\rm span}V_{0} \subseteq {\rm Kr}_{d,I}$ irrespective of $p$, the variational principle gives 
\begin{align}
 \hat{E}_{n}(d) \leq \tilde{E}_{n}(d) \leq E_n \implies 0 \leq \delta E_n(d) \leq E_{n} - \hat{E}_{n}(d), \quad n_{1} \leq n \leq n_{2}.
 \label{eq:thm2_spectral_error1}
\end{align}
The variational characterization, as we shall see, leads to an upper bound on the spectral error 
\begin{align}
    \bDel_{I,n_1,n_2} = {\rm diag}\begin{bmatrix}
        \delta E_{n_1} && \\
        &\ddots& \\
        && \delta E_{n_2}
    \end{bmatrix},
\end{align}
which generalizes $\bDel_{I}$ in Theorem 2 ($\bDel_{I,n_1,n_2} = \bDel_{I}$ when $n_1 = N-I$ and $n_2 = N-1$). With our previous assumption $E_{n_2} = 0$, we have $\hat{E}_n \leq 0$ for $n_1 \leq n \leq n_2$. Consequently, $\forall x \in \mathbb{C}^{I}$ 
\begin{align}
    -\frac{1}{\Dt} \theta(V_0 x) =  -\frac{1}{\Dt} {\rm arg} \frac{ \braket{ V_{0}x, \Ko V_{0} x}_{\mu} }{ \braket{ V_{0}x, V_{0} x}_{\mu} } \leq 0,
\end{align}
where we adopt the same notation as in Theorem 1 (c.f. \cref{eq:angle_expect}). By unfolding $V_0$ via \cref{eq:R_FdaggerRinv}, we arrive at
\begin{align}
   V_{0}^{\dagger} \Ko V_{0} &= G_{a}^{\dagger} \Lambda_{a} G_{a} + \Lambda_{b} + G_{c}^{\dagger} \Lambda_{c} G_{c} \label{eq:Vdagger_K_V},
\end{align}
and
\begin{align}
    V_{0}^{\dagger} V_{0} &= G_{a}^{\dagger} G_{a} + {\rm Id}_{I} + G_{c}^{\dagger} G_{c} \label{eq:Vdagger_V},
\end{align}
for $G_{a} = p(\Lambda_a) F_{a}^{\dagger} B_I [ p(\Lambda_b) F_{b}^{\dagger} B_I ]^{-1}$ and $G_{c} = p(\Lambda_c) F_{c}^{\dagger} B_I [ p(\Lambda_b) F_{b}^{\dagger} B_I ]^{-1}$. Substituting \cref{eq:Vdagger_K_V,eq:Vdagger_V} into the variational inequality, we can derive
\begin{align}
    -\frac{1}{\Dt}{\rm arg} \frac{x^{\dagger} (G^{\dagger}_a \Lambda_a G_a + \eta_b \lambda_{n_2} {\rm Id}_{I} ) x}{
 x^{\dagger}  x} -\frac{1}{\Dt} {\rm arg}\frac{x^{\dagger} \Lambda_b x}{ x^{\dagger} x} + \frac{1}{\Dt} {\rm arg} \lambda_{n_2}  &\leq -\frac{1}{\Dt} {\rm arg} \frac{ x^{\dagger} ( G_{a}^{\dagger} \Lambda_{a} G_{a} + \Lambda_{b} ) x }{ x^{\dagger} x }, \label{ineq:thm2_arg1} \\
    &\leq -\frac{1}{\Dt} {\rm arg} \frac{ x^{\dagger} ( G_{a}^{\dagger} \Lambda_{a} G_{a} + \Lambda_{b} + G_{c}^{\dagger}\Lambda_{c}G_{c}) x }{ x^{\dagger} (G_{a}^{\dagger} G_{a} + {\rm Id}_{I} + G_{c}^{\dagger}G_{c}) x}, \\
    &= -\frac{1}{\Dt} {\rm arg} \frac{ \braket{ V_{0}x, \Ko V_{0} x}_{\mu} }{ \braket{ V_{0}x, V_{0} x}_{\mu} } \leq 0 \label{ineq:thm2_arg2},
\end{align}
where $\eta_b = \inf_{x \in \mathbb{C}^{I}: x^{\dagger}x =1} \lVert x^{\dagger} \Lambda_b x \rVert$ is a constant of $\mathcal{O}(1)$ when $\vartheta_{n_2} - \vartheta_{n_1} < \pi/2$, and we recall that $E_{n_2} = 0 \implies \lambda_{n_2}=1$ in our current setting. In particular, the inequality \cref{ineq:thm2_arg1} above holds since for any unit vector $x \mapsto \frac{x}{\sqrt{x^{\dagger}x}}$,
\begin{align}
     {\rm arg} \frac{\displaystyle \lambda_{n_2} x^{\dagger} (G^{\dagger}_a \Lambda_a G_a + \Lambda_b ) x }{\displaystyle x^{\dagger} (G^{\dagger}_a \Lambda_a G_a + \eta_b \lambda_{n_2} {\rm Id}_{I} ) x \cdot x^{\dagger} \Lambda_b x} 
     = {\rm arg} \frac{\displaystyle \frac{\lambda_{n_2}}{x^{\dagger} \Lambda_b x} + \frac{\lambda_{n_2}}{x^{\dagger} G^{\dagger}_a \Lambda_a G_a x} }{\displaystyle 1 + \frac{ \eta_b \lambda_{n_2}}{x^{\dagger} G^{\dagger}_a \Lambda_a G_a x} } \leq {\rm arg} \frac{\displaystyle \frac{\lambda_{n_2}}{x^{\dagger} \Lambda_b x} + \frac{\lambda_{n_2}}{x^{\dagger} G^{\dagger}_a \Lambda_a G_a x} }{\displaystyle 1 + \frac{ \lVert x^{\dagger} \Lambda_b x \rVert \lambda_{n_2}}{x^{\dagger} G^{\dagger}_a \Lambda_a G_a x} } \leq 0,
\end{align}
where the numerator has a smaller phase than the denominator. The inequality \cref{ineq:thm2_arg2}, on the other hand, follows from our basic assumption that $E_{n} \leq 0$ for $n \leq n_2$. The inequalities indicate that the $G_{a}^{\dagger} \Lambda_{a} G_{a}$-eigenvalues, denoted as $\bar{\lambda}_{n}$ for $n_1 \leq n \leq n_2$ (with the eigenvalues ordered by decreasing arguments), are related to the $V^{\dagger}_{-} \Ko V_{-}$-eigenvalues through
\begin{align}
   -\frac{1}{\Dt} {\rm arg} ( \bar{\lambda}_{n} + \eta_b \lambda_{n_2} ) + E_n - E_{n_2} \leq  \hat{E}_n, \quad n_{1} \leq n \leq n_{2}.
\end{align}
Consequently, we may relax the error bound in \cref{eq:thm2_spectral_error1},
\begin{align}
     \delta E_n \leq E_n - \hat{E}_n &\leq \frac{1}{\Dt} {\rm arg}( \bar{\lambda}_{n}  + \eta_b \lambda_{n_2} ) + E_{n_2}, \\
     &\leq \frac{1}{\Dt} {\rm arg} \left[ \lambda_0 \frac{e^{\dagger}_n G^{\dagger}_a G_a e_n}{\eta_b e^{\dagger}_n e_n} + \lambda_{n_2} \right] + E_{n_2}, \\
     &\leq \frac{1}{\Dt} {\rm arg}\left[ \lambda_{0} \frac{y^{\dagger}_n G^{\dagger}_a G_a y_n}{\eta_b y^{\dagger}_n y_n} + \lambda_{n_2} \right] + E_{n_2},
     \label{ineq:thm2_arg3}
\end{align}
where $e_n$ denotes the eigenvector of $G^{\dagger}_a \Lambda_a G_a$ corresponding to $\bar{\lambda}_n$ and $y_{n}$ the eigenvector of $G^{\dagger}_a G_a$ corresponding to the $(n-n_1+1)$th largest eigenvalue. Importantly, we remark that \cref{ineq:thm2_arg1,ineq:thm2_arg3} can be established without phase ambiguity of $2\pi\mathbb{Z}$ by choosing suitable time step size $\Dt$.

We therefore seek a tight bound on the eigenvalues of $G_a^{\dagger} G_a$ (or equivalently the singular values of $G_a$). Notice that the singular values of $p(\Lambda_a)^{-1} \cdot G_a \cdot p(\Lambda_b) = p(\Lambda_a)^{-1} \cdot p(\Lambda_a) F_{a}^{\dagger} B_I [ p(\Lambda_b) F_{b}^{\dagger} B_I ]^{-1} \cdot p (\Lambda_b) = F_{a}^{\dagger} B_I [ F_{b}^{\dagger} B_I ]^{-1}$ are bounded above by the singular values of the canonical tangents (c.f. \cref{ineq:eigvec_tan_convg})
\begin{align}
    \tan \Theta[{\rm span}F_b, {\rm span}B_I] = \begin{bmatrix}
       F_a^{\dagger} B_I ( F_{b}^{\dagger} B_I)^{-1} \\
       F_c^{\dagger} B_I ( F_{b}^{\dagger} B_I )^{-1}
    \end{bmatrix}.
\end{align}
This allows us to effectively control the spectral radius of $G_a^{\dagger} G_a$,
\begin{align}
    \lVert G^{\dagger}_a G_a \rVert &\leq \lVert p(\Lambda_a) \rVert^2  \lVert p(\Lambda_b)^{-1} \rVert^2 \left \lVert \tan^2 \Theta[{\rm span}F_b, {\rm span}B_I] \right \rVert, \\ 
    &= \frac{\displaystyle \max_{n < n_1} \lvert p(\lambda_n) \rvert^2 }{\displaystyle \min_{n_1 \leq n \leq n_2} \lvert p(\lambda_n) \rvert^2 } \left \lVert \tan^2 \Theta[{\rm span}F_b, {\rm span}B_I] \right \rVert,
\end{align}
where it suffices to identify a degree-$(d-1)$ polynomial $p$ that tightly bounds the fraction on the RHS of the
expression above. In addition, here $p$ must satisfy the key orthogonality constraints,
\begin{align}
    \tilde{f}_n^{\dagger} R_I = [0 ~0~\ldots~0], \quad n_{2} + 1 \leq n \leq N-1,
\end{align}
where $\tilde{f}_n \in {\rm Kr}_{d,I}$ denotes the approximate eigenfunction corresponding to the Krylov eigenvalue $\tilde{\lambda}_n = \lvert \tilde{\lambda}_n \rvert e^{-i\tilde{E}_n\Dt}$. The set of $(N-n_2)$ constraints ensure that ${\rm span}R_I = p(\Ko){\rm span}B_I = p(\Ko){\rm span}\{g_1, \ldots, g_I\}$ belongs to the correct eigensector of the Krylov subspace.
Let us consider the familiar polynomial $p= p_1 \cdot p_2$ with $p_1$ of degree-$(d-N+n_2)$ and $p_2$ of degree $(N-n_2-1)$ given by 
\begin{align}
     p_1(z) = \frac{W_{d-N+n_{2}}(e^{-i \varphi}z)}{W_{d-N+n_{2}}(e^{-i \varphi}\lambda_{n_1})},
\end{align}
and 
\begin{align}
    p_2(z) = \prod_{n=n_{2}+1}^{N-1} \frac{z - \tilde{\lambda}_n}{\lambda_{n_{2}} - \tilde{\lambda}_n},
\end{align}
as introduced in \cref{eq:thm2_p1} and \cref{eq:thm2_p2} respectively. Accordingly, we can generalize \cref{eq:bound2} from Theorem 1,
\begin{align}
    \delta E_n &\leq -\frac{1}{\Dt} \arctan \left[ \frac{\zeta_{n_1, n_2} \sin\vartheta_{0} +
    \sin\vartheta_{n_2}  }{\zeta_{n_1, n_2} \cos\vartheta_{0} +
    \cos\vartheta_{n_2}  }  \right] + E_{n_2},\\
    &= \frac{ \sin(\vartheta_{n_2} -\vartheta_{0}) }{\Dt} \zeta_{n_1, n_2} + \mathcal{O}(\zeta^2_{n_1, n_2}),
\end{align}
where $\zeta_{n_1, n_2} = \eta_b^{-1} \iota_{n_1, n_2}^2 \Tilde{\epsilon}_{n_1 - 1 \rightarrow n_1}^{-2(d-N+n_{2})} \lVert \tan^2{\Theta} \rVert$ for
\begin{align}
    \iota_{n_1, n_2}  = \begin{cases}
        \displaystyle 1 &{\rm when}~ n_2 = N-1, \\
        \frac{\displaystyle \max_{n < n_{1}} \prod_{m=n_{2}+1}^{N-1} \left \lvert \frac{\lambda_n - \tilde{\lambda}_m}{\lambda_{n_{2}} - \tilde{\lambda}_m} \right \rvert}{\displaystyle \min_{n_{1} \leq n \leq n_{2}} \prod_{m=n_{2}+1}^{N-1} \left \lvert \frac{\lambda_n - \tilde{\lambda}_m}{\lambda_{n_{2}} - \tilde{\lambda}_m} \right \rvert} &{\rm otherwise}.
    \end{cases}
\end{align}
This immediately implies 
\begin{align}
    \lVert \bDel_{I,n_1,n_2} \rVert \leq \frac{\iota_{n_1, n_2}^2 
 \abs{\sin[(E_{n_2} - E_{0})\Dt]} \lVert \tan^2{\Theta} \rVert}{\eta_b \Tilde{\epsilon}_{n_1 - 1 \rightarrow n_1}^{2(d-N+n_{2})} \Dt},
\end{align}
which simplifies to the desired result when $n_2 = N-1$ as claimed. Finally, the dependence on conditioning, ${\rm cond}(\mathcal{B})^{1/2}$, can be reinstated to accommodate the nonorthogonal eigenfunction cases in analogy with Theorem 1. $\square$

\section{Conditioning of Koopman eigenfunctions in degenerate energy subspaces}
\label{app:conditioning_degenerate}

\subsection{Single-observable case}
Let us first consider the case when the ground state energy is degenerate. All Koopman eigenfunctions corresponding to $E_0$ acquire the same phase factor at each time step. Consequently, the single-observable signal collapses a degenerate eigenspace into a single spectral mode: while the energy $E_0$ remains identifiable, the individual Hamiltonian eigenstates do not. The mode $f_0(\ket{\phi}) = \braket{\psi_0|\phi} $ is replaced by an effective eigenfunction:
\begin{align}
   \tilde{f}_0(\ket{\phi}) = \sum_{n:E_n = E_0} w_n \braket{\psi_n|\phi},~\sum_{n:E_n = E_0} \lvert w_n \rvert^2 = 1,
\end{align}
which represents the ground state subspace. In the presence of degeneracies, the condition number ${\rm cond}(\mathcal{B})$ in Theorem 1 should be interpreted as ${\rm cond}(\tilde{\mathcal{B}})$, where $\tilde{\mathcal{B}}$ is the smaller $L^2(\mu)$ Gram matrix formed from the effective eigenfunctions associated with distinct energies. In other words, degenerate ground states (in fact any degenerate eigenstates) become linearly dependent under our $L^2(\mu)$ inner product and can never be resolved by a single-observable signal.

\subsection{Multi-observable case}
In comparison, degeneracies can be resolved by multi-observable signals. To see this, we revisit the example above of the degenerate ground state energy $E_0$ with some multiplicity $r_0$. Our signals $\vec g(\ket{\phi_k})$ contain, up to the common phase $e^{-iE_0 k\Dt}$, a collection of amplitudes $\mathfrak{A}_{i \mathfrak{a}} = \braket{\phi_0 | O_i | \psi_{0,\mathfrak{a}}}$ across observables $O_i$ and degenerate states $\{\ket{\psi_{0, \mathfrak{a}}}\}_{\mathfrak{a}=1}^{r_0}$. The amplitudes span a data-identifiable subspace, whose dimension is the rank of the $I \times r_0$ amplitude matrix $\mathfrak{A}$. The ground state mode thereby decomposes into ${\rm rank}(\mathfrak{A})$ effective eigenchannels.:
\begin{align}
    \tilde{f}_{E_0, \mathfrak{b}}(\ket{\phi}) = \sum_{\mathfrak{a} =1}^{r_0} w_{\mathfrak{b},(0,\mathfrak{a})} \braket{\psi_{0,\mathfrak{a}}|\phi}, \qquad
    \mathfrak{b} = 1,\ldots,{\rm rank}(\mathfrak{A}),
\end{align}
with the weight vectors $w_{\mathfrak{b}} \in \mathbb{C}^{r_0}$ forming an orthonormal basis of the row space ${\rm row}(\mathfrak{A})$.

More generally, the conditioning factor in Theorem 2 should be interpreted as the condition number of a \emph{block Gram matrix} $\mathcal{B}_{\mathrm{block}}$, constructed from Koopman eigenchannels $\tilde{f}_{E,\mathfrak{b}}$ that are distinguishable under the $L^2(\mu)$ inner product. With this replacement, ${\rm rank}(\mathfrak{A})$ matches the ground state multiplicity whenever the chosen observables resolve ground state subspace and ${\rm cond}(\mathcal{B}_{\mathrm{block}})$ becomes ${\rm cond}(\tilde{\mathcal{B}})$ in the single-observable limit. Therefore, the block Krylov structure underlying MODMD conceptually points to its ability to extract degenerate energies, and, with suitable observables, the associated eigenspaces.

\section{Additional simulation details}
\label{app:simulation details}
\subsection{Reference states}
\label{app:reference state}
Below we provide the exact reference states $\ket{\phi_0}$ employed for the TFIM and LiH calculations. The TFIM reference is an equal superposition of $6$ computational basis states, 
\begin{align*}
       \ket{\phi_0}_{\text{TFIM}}=\frac{1}{\sqrt 6}(\ket{000000000000000} + \ket{111111111111111} + \ket{100000000000000} \\+ \ket{000000001111111} + \ket{000000011111111} + \ket{000000111111111}), 
\end{align*}
while the LiH reference is a superposition of 
$6$ Slater determinant states,
\begin{align*}
    \ket{\phi_0}_{\text{LiH}}=\frac13(\ket{0000100001} + 2\ket{0001100001} + \ket{0000100011} + \ket{0111100001} + \ket{0000101111} + \ket{0000100111}),
\end{align*} where we work with the STO-3G basis set (2 core and 2 valence electrons) to represent the second-quantized molecular Hamiltonian. Note that both reference states are superpositions of a small number of computational basis states and thus have efficient classical representation.

The reference states for the additional simulations in \cref{sec: Additional Quantum Systems} are
\begin{align*}
    \ket{\phi_0}_\text{Heisenberg} = \frac12(\ket{000000000000000} + \ket{100000000000000} + \ket{110000000000000} + \ket{111000000000000})
\end{align*}
\begin{align*}
    \ket{\phi_0}_{\text{BeH}_2}=\frac13(\ket{000010111101} + \ket{001110111101} + \ket{000110111101} \\+ \ket{000010110001} + 2\ket{000010111001} + \ket{001100111101})
\end{align*}
\begin{align*}
    \ket{\phi_0}_{\text{N}_2} =\frac13(\ket{00010101110101} + \ket{00010101000101} + 2\ket{00010101100101} \\ + \ket{01110101110101}+\ket{00110101110101}+\ket{00010101001001})
\end{align*}
The unequal superposition for all three molecular Hamiltonians (LiH, $\text{BeH}_2$, and $\text{N}_2$) was necessary to ensure nonzero overlap with the first four eigenstates of each system.

For our comparative study between MODMD and QMEGS in \cref{subsec:QMEGS}, we employ the reference state,
\begin{align}
    \nonumber
    \ket{\varphi_{0}}_{\rm TFIM}= \frac{1}{\sqrt 6}(\ket{000000000000}+\ket{000000000001}+\ket{000111111111} \\
    +\ket{100000000000}+\ket{111111111110}+\ket{111111111111}).
\end{align} 

For our comparative study between MODMD and (U)VQPE in \cref{subsec:(U)VQPE}, we employ the reference state,
\begin{align}
    \nonumber
    \ket{\varphi_{0}}_{\rm TFIM} = \frac{1}{\sqrt{18}}(\ket{000000000000000}+\ket{111111111111111}+\ket{100000000000000}+\ket{000000001111111} \\ \nonumber
    +\ket{000000011111111}+\ket{000000111111111}+\ket{000001111111111}+\ket{000011111111111}\\ \nonumber
    +\ket{000111111111111}+\ket{001111111111111}+\ket{111111111111110}+\ket{000000000000001}\\ \nonumber
    +\ket{110000000000000}+\ket{111000000000000}+\ket{111100000000000}+\ket{111110000000000}\\ 
    +\ket{111111000000000}+\ket{111111100000000}),
\end{align}
which comprises a superposition of 18 computational basis states and is less structured than $\ket{\phi_0}_{\rm TFIM}$ in \cref{subsec:TFIM}.

\subsection{Convergence of eigenenergies}
\label{app:crossover of convergence}

\begin{figure*} [h]
    \centering
    \begin{subfigure}{0.49\textwidth}
        \centering
        \includegraphics[scale = .55]{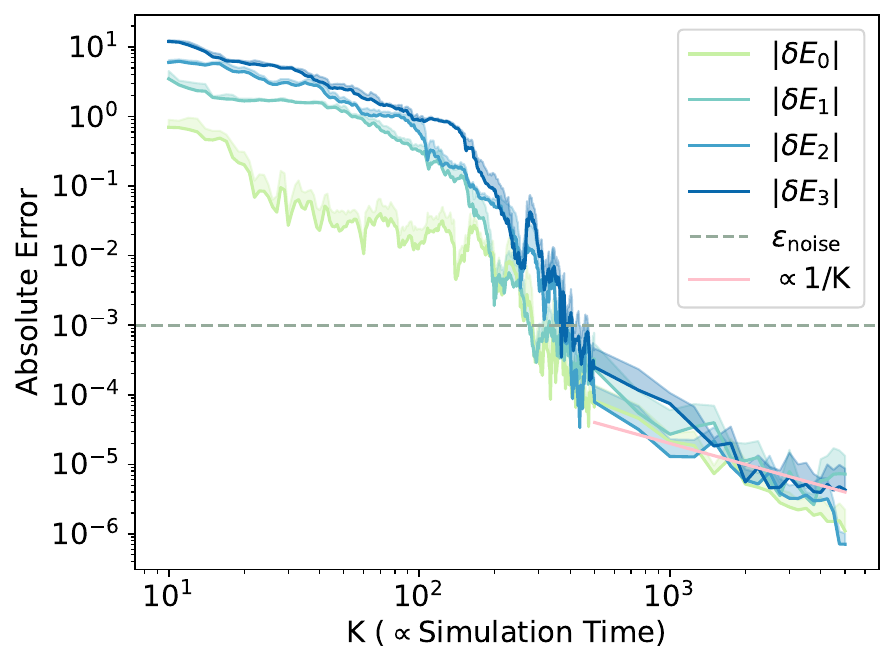}
        \caption{TFIM MODMD convergence}
        \label{fig: TFIM K convergence LogLog}
    \end{subfigure}%
    ~ 
    \begin{subfigure}{0.49\textwidth}
        \centering
        \includegraphics[scale = .55]{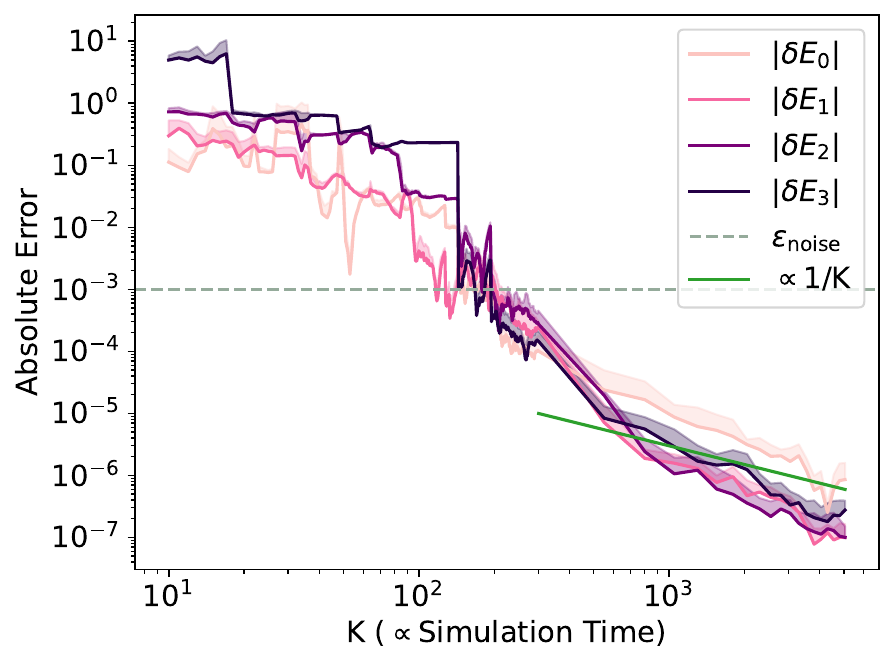}
        \caption{LiH MODMD convergence}
        \label{fig: LogLog MODMD LiH}
    \end{subfigure}
    \caption{Convergence of eigenenergies for TFIM and LiH. To obtain eigenenergy estimates $\tilde{E}_n$, we fix the MODMD parameters $\frac{K}{d}=\frac{5}{2}$ and $\Tilde{\delta}=10^{-2}$ for constructing and thresholding the data matrix pair $\mathbf{X},\mathbf{X}' \in \mathbb{R}^{d I \times (K+1)}$. Gaussian $\mathcal N(0,\varepsilon_{\rm noise}^2)$ noise with $\varepsilon_{\rm noise} = 10^{-3}$ is added independently to distinct matrix elements. The absolute errors, $\lvert \tilde{E}_n - E_n \rvert$, in the first four lowest target eigenenergies of the TFIM and LiH Hamiltonians are shown with respect to $K$ proportional to the non-dimensional maximal simulation time. The reference states $\ket{\phi_0}$ are sparse in the computational basis as prescribed in \cref{app:reference state}. The shading above the solid lines indicates standard deviation across repeated trials for each quantity. \textbf{Left}. TFIM absolute errors from the MODMD algorithm with $\Dt \approx 0.08$ and $I = 7$ (plotted on a log-log scale), where the convergence results are averaged over 5 trials, each involving a Gaussian noise realization and a random selection of $I-1$ observables. \textbf{Right}. LiH absolute errors from the MODMD algorithm with $\Dt \approx 0.39$ and $I=7$ (on a log-log scale), where the results are averaged over 20 trials, each involving a Gaussian noise realization.}
    \label{fig: LogLog MODMD Convergence}
\end{figure*}

In order to better illustrate the regime division in error convergence, as explained in \cref{sec:applications}, we provide alternative visualizations of \cref{fig: TFIM K Convergence MODMD} and \cref{fig: LiH K Convergence MODMD}. \cref{fig: LogLog MODMD Convergence} shows the absolute energy error as a function of time steps on the log-log scale for a larger range of $K$ values, highlighting the transition from the exponential to algebraic convergence. As the number of time steps increases, the log-log plot demonstrates the linear behavior expected for algebraic error decay.

\section{Additional Benchmarks on Quantum Systems}
\label{sec: Additional Quantum Systems}
\begin{figure*} 
    \centering
    \begin{subfigure}[t]{0.49\textwidth}
        \centering
        \includegraphics[scale = .55]{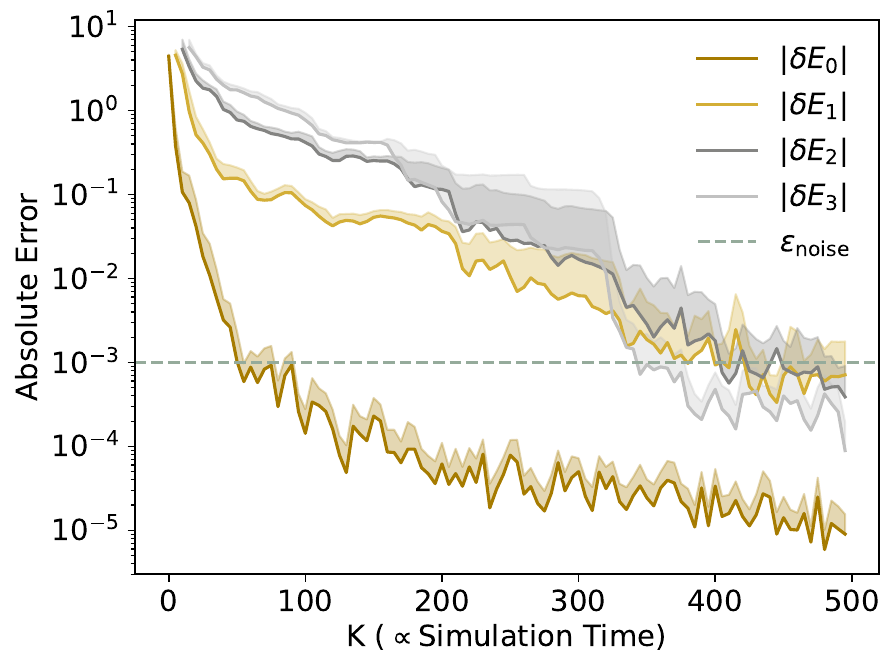}
        \caption{MODMD algorithm}
        \label{fig: Heisenberg K Convergence MODMD}
    \end{subfigure}%
    ~ 
    \begin{subfigure}[t]{0.49\textwidth}
        \centering
        \includegraphics[scale = .55]{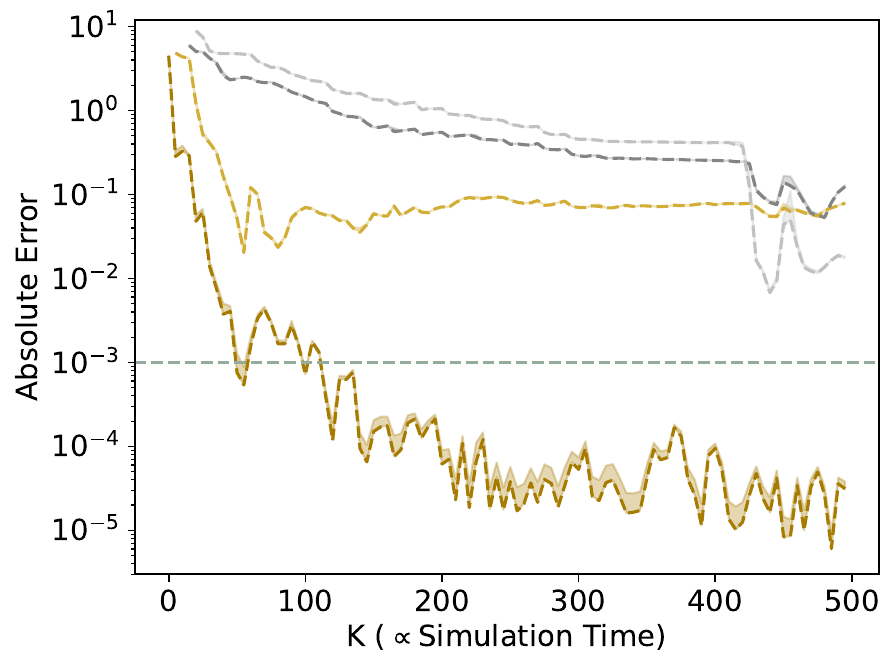}
        \caption{ODMD algorithm}
        \label{fig: Heisenberg K Convergence ODMD}
    \end{subfigure}
    \caption{Convergence of eigenenergies for the isotropic Heisenberg model. To obtain eigenenergy estimates $\tilde{E}_n$, we fix the (M)ODMD parameters $\frac{K}{d}=\frac{5}{2}$ and $\Tilde{\delta}=10^{-2}$ for constructing and thresholding the pair of data matrices $\mathbf{X},\mathbf{X}' \in \mathbb{R}^{d I \times (K+1)}$. Gaussian $\mathcal N(0,\varepsilon_{\rm noise}^2)$ noise with $\varepsilon_{\rm noise} = 10^{-3}$ is added independently to the real or/and imaginary parts of the matrix elements. Absolute errors, $\lvert \tilde{E}_n - E_n \rvert$, in the first four eigenenergies of the Heisenberg Hamiltonian are shown with respect to $K$ proportional to the non-dimensional maximal simulation time. The reference state $\ket{\phi_0}$ is an equal superposition of four computational basis states (see \cref{app:reference state}) and we employ a time step of $\Dt \approx 0.09$. The shading above the solid/dashed lines shows the standard deviation across trials for each quantity. For the Heisenberg model, the Hamiltonian is of size $32768 \times 32768$ and the largest linear system in the corresponding MODMD LS problem has size $1400 \times 501$. \textbf{Left}. Absolute errors from the multi-observable (MODMD) algorithm with $I = 7$ distinct observables, where the convergence results are averaged over 20 trials, each involving a Gaussian noise realization and a selection of $I-1$ random observables. \textbf{Right}. Absolute errors from the single-observable (ODMD) algorithm where the convergence results are averaged over 20 trials, each involving a Gaussian noise realization.}
    \label{fig: Heisenberg K Convergence}
\end{figure*}

\begin{figure*}[t]
    \centering
    \begin{subfigure}[t]{0.49\textwidth}
        \centering
        \includegraphics[scale = .55]{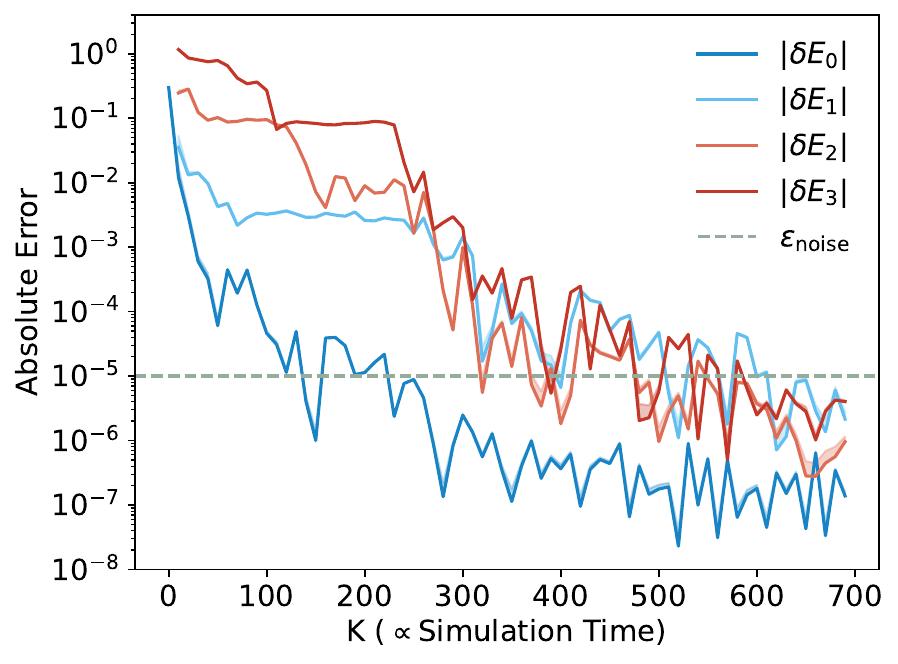}
        \caption{MODMD algorithm}
        \label{fig: BeH2 K Convergence MODMD}
    \end{subfigure}%
    ~ 
    \begin{subfigure}[t]{0.49\textwidth}
        \centering
        \includegraphics[scale = .55]{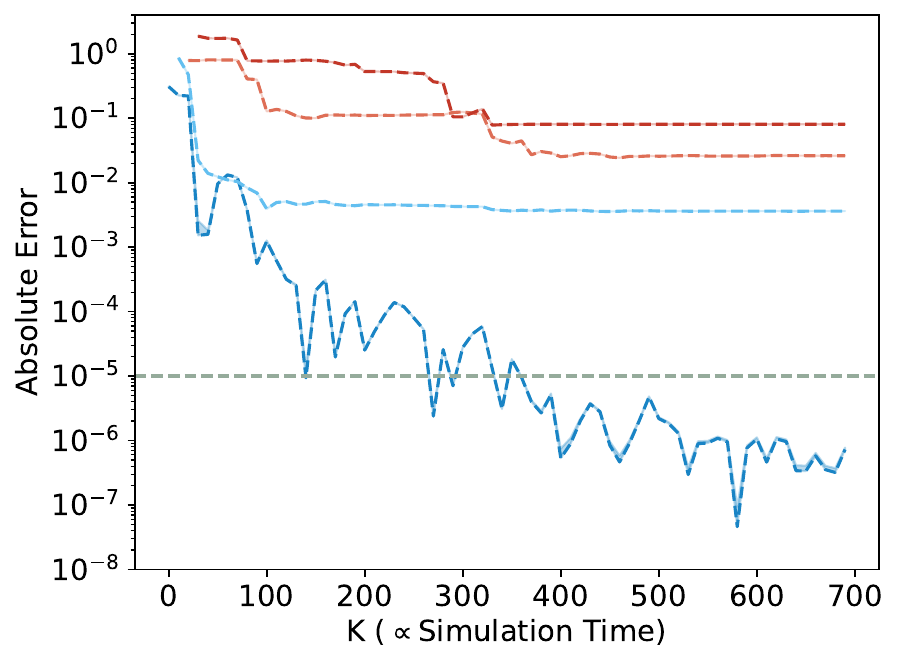}
        \caption{ODMD algorithm}
        \label{fig: BeH2 K Convergence ODMD}
    \end{subfigure}
    \caption{Convergence of eigenenergies for the beryllium hydride ($\text{BeH}_2$) molecule. To obtain eigenenergy estimates $\tilde{E}_n$, we fix the (M)ODMD parameters $\frac{K}{d}=\frac{5}{2}$ and $\Tilde{\delta}=10^{-4}$ for constructing and thresholding the pair of data matrices $\mathbf{X},\mathbf{X}' \in \mathbb{R}^{d I \times (K+1)}$. Gaussian $\mathcal N(0,\varepsilon_{\rm noise}^2)$ noise with $\varepsilon_{\rm noise} = 10^{-5}$ is added independently to the real or/and imaginary parts of the matrix elements. The reference state $\ket{\phi_0}$ is a superposition of six Slater determinant states (see \cref{app:reference state}) and we employ a time step of $\Delta t=0.20$. The absolute errors, $\lvert \tilde{E}_n - E_n \rvert$, in the first four eigenenergies of the $\text{BeH}_2$ Hamiltonian are shown with respect to $K$ proportional to the non-dimensional maximal simulation time. Convergence results are averaged over 20 trials of Gaussian noise realizations with shading  above the solid/dashed lines showing the standard deviation across trials for each quantity. For $\text{BeH}_2$, the model Hamiltonian is of size $4096 \times 4096$ and the largest linear system in the corresponding MODMD LS problem has size $1960 \times 701$. \textbf{Left}. Absolute errors from the multi-observable (MODMD) algorithm with $I = 7$ observable types. \textbf{Right}. Absolute errors from the single-observable (ODMD) algorithm.}
    \label{fig: BeH2 K Convergence}
\end{figure*}

\begin{figure*}[h!]
    \centering
    \begin{subfigure}[t]{0.49\textwidth}
        \centering
        \includegraphics[scale = .55]{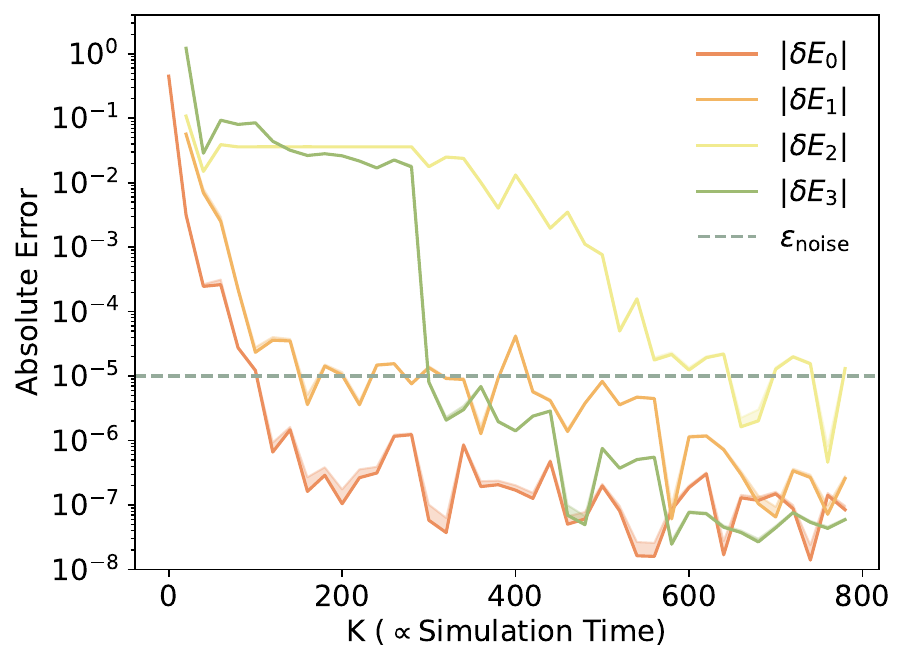}
        \caption{MODMD algorithm}
        \label{fig: N2 K Convergence MODMD}
    \end{subfigure}%
    ~ 
    \begin{subfigure}[t]{0.49\textwidth}
        \centering
        \includegraphics[scale = .55]{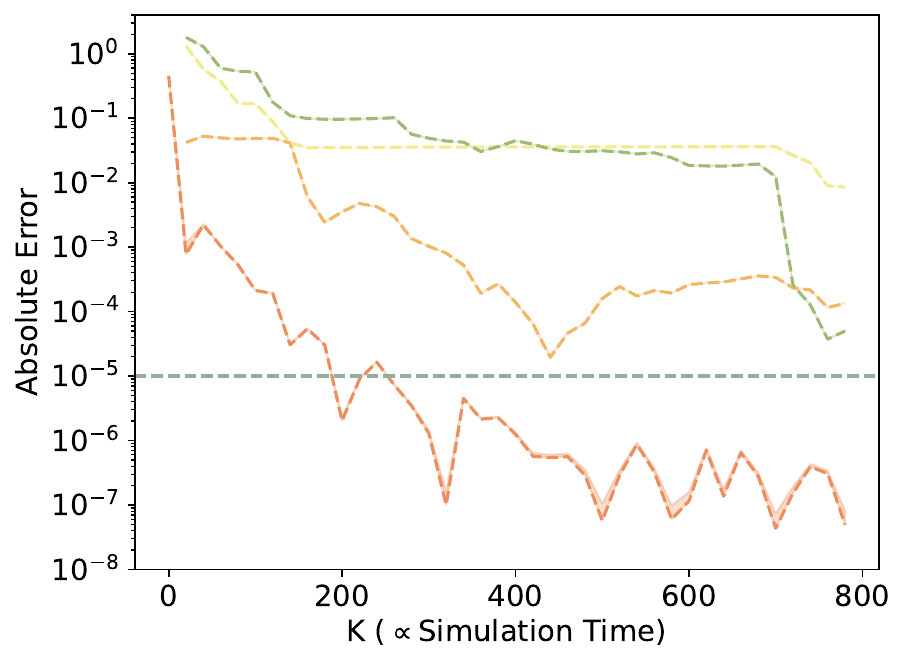}
        \caption{ODMD algorithm}
        \label{fig: N2 K Convergence ODMD}
    \end{subfigure}
    \caption{Convergence of eigenenergies for molecular nitrogen ($\text{N}_2$). To obtain eigenenergy estimates $\tilde{E}_n$, we fix the (M)ODMD parameters $\frac{K}{d}=\frac{5}{2}$ and $\Tilde{\delta}=10^{-4}$ for constructing and thresholding the pair of data matrices $\mathbf{X},\mathbf{X}' \in \mathbb{R}^{d I \times (K+1)}$. Gaussian $\mathcal N(0,\varepsilon_{\rm noise}^2)$ noise with $\varepsilon_{\rm noise} = 10^{-5}$ is added independently to the real or/and imaginary parts of the matrix elements. The reference state $\ket{\phi_0}$ is a superposition of six Slater determinant states (see \cref{app:reference state}) and we use a time step of $\Delta t=0.37$. The absolute errors, $\lvert \tilde{E}_n - E_n \rvert$, in the first four eigenenergies of the $\text{N}_2$ Hamiltonian are shown with respect to $K$ proportional to the non-dimensional maximal simulation time. Convergence results are averaged over 20 trials of Gaussian noise realizations with shading  above the solid/dashed lines showing the standard deviation across trials for each quantity. For $\text{N}_2$, the model Hamiltonian is of size $16384 \times 16384$ and the largest linear system in the corresponding MODMD LS problem has size $2240 \times 801$. \textbf{Left}. Absolute errors from the multi-observable (MODMD) algorithm with $I = 7$ observable types. \textbf{Right}. Absolute errors from the single-observable (ODMD) algorithm.}
    \label{fig: N2 K Convergence}
\end{figure*}

In this section we present additional benchmark results similar to those presented in \cref{sec:applications}, but for new quantum systems. The behaviors illustrated in \cref{fig: Heisenberg K Convergence}, \cref{fig: BeH2 K Convergence}, and \cref{fig: N2 K Convergence} are consistent with those highlighted in the main text. While the ground state energy converges at a somewhat comparable rate for both the single- and multi-observable algorithms, MODMD achieves high accuracies for the excited state energies significantly more rapidly, reaching errors below the noise level for relatively small $K$.

\begin{figure} [t!]
    \centering
    \includegraphics[scale=.55]{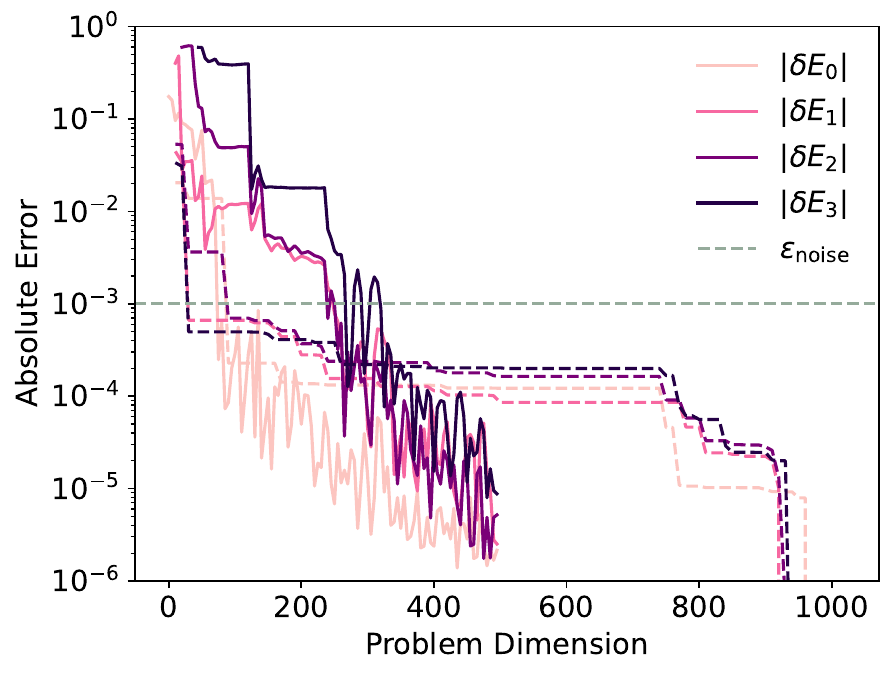}
    \caption{Performance comparison of MODMD and truncated configuration interaction for the lithium hydride (LiH) molecule. To obtain eigenenergy estimates $\tilde{E}_n$, we fix the MODMD parameters $\frac{K}{d}=\frac{5}{2}$ and $\Tilde{\delta}=10^{-2}$ for constructing and thresholding data matrices $\mathbf{X},\mathbf{X}' \in \mathbb{R}^{d I \times (K+1)}$. Gaussian $\mathcal N(0,\varepsilon_{\rm noise}^2)$ noise with $\varepsilon_{\rm noise} = 10^{-3}$ is added independently to the real or/and imaginary parts of the matrix elements. The reference state $\ket{\phi_0}$ is a superposition of six Slater determinants (see \cref{app:reference state}) and we employ a time step of $\Delta t\approx0.39$. As a baseline, eigenenergy estimates are also computed through exact (and therefore classical) diagonalization of truncated configuration interaction Hamiltonian, where we project the full problem onto a subspace spanned by lowest-energy Slater determinants. The absolute errors, $\lvert \tilde{E}_n - E_n \rvert$, in the first four eigenenergies of the LiH Hamiltonian are shown with respect to $K$ proportional to the non-dimensional maximal simulation time. MODMD convergence results are averaged over 20 trials of Gaussian noise realizations. Solid and dashed lines indicate the MODMD and truncated configuration interaction estimates. For LiH, the full Hamiltonian is of size $1024 \times 1024$ whereas the largest linear system in the corresponding MODMD LS problem has size $1400 \times 501$. }
    \label{fig: Truncated CI Comparison}
\end{figure}

\subsection{Lithium Hydride Molecule}
While increasing the number of observables enhances richness of the signal subspace and thus improves accuracy at a favorable quantum cost, it unavoidably raises the classical cost of solving the relevant matrix LS problem in MODMD. To more quantitatively assess the trade-off between quantum and classical resources, we compare the performance of MODMD with that of the classical truncated configuration interaction method on the LiH molecule.

In particular we can construct the truncated configuration interaction (CI) Hamiltonian~\cite{Zgid2012} in a natural way: for a given subspace dimension, we select the Slater determinants with the lowest diagonal energies and project the full CI Hamiltonian onto the corresponding low-energy subspace. For each linear system solved in the MODMD LS problem, we solve a truncated CI problem of the same subspace dimension, therefore enabling a direct comparison between the two approaches in terms of accuracy and computational cost.

In \cref{fig: Truncated CI Comparison}, we observe that the classical truncated CI method outperforms MODMD when the number of time steps is either relatively small -- due to the greedy selection of initial Slater determinant(s) -- or relatively large, where the CI subspace approaches exact diagonalization. For intermediate values of time step number, however, MODMD achieves measurably higher accuracies, outperforming truncated CI by one to two orders of magnitude. Such a non-monotonic dependence points to a regime in which MODMD offers a favorable balance of accuracy and resource efficiency.

\subsection{Quantum Heisenberg Model}
As another example from condensed matter physics, we examine the convergence of the ground and excited
state eigenenergies of the 15-qubit isotropic (also known as XXX) Heisenberg model subject to open boundary conditions. The system Hamiltonian is given by 
\begin{align}
    H_{\text{Heisenberg}}=- J\sum_{i=1}^{L-1}  X_i X_{i+1}+Y_i Y_{i+1}+Z_i Z_{i+1} - h \sum_{i=1}^{L} Z_i,
    \label{eq:H_Heisenberg}
\end{align}
The strategy used to select the observable pool is the same as in the transverse-field Ising model calculations, and the reference state is an equal superposition of four computational basis states.

\subsection{Beryllium Hydride Molecule}
The second chemical Hamiltonian for which we evaluate the performance of MODMD is that of beryllium hydride ($\text{BeH}_2$). The 12-qubit Hamiltonian is generated in an equilibrium linear geometry with a bond length of  $\approx 2.67$\AA \text{ }using the STO-3G basis set and parity fermion-to-qubit mapping. The observable pool was constructed in a similar fashion to the lithium hydride calculations and the reference state is a superposition of six Slater determinants, each defining a Hartree-Fock-like configuration with single-particle excitations. Here, the simulated $\varepsilon_\text{noise}$ and defined $\tilde\delta$ values were set smaller than those for LiH in the main text since there is a near-degeneracy between two of the lowest eigenstates, requiring less severe SVD truncation in order to accurately resolve the spectrum.
\subsection{Molecular Nitrogen}
As a third chemical example, we evaluate the performance of MODMD on the molecular nitrogen ($\text{N}_2$) Hamiltonian. The 14-qubit Hamiltonian is generated in an equilibrium geometry with a bond length of  $\approx 1.10$\AA \text{ }using the STO-3G basis set and parity fermion-to-qubit mapping. The observables were chosen following a procedure similar to that used in the lithium hydride calculations and the reference state is a superposition of six Slater determinants, each defining a Hartree-Fock-like configuration with single-particle excitations. Here, the simulated $\varepsilon_\text{noise}$ and defined $\tilde\delta$ values were set smaller than those for LiH in the main text, due to a near-degeneracy between two of the lowest eigenstates.

\section{Parameter Sweeps}
\label{app:parameter_sweeps}
\subsection{MODMD: Lithium Hydride Molecule}
\begin{figure*}[t!]
    \centering
    \begin{subfigure}[t]{\textwidth}
        \centering
        \includegraphics[width=\textwidth]{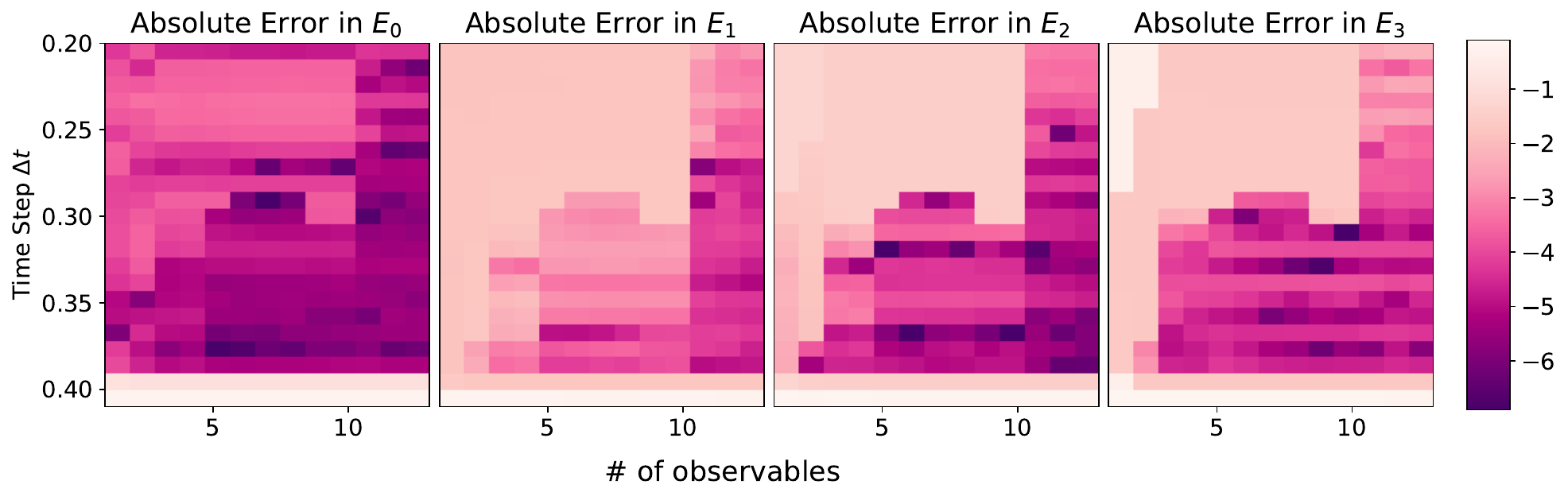}
        \caption{Time step and observable pool size sweep}
        \label{fig: LIH IDeltaT Sweep}
    \end{subfigure}%
    \hfill
    ~
    \begin{subfigure}[t]{\textwidth}
        \centering
        \includegraphics[width=\textwidth]{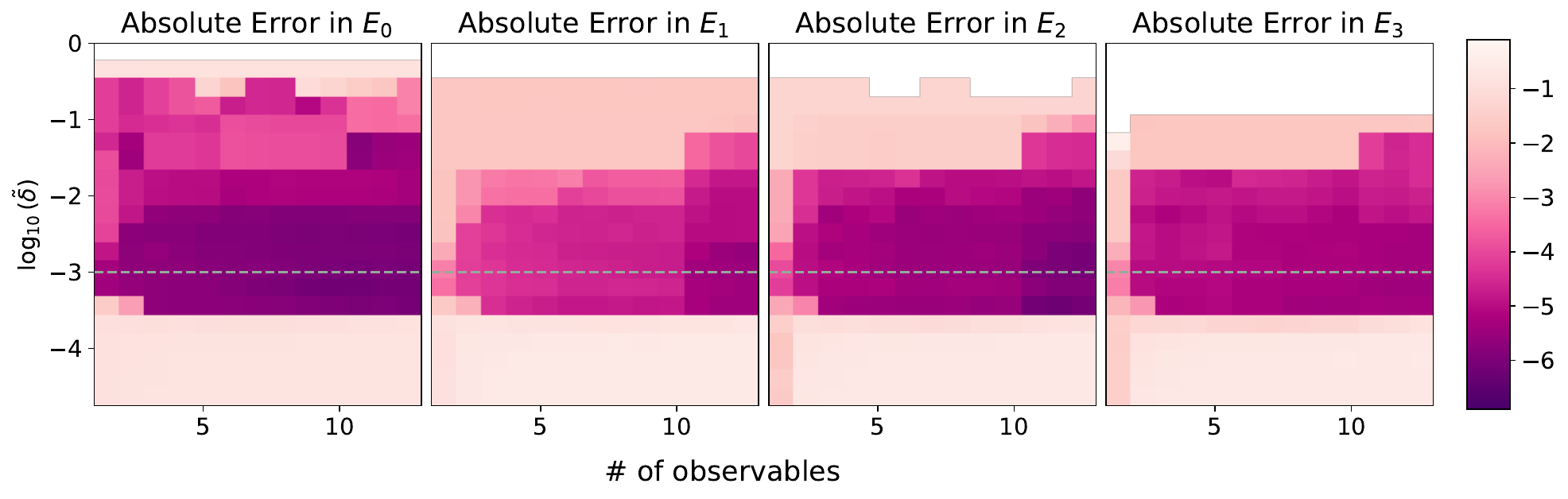}
        \caption{SVD threshold and observable pool size Sweep}
        \label{fig: LiH IThrshold Sweep}
    \end{subfigure}
    \caption{Parameter sweeps for convergence of eigenenergies for the lithium hydride (LiH) molecule. To obtain eigenenergy estimates $\tilde{E}_n$, we fix the (M)ODMD parameters $\frac{K}{d}=\frac{5}{2}$ and $K = 500$ for constructing and thresholding the pair of data matrices $\mathbf{X},\mathbf{X}' \in \mathbb{R}^{d I \times (K+1)}$. Gaussian $\mathcal N(0,\varepsilon_{\rm noise}^2)$ noise with $\varepsilon_{\rm noise} = 10^{-3}$ is added independently to the real or/and imaginary parts of the matrix elements. The reference state $\ket{\phi_0}$ is a superposition of six  Slater determinants (see \cref{app:reference state}). The absolute errors, $\lvert \tilde{E}_n - E_n \rvert$, in the first four eigenenergies of the LiH Hamiltonian are shown where convergence results are averaged over 20 trials of Gaussian noise realizations. \textbf{Left}. Absolute errors from the MODMD algorithm for varying $\Dt$ and $I$ values. The singular value threshold is fixed at the empirically near-optimal value of $\tilde\delta=10^{-2}$. \textbf{Right}. Absolute errors from the MODMD algorithm for varying $\tilde\delta$ and $I$ values. The time step is fixed at the empirically near-optimal value of $\Dt\approx0.39$.}
\end{figure*}

Here we present additional results for the eigenenergy convergence of the LiH molecule. We examine the performance of MODMD under different parameter regimes upon sweeping the time step $\Dt$ against the size of the observable pool $I$ in one set of calculations (\cref{fig: LIH IDeltaT Sweep}) and the singular value threshold $\tilde\delta$ against $I$ in another (\cref{fig: LiH IThrshold Sweep}).

As discussed in \cref{subsec:selection of hyperparameters}, we expect that an excessively large time step will cause an undesired aliasing ambiguity in eigenphases, independent of the number of observables, as is evident in the edge behavior of \cref{fig: LIH IDeltaT Sweep}. Conversely, an overly small time step also slows down the convergence. We find that using a richer subspace, constructed from our multi-observable signal, can overcome this limitation. For all four eigenenergies, increasing the number of observables generally leads to reliable convergence in a larger range of $\Dt$ values, particularly for the excited state energies. This robustness can be advantageous in an experimental setting where the achieved evolution time is relatively imprecise. Additionally, a smaller time step reduces the overall Trotter error~\cite{Zhu_2021}. 

As mentioned in \cref{subsec:selection of hyperparameters}, using a $\tilde\delta$ value that is too small (for instance, below the noise level) retains excessive noise, preventing proper eigenenergy convergence regardless of how many observables we employ. Conversely, a $\tilde\delta$ value that is too large truncates too much of the information recovered from MODMD measurements. This, however, can be overcome to some extent by increasing the number of observables in the observable pool. We observe that increasing $I$ generally increases the range of $\tilde\delta$ values at which the lowest four eigenenergies achieve good convergence. Measuring several observables thus allows for a harsher truncation (larger $\tilde{\delta})$ in the post-processing stage, better ensuring that the algorithm $(i)$ does not fall into excessive noise retention regime and $(ii)$ is less sensitive to hyperparameter tuning.

\subsection{QMEGS: Transverse-Field Ising Model}
\label{app: QMEGS sweep}
\begin{figure} [t!]
    \centering
    \includegraphics[scale=.55]{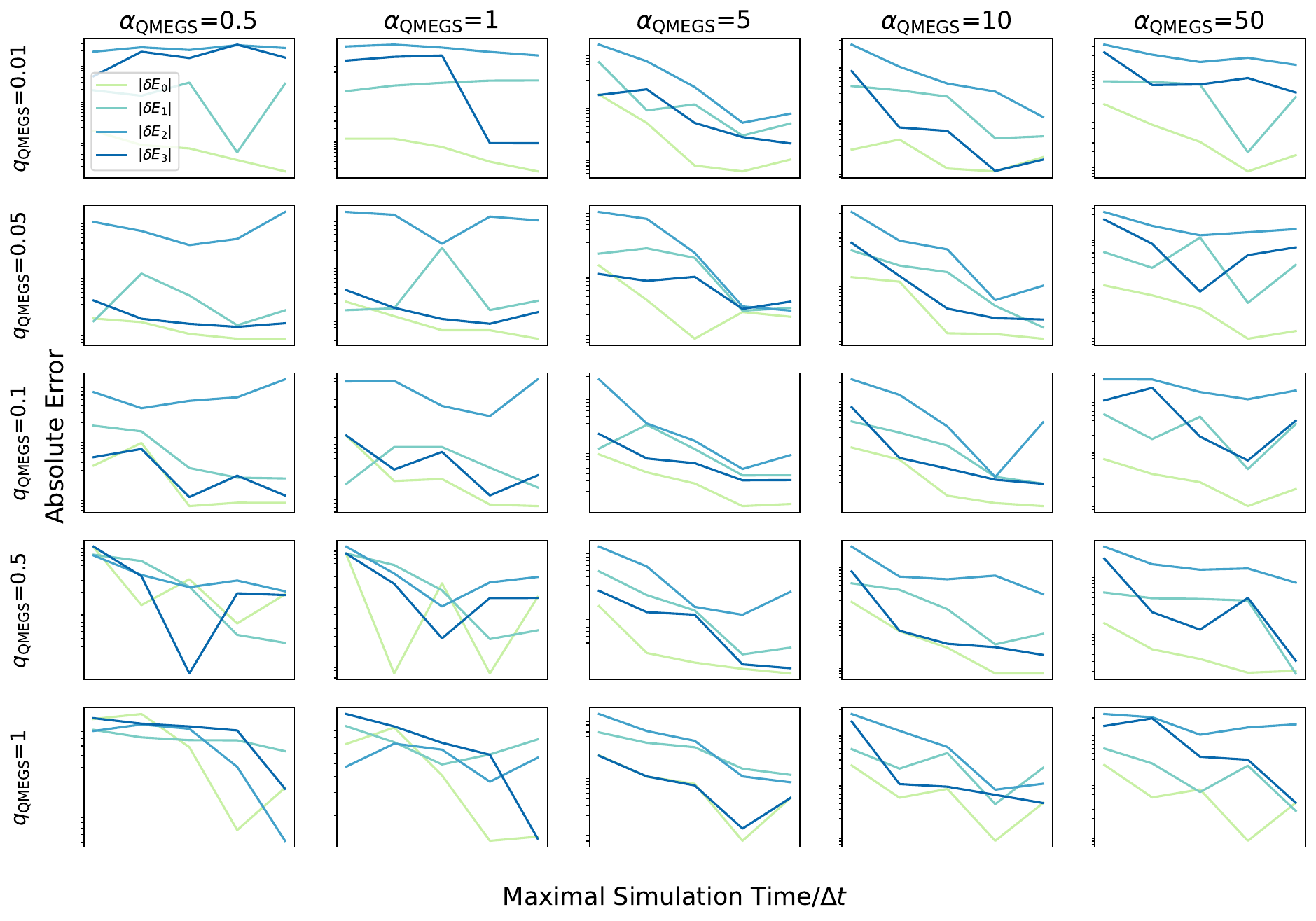}
    \caption{Convergence of QMEGS eigenenergy approximations on the 12-qubit transverse-field Ising model (TFIM) for varying values of the QMEGS hyperparameters $\alpha_{\rm QMEGS}$ and $q_{\rm QMEGS}$.}
    \label{fig: QMEGS hyperparameters}
\end{figure}

QMEGS features two hyperparameters in its classical search-and-block post-processing: the block size $\alpha_{\rm QMEGS}$ and the search resolution $q_{\rm QMEGS}$. Following~\cite{ding2024quantummultipleeigenvaluegaussian}, we adopt $q_{\rm QMEGS} < \alpha_{\rm QMEGS} \ll t_{\rm max}$. To maximize performance, we also conduct a systematic hyperparameter sweep around the values reported in the original QMEGS work, $\alpha_{\rm QMEGS} = 5$ and $q_{\rm QMEGS} =0.05$. The optimal pair is selected by minimizing the total QMEGS error defined as the sum of errors over the four lowest energy levels and simulations of increasing maximal runtime:
\begin{align}
    (\alpha_{\rm QMEGS}^*,q_{\rm QMEGS}^*) =\argmin_{(\alpha_{\rm QMEGS},q_{\rm QMEGS}) } \sum_{t_{\rm max}} \sum_{n=0}^{N_{\rm eig}-1}\ \lvert \delta E_n(t_{\rm max},\alpha_{\rm QMEGS},q_{\rm QMEGS}) \rvert,
    \label{eq:qmegs_hyperparameters}
\end{align}  
where \cref{fig: QMEGS hyperparameters} displays the results of this hyperparameter sweep, including relative performance plots similar to \cref{fig: QMEGS Comparison} for varying values of $\alpha_{\rm QMEGS}$ and $q_{\rm QMEGS}$. As explained in the main text, the error $\delta E_n$ given in \cref{eq:qmegs_hyperparameters} is optimized over the set of observables $\{O_i\}_{i=1}^{I}$ such that $\displaystyle|\delta E_n^{\rm{QMEGS}}|=\min_{1\leq i\leq I}|\delta E_n^{\rm{QMEGS}}(O_i)|$.

\begin{figure} [t!]
    \centering
    \includegraphics[scale=.55]{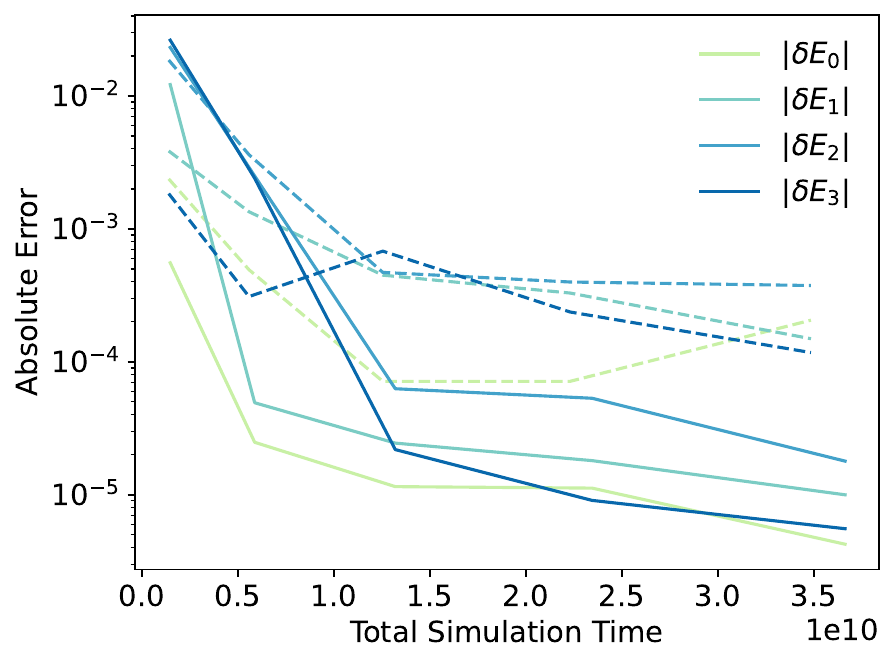}
    \caption{Comparison of MODMD and QMEGS on the transverse-field Ising model (TFIM). Absolute errors in the first four eigenenergies (obtained via QMEGS and MODMD with $I =7$ distinct observables) are shown with respect to total simulation time for each algorithm. MODMD errors (solid) are averaged over a total of 20 trials, each involving a Gaussian noise realization and a set of $I-1$ observables randomly selected within a pool of candidate operators. QMEGS errors (dashed) are calculated using the hyperparameter $(\alpha_{\rm QMEGS}, q_{\rm QMEGS}) = (5, 0.05)$. We search for a total of 20 peaks in QMEGS, where the four lowest-lying eigenstates have the 1st, 7th, 10th, and 5th highest initial state overlap respectively.}
    \label{fig: QMEGS total time}
\end{figure}

Fixing the pair $(\alpha_{\rm QMEGS}^*,q_{\rm QMEGS}^*)$, we also examine the energy error with respect to the total runtime defined as
\begin{align}
    t_{\rm tot}^{\rm QMEGS}=\sum_{k=0}^{K_{\rm QMEGS}-1} |t_k^{\rm QMEGS}|, \qquad 
    t_{\rm tot}^{\rm MODMD}= \sum_{k=0}^{K+d} |t_k^{\rm MODMD}| = \sum_{k=0}^{K+d}\frac{k\Dt}{\epsilon^2}=\frac{(K+d)(K+d+1)}{2\epsilon^2},
\end{align}
as shown in \cref{fig: QMEGS total time}.

\section{Eigenstate Convergence}
\label{app:eigenstate_convg}
As discussed in \cref{subsec:main algorithm}, the MODMD algorithm can approximate the low-lying eigenstates of the Hamiltonian in addition to the eigenenergies via \cref{eq:eigstate_2}. Here we demonstrate the convergence of the TFIM and LiH eigenstates with respect to the number of time steps $K$. We note that both the noise and SVD threshold in the TFIM simulation were set differently (smaller by an order of magnitude) compared to those in the main text.
\begin{figure*}[h]
    \centering
    \begin{subfigure}[t]{0.49\textwidth}
        \centering
        \includegraphics[scale=0.55]{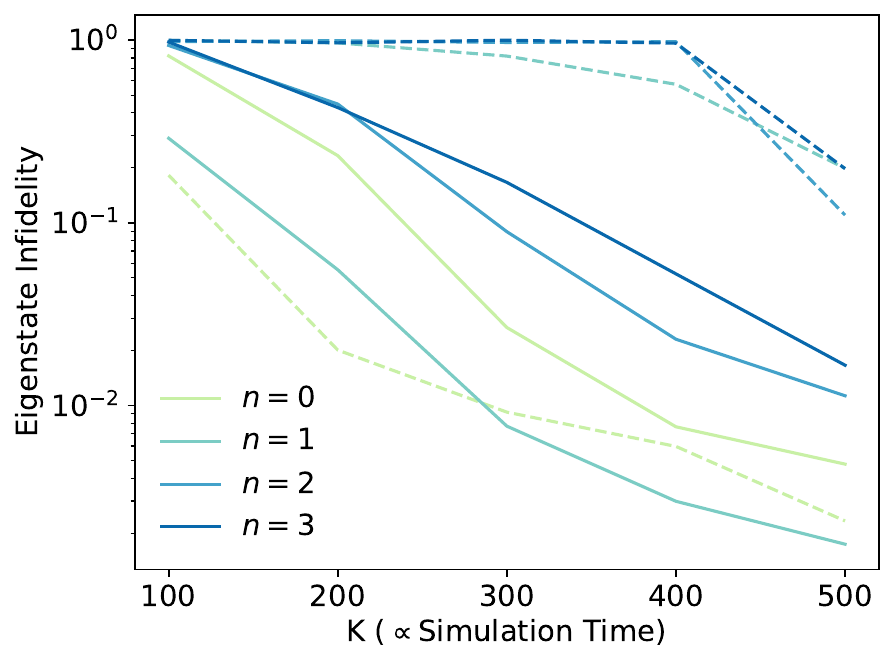}
        \caption{Eigenstate infidelity}
        \label{fig: TFIM Eigenstate Fidelity}
    \end{subfigure}%
    \hfill
    ~
    \begin{subfigure}[t]{0.49\textwidth}
        \centering
        \includegraphics[scale=0.55]{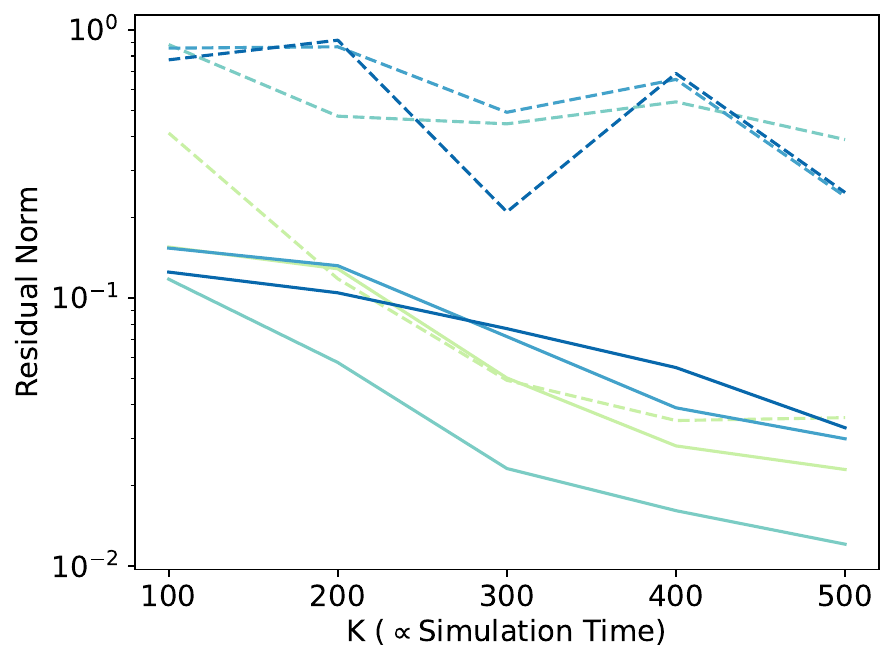}
        \caption{Eigenstate residual norm}
        \label{fig: TFIM Eigenstate Residual}
    \end{subfigure}
    \caption{Convergence of eigenstates for the transverse-field Ising model. To obtain eigenstate estimates $\tilde{\psi}_n$, we fix the (M)ODMD parameters $\frac{K}{d}=\frac{5}{2}$ and $\Tilde{\delta}=10^{-3}$ for constructing and thresholding the pair of data matrices $\mathbf{X},\mathbf{X}' \in \mathbb{R}^{d I \times (K+1)}$. Gaussian $\mathcal N(0,\varepsilon_{\rm noise}^2)$ noise with $\varepsilon_{\rm noise} = 10^{-4}$ is added independently to the real or/and imaginary parts of the matrix elements. The reference state $\ket{\phi_0}$ is an equal superposition of six computational basis states and we use a time step of $\Delta t \approx 0.08$. The state infidelity $1-|\langle\tilde\psi_n|\psi_n\rangle|^2$ and normalized residual norm $\|H\ket{\tilde\psi_n}-\tilde {E_n}\ket{\tilde\psi_n}\|_2/\|H\|_2$ are shown with respect to $K$ proportional to the non-dimensional maximal simulation time. Convergence results are averaged over 20 trials of Gaussian noise realizations and (for MODMD) random one-local Pauli observables. The solid lines depict the results for the multi-observable (MODMD) algorithm with $I = 7$ and the dashed lines depict the results for the single-observable (ODMD) algorithm. \textbf{Left}. Approximate eigenstate fidelity $1-|\langle\tilde\psi_n|\psi_n\rangle|^2$. \textbf{Right}. Normalized residual norm $\|H\ket{\tilde\psi_n}-\tilde E_n\ket{\tilde\psi_n}\|_2/\|H\|_2$.}
    \label{fig: TFIM Eigenstate Convergence}
\end{figure*}


\begin{figure*}[h]
    \centering
    \begin{subfigure}[t]{0.49\textwidth}
        \centering
        \includegraphics[scale=0.55]{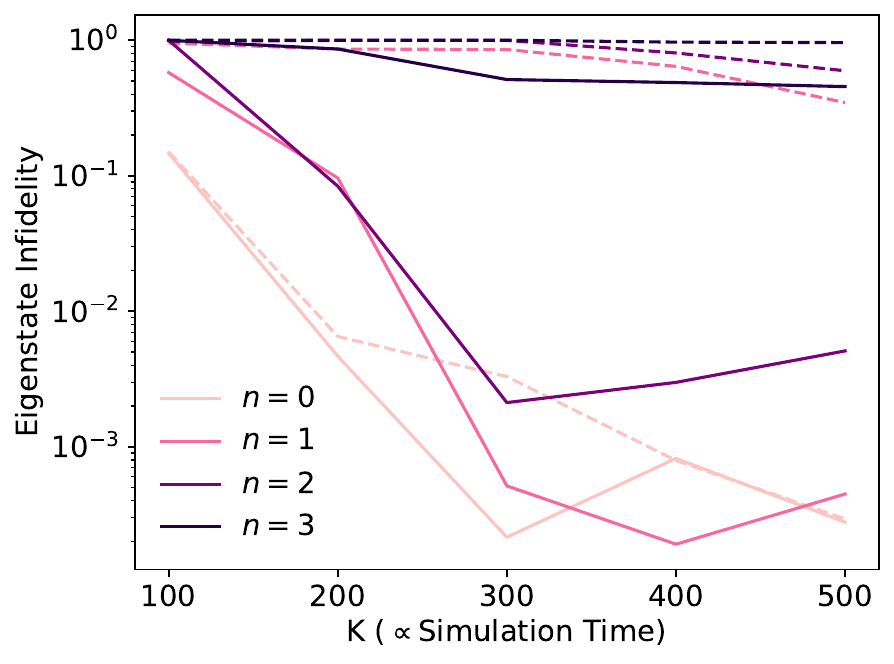}
        \caption{Eigenstate infidelity}
        \label{fig: LiH Eigenstate Fidelity}
    \end{subfigure}%
    \hfill
    ~
    \begin{subfigure}[t]{0.49\textwidth}
        \centering
        \includegraphics[scale=0.55]{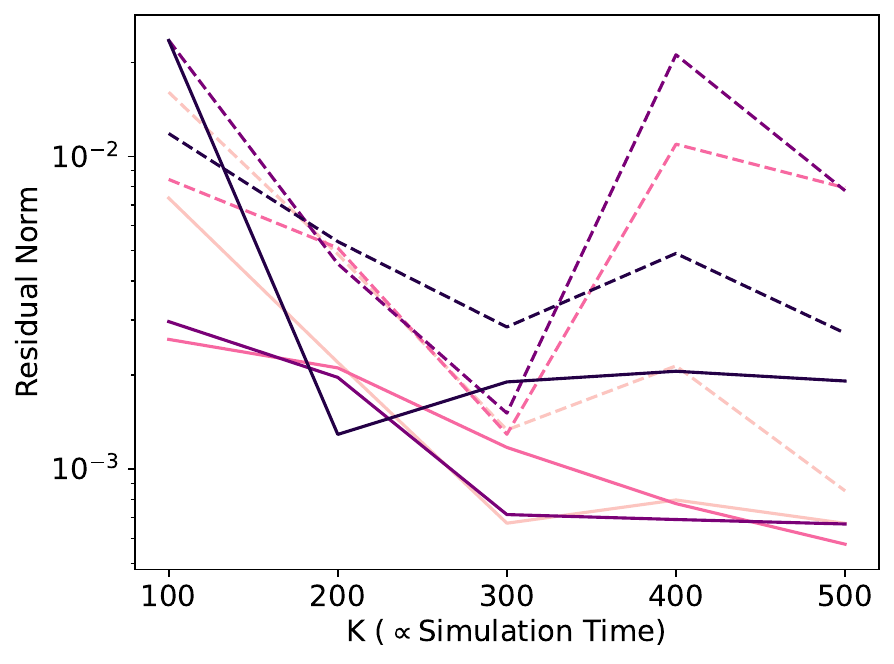}
        \caption{Eigenstate residual norm}
        \label{fig: LiH Eigenstate Residual}
    \end{subfigure}
    \caption{Convergence of eigenstates for the lithium hydride (LiH) molecule. To obtain eigenstate estimates $\tilde{\psi}_n$, we fix the (M)ODMD parameters $\frac{K}{d}=\frac{5}{2}$ and $\Tilde{\delta}=10^{-2}$ for constructing and thresholding the pair of data matrices $\mathbf{X},\mathbf{X}' \in \mathbb{R}^{d I \times (K+1)}$. Gaussian $\mathcal N(0,\varepsilon_{\rm noise}^2)$ noise with $\varepsilon_{\rm noise} = 10^{-3}$ is added independently to the real or/and imaginary parts of the matrix elements. The reference state $\ket{\phi_0}$ is a superposition of six computational basis states and we employ a time step of $\Delta t \approx 0.39$. The state infidelity, $1-|\langle\tilde\psi_n|\psi_n\rangle|^2$, and normalized residual norm, $\|H\ket{\tilde\psi_n}-\tilde {E_n}\ket{\tilde\psi_n}\|_2/\|H\|_2$, are shown with respect to $K$ proportional to the non-dimensional maximal simulation time. Convergence results are averaged over 20 trials of Gaussian noise realizations. The solid lines depict the results for the multi-observable (MODMD) algorithm with $I = 7$ and the dashed lines depict the results for the single-observable (ODMD) algorithm. \textbf{Left}. Approximate eigenstate fidelity $1-|\langle\tilde\psi_n|\psi_n\rangle|^2$. \textbf{Right}. Normalized residual norm $\|H\ket{\tilde\psi_n}-\tilde E_n\ket{\tilde\psi_n}\|_2/\|H\|_2$.}
    \label{fig: LiH Eigenstate Convergence}
\end{figure*}

Here, we use two metrics to evaluate the quality of our eigenstate estimates $\ket{\tilde\psi_n}$, which are obtained via MODMD as linear combinations of time-evolved states acted on by selected operators. First, we examine the infidelity $1-|\langle\tilde\psi_n|\psi_n\rangle|^2$ determined by the squared overlap of the approximate and exact eigenstates. Clearly, the (in)fidelity provides a direct measure of how well the approximate eigenstate represents the true wavefunction. As a second alternative, we consider the normalized residual norm, $\|H\ket{\tilde\psi_n}-\tilde E_n\ket{\tilde\psi_n}\|_2/\|H\|_2$, which is indicative of the energy variance of the approximate eigenstate and is commonly monitored in numerical linear algebra tasks to track eigenspace convergence. 

In both metrics, MODMD captures most of the first three excited states with accuracies close to that of the ground state -- a remarkable behavior not observed in ODMD. This uniformity in performance across the low-lying eigenstates underscores the effectiveness of our multi-observable approach in stabilizing recovery of essential spectral information.

\clearpage
\twocolumngrid

\bibliographystyle{apsrev4-2}
\bibliography{mainbib}
\end{document}